%% file: Iax.tex
\documentclass{aa}
\usepackage{natbib}
\usepackage{multicol}
\usepackage{siunitx}
\bibpunct{(}{)}{;}{a}{}{,}
\usepackage{graphicx}
\usepackage{grffile}
\usepackage[varg]{txfonts}
\usepackage{pdfpages}
\usepackage{booktabs}
\usepackage{placeins}
\usepackage{tabularx}
\usepackage{xspace}
\usepackage{xcolor}
\usepackage{rotating}
\usepackage{stfloats}
\usepackage{float}
\usepackage{hyperref}

\newcommand{\msun}{\ensuremath{\mathrm{M}_\odot} }
\newcommand{\mni}{\ensuremath{M(^{56}\mathrm{Ni})} }
\newcommand{\nifs}{\ensuremath{^{56}\mathrm{Ni}} }
\newcommand{\mch}{\ensuremath{M_\mathrm{Ch}} }

\input{./journals.tex}

\titlerunning{Type Iax Supernovae}
\authorrunning{F. Lach et al.}

\begin{document}

\title{Type Iax supernovae from deflagrations in Chandrasekhar mass
  white dwarfs}

\author{F.~Lach\inst{1,2}\thanks{E-mail: florian.lach@h-its.org}
\and F.~P.~Callan\inst{3}
\and D.~Bubeck
\and F.~K.~Roepke\inst{1,2}
\and S.~A.~Sim\inst{3}
\and M.~Schrauth\inst{4}
\and S.~T.~Ohlmann\inst{5}
\and M.~Kromer}

\institute{ Heidelberger Institut f\"{u}r Theoretische Studien, 
            Schloss-Wolfsbrunnenweg 35, D-69118 Heidelberg, Germany
  \and Zentrum f{\"u}r Astronomie der Universit{\"a}t Heidelberg, 
       Institut f{\"u}r Theoretische Astrophysik, Philosophenweg 12, 
       D-69120 Heidelberg, Germany
  \and School of Mathematics and Physics, Queen’s University
       Belfast, Belfast BT7 1NN, UK 
  \and Institute of Theoretical Physics and Astrophysics, University of
  Würzburg, D-97074 Würzburg, Germany
  \and Max Planck Computing and Data Facility, Gießenbachstr. 2, 85748
  Garching, Germany
}

\date{22 November 2021/ \today}

\abstract{}{}{}{}{}
 \abstract
 {Due to the ever increasing number of observations during the past
   decades, Type Ia supernovae are nowadays regarded as a heterogeneous
   class of optical transients consisting of several subtypes. One of
   the largest of these subclasses is the class of Type Iax supernovae.
   They have been suggested to originate from pure deflagrations in
 carbon-oxygen Chandrasekhar mass white dwarfs because the outcome of
 this explosion scenario is in general agreement with their subluminous
 nature.} {Although a few deflagration studies have already been carried
   out, the full diversity of the class has not been captured yet. This, in
   particular, holds for the faint end of the subclass. We therefore
   present a parameter study of single-spot ignited deflagrations in
   Chandrasekhar mass white dwarfs varying the location of the ignition
   spark, the central density, the metallicity, and the composition of
   the white dwarf. We also explore a rigidly rotating progenitor to
 investigate whether the effect of rotation can spawn additional
 trends.} {We carried out three-dimensional hydrodynamic simulations
   employing the \textsc{leafs} code. Subsequently, detailed
   nucleosynthesis results were obtained with the nuclear network code
   \textsc{yann}. In order to compare our results to observations, we
   calculated synthetic spectra and light curves with the \textsc{artis}
 code.} {The new set of models extends the range in brightness covered
   by previous studies to the lower end. Our single-spot ignited
   explosions produce \nifs masses from $5.8 \times 10^{-3}$ to $9.2
   \times 10^{-2}\,\mathrm{M}_\odot$. In spite of the wide exploration of the
   parameter space, the main characteristics of the models are primarily
   driven by the mass of \nifs and form a one-dimensional sequence.
   Secondary parameters seem to have too little impact to explain the
   observed trend in the faint part of the Type Iax supernova class. We
   report kick velocities of the gravitationally bound explosion
   remnants from $6.9$ to $369.8\,\si{km.s^{-1}}$. The magnitude as well
 as the direction of the natal kick is found to depend on the strength
 of the deflagration.} {This work corroborates the results of previous
   studies of deflagrations in Chandrasekhar mass white dwarfs. The wide
   exploration of the parameter space in initial conditions and
   viewing angle effects in the radiative transfer lead to a significant
   spread in the synthetic observables. The trends in observational
   properties toward the faint end of the class are, however, not
   reproduced. This motivates a quantification of the systematic
   uncertainties in the modeling procedure and the influence of the
   $^{56}$Ni-rich bound remnant to get to the bottom of these discrepancies.
   Moreover, while the pure deflagration scenario remains a
     favorable explanation for bright and intermediate luminosity Type
     Iax supernovae, our results suggest that other mechanisms
     also contribute to this class of events.}

\maketitle

%--------------------------------------------------------------------
%-------------          INTRODUCTION    -----------------------------
%--------------------------------------------------------------------

\section{Introduction}
\label{sec:introduction}

Type Ia supernovae (SNe~Ia) have, for a long time, been thought to form
a rather homogeneous class of cosmic explosions. In fact, the majority
of events show similar photometric and spectroscopic behavior. Their
light curves follow an empirical width-luminosity relation (WLR,
Phillips relation, \citealp{phillips1993a}), associating broad shapes
with bright events. This property makes these so-called Branch-normal
supernovae \citep{branch1993a, branch2006a} standardizable candles,
widely employed for distance measurements
\citep{riess1996a,schmidt1998a,perlmutter1999a,jha2007a}.  However, the
increasing number of observations over the past decades led to the
detection of outliers and eventually called for the introduction of
several subclasses of SNe~Ia. These subtypes are believed to share the
thermonuclear origin with normal SNe~Ia, that is, the disruption of a
carbon-oxygen (CO, in some cases also oxygen-neon (ONe),
\citealp{marquardt2015a,kashyap2018a}) white dwarf (WD) due to explosive
nuclear burning, but they exhibit clear differences in some
characterizing properties (see \citealp{taubenberger2017a} for a
review).

The largest of these subclasses of SNe~Ia is the subluminous supernovae
of Type Iax (SNe~Iax), with more than ${\sim}\,50$ members observed.
According to \citet{foley2013b}, SNe~Iax might account for about 31\% of
all SNe~Ia while \citet{li2011a} and \citet{white2015a} estimated their
rate to be $\sim 5\%$.  \citep{jha2017type}. The typical object for this
class is SN~2002cx \citep{li2003a}, and, therefore, at least the
brighter objects are also called 02cx-like events.  \citet{foley2013b}
named these transients SNe~Iax and defined the key observational
features of this class as follows: (i) no evidence of hydrogen in any
spectrum, (ii) lower maximum-brightness photospheric velocity than any
normal SN~Ia, (iii) low absolute magnitude near maximum light in
relation to the width of its light curve (i.e., they fall below the
WLR), and (iv) spectra similar to SN~2002cx at comparable epochs.

While the first criterion represents the usual distinction between
SNe~Ia and supernovae of Type II, the second and third requirements
suggest the production of only a small amount of energy. The optical
emission from SNe~Ia is powered by the radioactive decay of \nifs to
$^{56}$Co in the early phase, and, therefore, their brightness is
directly linked to the ejected mass of \nifs. Peak absolute brightnesses
derived from observations range between $-14\ge M_V \ge -19$, inferred
$^{56}$Ni masses lie between $0.003$ and $0.27\,\mathrm{M}_\odot$
\citep{foley2013b,jha2017type}, and ejected masses $M_\mathrm{ej}$
between $0.15\,\mathrm{M}_\odot$ for the faintest members of this class,
such as SN~2008ha \citep{foley2009a}, and up to approximately \mch for
the brightest events (e.g., SN~2012Z, \citealp{stritzinger2015a}).
Normal SNe~Ia, in contrast, eject between $0.3$ and $0.9\,$\msun of
\nifs \citep{hillebrandt2013a}. In addition, ejecta velocities are
significantly lower than those of normal SNe~Ia and usually fall below
$8000\,\si{km.s^{-1}}$ down to $\lesssim 2000\,\si{km.s^{-1}}$ for the
faintest SNe~Iax \citep{foley2009a}. However, B-band light curve decline
rates span a similar range compared to normal events, but they cannot be
connected to the corresponding brightness via the WLR, and, thus,
SNe~Iax are not employed as distance indicators. The shape of their
light curves in the optical regime resembles normal SNe~Ia while the
prominent second peak in the near-infrared (NIR) bands (IJHK)
\citep{hamuy1996a,kasen2006b} is absent in SNe~Iax
\citep{foley2013a,magee2016a,jha2017type}. Also the redder bands tend to
decline slower compared to lower wavelengths \citep{tomasella2016a} and
show a larger variation in decline rates compared to normal SNe~Ia.
There is also a loose corellation between photosperic velocities and
peak magnitudes \citep{mcclelland2010a,foley2013a}, that is, brighter
events also show higher expansion velocities, challenged only by some
outliers such as SN~2009ku \citep{narayan2011a} and SN~2014ck
\citep{tomasella2016a}.  In addition, \citet{magee2016a} study the rise
times of SNe~Iax ($t_\mathrm{rise} = 9 - 28\,\si{d}$) and find an
increasing trend with peak r-band magnitude, but again with a clear
outlier, namely SN~2009eoi. 

The near maximum spectrum of SN~2002cx, and, by definition those of
other SNe~Iax, is characterized by prominent Fe III lines and relatively
weak signs of IMEs, for example, Si II and S II \citep{branch2004a}.
Surprisingly, this is quite similar to the overluminous and rapidly
expanding SN~1991T-like events. The photospheric velocity, however, is
very low ($\sim 7000\,\si{km.s^{-1}}$) making it easier to identify
individual species since absorption and emission features are broadened
less. About two weeks after maximum the Si~II feature has disappeared
and the spectrum is dominated by singly ionized iron group elements
(IGEs). Until $56\,\si{d}$ the spectrum resembles those of normal SNe~Ia
at slightly later times and has not turned nebular yet
\citep{branch2004a}. The late-time spectra of SN~2002cx (227 and
$277\,\si{d}$) were examined by \citet{jha2006b}.  They found that the
spectra did not change significantly after $56\,\si{d}$ post maximum and
deviate significantly from those of normal SNe~Ia. Instead of being
dominated by forbidden lines of Fe and Co typical for nebular spectra
they exhibit P Cygni features of Fe II but also IMEs and potentially
signs of oxygen. \citet{foley2016a} also state that late-time spectra of
SNe~Iax only show minor changes over time and that differences among the
subclass originate primarily from different expansion velocities.
Finally, they find clear signs of Ni II pointing toward a significant
amount of stable Ni in the ejecta.

Several different progenitor systems and explosion mechanisms have been
proposed for thermonuclear supernovae during the past decades. The
explosion of the WD is triggered by the interaction with a hydrogen or
helium-rich companion (single-degenerate scenario,
\citealp{whelan1973a}) or another WD (double-degenerate scenario,
\citealp{iben1984a}). Furthermore, the explosion is caused by a
thermonuclear runaway in degenerate matter resulting in either a
subsonic deflagration or a supersonic detonation. The propagation of a
deflagration is mediated by heat conduction while a detonation proceeds
as a shock wave. Moreover, a spontaneous transition from a deflagration
to a detonation has been proposed by
\citet{blinnikov1986a,khokhlov1989a} and was studied by
\citet{khokhlov1991a,gamezo2005a,roepke2007d}, and
\citet{seitenzahl2013a}, for instance. Comprehensive summaries of the
whole ``zoo'' of possible SN~Ia explosion scenarios can be found in
reviews, for example, \citet{wang2012b}, \citet{hillebrandt2013a},
\citet{livio2018a}, \citet{roepke2018a}, and \citet{soker2019a}.

One promising scenario for explaining SNe~Iax is a \mch pure
deflagration as proposed by \citet{branch2004a,branch2006a} for
SN~2002cx and \citet{chornock2006a} and \citet{phillips2007a} for
SN~2005hk. Recently, deflagrations in hybrid carbon-oxygen-neon (CONe)
instead of pure CO WDs have also been considered as progenitors for
SNe~Iax \citep{denissenkov2015a,kromer2015a,bravo2016a}. This scenario
naturally accounts for many characteristics of SNe~Iax. Most notably,
the simulations of these explosions do not eject enough \nifs to reach
the absolute magnitudes of normal SNe~Ia, and, depending on the ignition
configuration, do not disrupt the whole WD, but leave behind a bound
remnant \citep{jordan2012b,kromer2013a,fink2014a}.  \citet{foley2014b}
speculate about the existence of such a bound remnant in SN~2008ha and
suggest that it emits a wind preventing the spectra from transitioning
to the nebular phase which is one of the most characterisitc properties
of SN~Iax spectra. Moreover, evidence for a bound remnant was also found
in the slowly declining late-time light curve of SN~2014dt by
\citet{kawabata2018a} and very recently in SN~2019muj
\citep{kawabata2021a}. The latter also employ their analytical,
phenomenological light curve model to other SNe Iax and conclude that
their late-time light curves are well matched assuming radiating burning
products at low velocities. The radiation of gravitationally bound or
slowly expanding \nifs is also suspected to contribute to the light
curves and spectra by \citet{foley2016a, magee2016a}, and
\citet{shen2017a}.  

These deflagrations lead to photospheric velocities in the observed
range below those of normal SNe~Ia. The detection of IGEs and IMEs in
their spectra at all times and the lack of a second maximum in the NIR
ligth curves (see \citealp{kasen2006b}) suggests a rather mixed ejecta
structure which is a natural outcome of a turbulent deflagration.
However, the assumption of completely mixed ejecta in some bright
SNe~Iax has been challenged by \citet{stritzinger2015a,barna2017a},
and \citet{barna2018a}.  Therefore, a pulsational delayed detonation
\citep{hoeflich1996a} scenario was proposed for the most energetic
SNe~Iax \citep{stritzinger2015a}.  However, the need for
stratification in the moderately bright SN~2019muj was not found to be
very strong for all elements except carbon in the outer layers
\citep{barna2021a}. It should be noted that their approach is only
sensitive to velocities between $3600$ and
${\sim}\,6500\,\si{km.s^{-1}}$. In this region, their stratified
template model does not differ significantly from well mixed
deflagration ejecta.

The single degenerate scenario suffers the problem of stripped hydrogen
or helium from the companion star: It is expected that the explosion of
the WD will remove material from its donor which should be seen in the
late-time spectra \citep{pakmor2008a,lundqvist2013a,bauer2019a}. Since
SNe~Iax are less energetic than normal SNe~Ia, \citet{liu2013c} and
\citet{zeng2020a} argue that the stripped hydrogen or helium might stay
below the detection limit. \citet{magee2019a} also look for helium
features in SNe~Iax and find that their NIR spectra are compatible with
only very small amounts of helium and that it is easier to detect in
fainter models. Another hint toward the \mch scenario for SNe~Iax is the
possible detection of a bright helium-star at the location of SN~2012Z
\citep{mccully2014a} because accretion of helium is a plausible way for
a WD to reach $M_\mathrm{Ch}$.  \citet{mccully2021a} present very late
time observations of SN~2012Z up to $1425\,\si{d}$ after maximum light.
They find that the event is still about a factor of two brighter than
pre-explosion and attribute this either to the shock-heated companion, a
bound remnant (see above), or radioactive decay of long-lived isotopes
other than $^{56,57}$Co. In addition, the recently detected
hypervelocity WDs with an unusual composition \citep{raddi2019a} have
also been associated with the bound remnant WD cores found in
simulations of \mch deflagrations. These characteristics make \mch
deflagrations a promising scenario for SNe~Iax and motivate a deeper
investigation of this model.  Moreover, their occurrence in star-forming
regions of late type galaxies are in line with a WD accreting helium
from its donor star \citep{lyman2018a}.

SNe~Ia contribute substantially to cosmic nucleosynthesis
\citep{nomoto2013a,thielemann2018a} and it is widely accepted that they
are required for the production of the neutron-rich element manganese
alongside core-collapse supernovae
\citep{seitenzahl2013a,kobayashi2006a,kobayashi2020a,lach2020a}.
Explosions of \mch WDs are able to produce supersolar amounts of Mn
relative to Fe in their innermost, that is, densest, parts in normal
freeze-out from nuclear statistical equilibrium (NSE, see
\citealp{woosley1973a,hix1999a,seitenzahl2009a,bravo2019a}) since the
rate of electron captures increases with density
\citep{chamulak2008a,piro2008c,brachwitz2000a}.  Therefore, if SNe~Iax
originate from deflagrations in \mch WDs they are promising candidates
for the enrichment of the Universe with Mn.  Recently, double-detonation
models for SNe~Ia came into the focus as a production site of Mn, but
they can probably not completely replace explosions of \mch WDs
\citep{lach2020a,gronow2021b}.

Simulations of deflagrations in \mch CO WDs with a clear relation to SNe
Iax have been carried out by \citet{jordan2012b}, \citet{long2014a},
\citet{kromer2013a}, \citet{fink2014a}, \citet{kromer2015a}, and
\citet{leung2020a}. We summarize the findings and shortcomings of this
previous work in more detail in Sect.~\ref{sec:previous}.

As an alternative scenario, SN~2008ha has also been speculated to be a
core-collapse SN \citep{valenti2009a} or even a failed detonation of an
ONe WD merging with a CO secondary \citep{kashyap2018a}.
\citet{fernandez2013a} also bring a detonation ignited in a WD-neutron
star merger event into play for SNe~Iax. While the historic SN~Iax
remnant SN~1181 is believed to originate from an ONe-CO WD merger
\citep{oskinova2020a,ritter2021a} the SN remnant Sgr A East has recently
been associated with a failed deflagration producing a SN~Iax
\citep{zhou2021a}. Hence, the question whether it is possible to explain
the entire class of SNe~Iax in the framework of the pure deflagration in
a \mch WD model remains. 

To address this question, we present an extensive set of
three-dimensional (3D) full-star simulations of deflagrations in \mch CO
WDs. Since the most realistic ignition configuration consists of only
one spot \citep{kuhlen2006a,zingale2009a,nonaka2012a} we restrict our
suite of simulations to single-spot ignition. The location of this
ignition spark and the central density are systematically varied to
investigate the dependence on these parameters. With this study, we aim
to extend the set of deflagration models toward the faintest events of
the SNe~Iax class and systematically determine expected properties of
bound remnants.  Furthermore, we aim to explore whether the restriction
to a single ignition spark but variation of other parameters can change
the general characteristics of pure deflagration models.

The paper is structured as follows: We give a summary of previous
deflagration studies in Sect.~\ref{sec:previous} followed by a short
overview of the numerical methods used to simulate the explosion in
Sect.~\ref{sec:numerics}. Subsequently, we describe our initial models
in Sect.~\ref{sec:setup} and discuss the results of the hydrodynamic
simulations in Sect.~\ref{sec:results}. The outcome of the radiative
transfer (RT) calculations is compared to observations in
Sect.~\ref{sec:rt}. Finally, we wrap up our findings and conclusions in
Sect.~\ref{sec:conclusion}.

%-----------------------------------------------------------------------
%----------------------  Previous Work  --------------------------------
%-----------------------------------------------------------------------

\section{Previous work}
\label{sec:previous}

In the following we summarize the results of various works concerning
the modeling of deflagrations in \mch WDs in connection to SNe~Iax.
These results help to understand the open questions and shortcomings of
the currently available simulations and also guide the choice of
parameters for the study carried out in our work.

\subsection{\citet{jordan2012b}}
\label{subsec:jordan12}

\citet{jordan2012b}, hereafter J12, present a set of models of pure
deflagrations in CO \mch WDs. Since their models fail to ignite a
delayed detonation and do not unbind the WD, they call these explosions
failed-detonation SNe. They carry out two-dimensional (2D) as well as 3D
hydrodynamic simulations, but do neither present detailed
nucleosynthesis yields nor radiative transport calculations. Their \mch
WD (1.365$\,\mathrm{M}_\odot$, $\rho_c=2.2\times 10^9\,\si{g.cm^{-3}}$)
is ignited in 63 spots of $16\,\si{km}$ radius contained inside a sphere
of $128\,\si{km}$ radius. This sphere is located slightly off-center at
distances of 48, 38, 28 and $18\,\si{km}$ from the center of the WD. The
2D run, however, uses only four ignition sparks inside a circle with a
radius of $64\,\si{km}$ displaced by $70\,\si{km}$ from the center of
the WD. 

As mentioned above, the explosion in the WD leaves behind a bound
remnant in all of their simulations although 89 to 167\% of the initial
binding energy $E_\mathrm{bind}$ are released during the deflagration
phase. This shows that the released energy is not necessarily
distributed homogeneously and that multidimensional effects need to be
taken into account. Their explosions eject between 0.23 and
$1.09\,\mathrm{M}_\odot$ of material including 0.07 to
$0.34\,\mathrm{M}_\odot$ of IGEs.  Unfortunately, values for the
production of $^{56}$Ni are not given. The ejecta velocities lie below
$10\,000\,\si{km.s^{-1}}$ and they find an asymmetric ejecta structure
characterized by a surplus of burning products at the ignition side and
CO fuel in the opposite direction.  These features lead them to the
conclusion that their models might be candidates for SNe~Iax, at least
for the bright SN~2002cx-like  members of this subclass. However, a
comparison of synthetic observables to observations is not part of this
study.  

Another interesting aspect investigated by J12 is the kick velocity
$v_\mathrm{kick}$ of the bound remnant. Since the explosion proceeds
asymmetrically the bound core receives a kickback between 119 and
$549\,\si{km.s^{-1}}$ which might be enough to unbind the remnant from
the binary system. Together with its composition enriched by burning
products such a puffed-up WD is a possible explanation for peculiar
iron-rich WDs (see e.g., \citealp{provencal1998a,catalan2008a} and for
more recent works
\citealp{vennes2017a,raddi2018a,raddi2018b,raddi2019a,neunteufel2020a}).

\subsection{\citet{long2014a}}
\label{subsec:long13}

\citet{long2014a}, hereafter L14, present a follow-up study of the work
by J12. They utilize the same progenitor model, but also specify that
they employ zero metallicity, a C/O ratio of 1 (i.e., 50\% C and 50\%
O), and a constant temperature of $T=3 \times 10^7\,\si{K}$.  The only
free parameter is  the ignition geometry. It consists again of bubbles
of $16\,\si{km}$ radius. These are confined to spheres of radius 128,
256 and $384\,\si{km}$ located at the center of the WD.  In addition,
the total number of ignition sparks is varied between 63 and 3500.  They
carry out six 3D full-star simulations including a postprocessing step
and one-dimensional (1D) RT calculations yielding synthetic light
curves.

Although the total energy $E_\mathrm{tot}$ (sum of gravitational energy
$E_\mathrm{grav}$, kinetic energy $E_\mathrm{kin}$ and internal energy
$E_\mathrm{I}$) is positive in all their models we cannot judge whether
the whole WD is unbound since it is not specified in the paper. They
arrive at \nifs masses \mni between 0.135 and $0.288\,\mathrm{M}_\odot$,
nuclear energies $E_\mathrm{nuc}$ between 6.78 and $9.60 \times
10^{50}\,\si{erg}$, and kinetic energies of the ejecta
$E_\mathrm{kin,ej}$ between 2.44 and $5.01 \times 10^{50}\,\si{erg}$.
The main finding is that the initial spatial density of the ignition
sparks as well as the outer radius of the confining sphere determine the
outcome of the simulation. As a first effect, a dense distribution leads
to a high burning rate in the early phase of the flame propagation.
Subsequently, the bubbles merge very rapidly, and, thus, the flame
surface decreases leading to a reduction in the burning rate.  Second,
the gravitational acceleration, and, therefore, the buoyant force
increases with radius.  This causes a rapid increase in the energy
production for large ignition radii but also an earlier quenching of the
flame as the deflagration reaches the outer part of the WD. There is
most probably an optimal choice for the number of ignition sparks per
volume and the maximum ignition radius in terms of \mni. The approximate
number and spatial density of ignition kernels has already been
estimated by \citet{roepke2006a}. In the study of L14 a modest number of
bubbles (128) confined to a $128\,\si{km}$ sphere yields the highest
mass of \nifs while very many sparks (3500) inside a $384\,\si{km}$
sphere set the lower limit. 

Finally, an examination of the synthetic light curves shows that models
ignited with ${\sim}\,100$ kernels show more similarities to SNe~Iax
than vigorously ignited models with ${\sim}\,1000$ sparks. However, also
in this study only the luminosities of bright SNe~Iax are reached. 

\subsection{\citet{fink2014a}}
\label{subsec:fink14}

The work of \citet{fink2014a} (hereafter, F14) was developed
simultaneously to the study of L14. They conduct 14 3D full-star
simulations of pure deflagrations in \mch CO WDs studying mainly the
impact of the ignition geometry. In detail, they vary the number $N$ of
ignition kernels from 1 to 1600, and, for two models ($N$=300,1600),
they also set up a denser distribution of the bubbles. Their initial WD
has a central density of $2.9\times10^9\,\si{g.cm^{-3}}$, a constant
temperature of $5\times10^5\,\si{K}$, and mass fractions of
$X(^{12}\mathrm{C})=0.475$, $X(^{16}\mathrm{O})=0.50$ and
$X(^{22}\mathrm{Ne})=0.025$ to account for solar metallicity. Moreover,
one model ($N$=100) has also been calculated at central densities of
$1.0\times10^9\,\si{g.cm^{-3}}$ and $5.5\times10^9\,\si{g.cm^{-3}}$.

Concerning the nuclear energy release and the production of \nifs F14
arrive at the same conclusion as L14 stating that models ignited in
many, densely located sparks lead to less powerful explosions compared
to moderately ignited WDs. The highest amount of \nifs is found in the
model with 150 initial kernels. Moreover, the energy production in the
high-density model does not differ significantly from their standard
model while the low-density simulation leads to noticeable less released
nuclear energy. However, the high-density model produces no more \nifs
than the low-density run because of the more neutron-rich
nucleosynthesis at high densities favoring the production of stable IGEs
instead. 

In contrast to L14, F14 also explicitly report the existence of bound
remnants for models ignited in less than 100 spots. These remnants are
enriched with burning products and reach masses up to
$1.32\,\mathrm{M}_\odot$.  In addition, they also calculate their
respective kick velocity, but arrive at values an order of magnitude
below those presented by J12 ranging between 4.4 and
$36\,\si{km.s^{-1}}$. They speculate that their use of an approximate
monopole solver for the gravitational force might account for the
differences. 

F14 also present synthetic light curves and spectra resulting from 3D RT
simulations concluding that models with less than 20 ignition sparks are
compatible with SNe~Iax.  However, even their faintest explosion ($N=1$)
does not reach the faintest members of the SN~Iax class, SN~2008ha and
SN~2019gsc.  The predicted spectra look very similar without a
significant viewing angle dependency.  \citet{kromer2013a} (hereafter,
K13) present an in-depth analysis of the explosion ignited in 5 bubbles
(Model N5). They compare N5 to SN~2005hk and state that peak luminosity,
the colors at maximum and the decline in UBV bands coincide very well.
Also the presence of IGEs at all times and the lack of a secondary
maximum matches the characteristics of SNe~Iax. However, the decline in
red bands (RIJH) is significantly too fast and also the B-band rise time
is too short to conform with SN~2005hk. One way to achieve a slower
decline and keeping the peak luminosity fixed is to increase the ejected
mass at constant \mni. In addition, the influence of the bound remnant
on the observables was not investigated yet. 

\subsection{\citet{kromer2015a}}
\label{subsec:kromer15}

In order to reach the faint end of the SN~Iax subclass,
\citet{kromer2015a} (hereafter, K15) carry out a deflagration simulation
inside a \mch hybrid CONe WD. These progenitors had been proposed in the
work of \citet{denissenkov2013c,denissenkov2015a}. The initial
conditions, that is,  density, temperature and ignition condition, were
chosen to be similar to N5 of F14.  The WD consists of a
$0.2\,\mathrm{M}_\odot$ CO core and a $1.1\,\mathrm{M}_\odot$ ONe
mantle.  The burning is assumed to quench as soon as the flame reaches
the ONe layer which leads to a very faint explosion ejecting only
$0.014\,\mathrm{M}_\odot$ in total and $3.4 \times
10^{-3}\,\mathrm{M}_\odot$ of $^{56}$Ni.  This provides a good fit to
the peak luminosity of one of the faintest SN~Iax, SN~2008ha. But still
the decline (red bands) and rise of the light curve is too fast.
Moreover, some line features around maximum cannot be reproduced. These
deficiencies also hint toward too little ejected mass. K15 point out
that the \nifs enriched bound remnant might also have an influence on
the light curve at late and possibly even early times depending on the
structure of the remnant. 

\citet{bravo2016a} also investigated the hybrid CONe WD scenario with 1D
simulations and varying CO core masses. Their study includes pure
detonations and delayed detonation models. They conclude that only
delayed detonations resemble bright SNe~Iax not reaching intermediate
and low-luminosity events. 

\subsection{\citet{leung2020a}}
\label{subsec:leung20}

An extensive deflagration study with reference to SNe~Iax was presented
by \citet{leung2020a} (hereafter, L20). They start with \mch CO and CONe
WDs and vary the central density between $0.5 \times
10^9\,\si{g.cm^{-3}}$ and $9.0 \times 10^9\,\si{g.cm^{-3}}$.
Furthermore, they use two different ignition conditions, that is, the
c3-ignition by \citet{reinecke1999b} and a single bubble. Unfortunately,
they do not give details on the radius and the location of the ignition
spark. Their simulations are conducted in 2D only and the emphasis in
their work lies on the nucleosynthesis products and not on optical
observables. 

They report \nifs masses between 0.20 and $0.36\,\mathrm{M}_\odot$ and
total ejected masses between 0.92 and $1.36\,\mathrm{M}_\odot$. This
indicates that these models can only account for the brightest SNe~Iax,
such as SN~2012Z.

\subsection{Summary}
\label{subsec:summary}

In summary, all the deflagration studies above claim to produce models
which can account for SNe~Iax although not all of them carry out RT
simulations and compare to observations. In these cases, the claim
originates from the low explosion energies, low ejecta velocities and
low masses of \nifs compared to normal SNe~Ia. It is also well known
that pure deflagrations do not produce a secondary maximum in infrared
bands. Studies including synthetic observables, L14, K13, F14, and K15,
corroborate this claim. However, there are some discrepancies regarding
the width of the model light curves (see above). Therefore, to explain
the decline rate and rise times of SNe~Iax and also the diversity among
this subclass a much wider exploration of the parameter space is
necessary. 

Moreover, these modeled events cover a reasonable range in brightness,
that is, ejected mass of \nifs, but they do not account for the full
diversity of the class of SNe~Iax and show a bias toward bright events
while the fainter objects, for instance, SN~2008ha \citep{foley2009a},
SN~2010ae \citep{stritzinger2014a}, and SN~2019gsc
\citep{srivastav2020a, tomasella2020a} are not reached.  One of the
major shortcoming of the works mentioned above is their use of the
ignition configuration, that is, the location, shape, size and number of
the initial ignition kernels, as a free parameter to control the
strength of the deflagration. This is in strong contrast to the works of
\citet{zingale2011a} and \citet{nonaka2012a} which suggest the ignition
in one spark off-center.

%-------------------------------------------------------------------
%---------------    Numerics, Codes --------------------------------
%-------------------------------------------------------------------

\section{Numerical methods}
\label{sec:numerics}

For the hydrodynamic simulations we employ the \textsc{leafs} code which
has already been successfully used for a large number of explosion
simulations (e.g.,
\citealp{roepke2005b,roepke2007c,seitenzahl2013a,fink2014a,ohlmann2014a,marquardt2015a,fink2018a}).
It is based on the Prometheus code by \citet{fryxell1989a} that has been
extended and adapted by \citet{reinecke1999b} to simulate SNe~Ia. For
solving the reactive Euler equations it takes a finite volume approach
using the piecewise parabolic scheme by \citet{colella1984a}. In order
to treat flame fronts as discontinuities, the level-set technique
\citep{osher1988a} has been implemented by \citet{reinecke1999a}.
Nuclear burning and, therefore, the production of energy at the flame
front is taken care of by the appropriate conversion of 5 pseudospecies
representing carbon, oxygen, IMEs, IGEs and $\alpha$-particles (see
\citealp{ohlmann2014a}).  To follow the explosion until the expelled
material expands homologously, \citet{roepke2005c} implemented two
nested expanding grids. The inner grid tracks the flame while the outer
one follows the expansion of the star. The most recent developments are
the implementation of the Helmholtz equation of state (EoS,
\citealp{timmes1999a}) including Coulomb corrections and a fast Fourier
transform based gravity solver which solves the full Poisson equation
and replaces the monopole solver of previous simulations.

In order to capture the detailed nucleosynthesis, we employ the tracer
particle method \citep{travaglio2004a}. Virtual particles are advected
passively with the flow and record thermodynamic quantities such as
temperature, density, pressure etc. In a postprocessing step the
nucleosynthesis results are calculated with the nuclear network code
\textsc{yann} \citep{pakmor2012b}. We employ the 384 species network of
\citet{travaglio2004a} based on work by
\citet{thielemann1996a,iwamoto1999a}, nuclear reaction rates (version
2009) from the \textsc{REACLIB} database \citep{rauscher2000a}, and weak
rates from \citet{langanke2001a}.

To compare synthetic observables (i.e., light curves and spectra) with
data, we carry out RT simulations using the 3D Monte Carlo RT code
\textsc{artis} \citep{sim2007b,kromer2009a}. \textsc{artis} follows the
propagation of $\gamma$-ray photons emitted by the radioactive decay of
the nucleosynthesis products and deposits energy in the supernova
ejecta.  It then solves the RT problem self-consistently enforcing the
constraint of energy conservation in the co-moving frame. Assuming a
photoionization-dominated plasma, the equations of ionization
equilibrium are solved together with the thermal balance equation
adopting an approximate treatment of excitation. Since a fully general
treatment of line formation is implemented, there are no free parameters
to adjust. This allows direct comparisons to be made between the
synthetic spectra and light curves and observational data. Line of sight
dependent spectra are calculated employing the method detailed by
\citet{bulla2015a} which utilizes "virtual packets".  This method
significantly reduces the Monte-Carlo noise of the viewing angle
dependent spectra.

%---------------------------------------------------------------------
%----------------  Setup ----------------------------------------
%---------------------------------------------------------------------

\section{Initial setup}
\label{sec:setup}

We carried out full-star simulations of the explosion of a CO WD on a
spatial grid with $528^{3}$ cells in the nested expanding grid approach.
This resulted in an initial resolution of $2.06\,\si{km}$ per cell in the
central part of the star for all initial models and guaranteed that the
flame resolution is similar across all models during the early phase of
the explosion. We distributed $4\,096\,000$ tracer particles representing
equal mass fractions of the material throughout the star.  This rather
large number was chosen to guarantee a sufficient representation of the
ejecta since only a small part of the star is expected to become
unbound. The metallicity in the hydrodynamic simulation was only
represented by the electron fraction $Y_e$, which is set to the solar
value $Y_e=0.499334658$ according to the solar composition published by
\citet{asplund2009a}. In the nucleosynthesis postprocessing step we
used
the \citet{asplund2009a} values with C, O, and N isotopes converted to
$^{22}$Ne and ignoring hydrogen and helium. For subsolar metallicities,
we did not simply scale all isotope mass fractions but fix the ratios of
some $\alpha$-elements to measurements in low-metallicity stars (i.e.,
[C/Fe]=0.18, [O/Fe]=0.47, [Mg/Fe]=0.27, [Si/Fe]=0.37, [S/Fe]=0.35,
[Ar/Fe]=0.35, [Ca/Fe]=0.33, [Ti/Fe]=0.23, \citealp{prantzos2018a}). This
choice reflects the fact that $\alpha$-elements were overabundant at
early times since they are primarily produced in core collapse
supernovae. Moreover, all simulations assumed equal amounts of C and O in
the WD material if not specified otherwise.

\begin{figure}[htbp] 
  \centering
  \resizebox{1.00\columnwidth}{!}{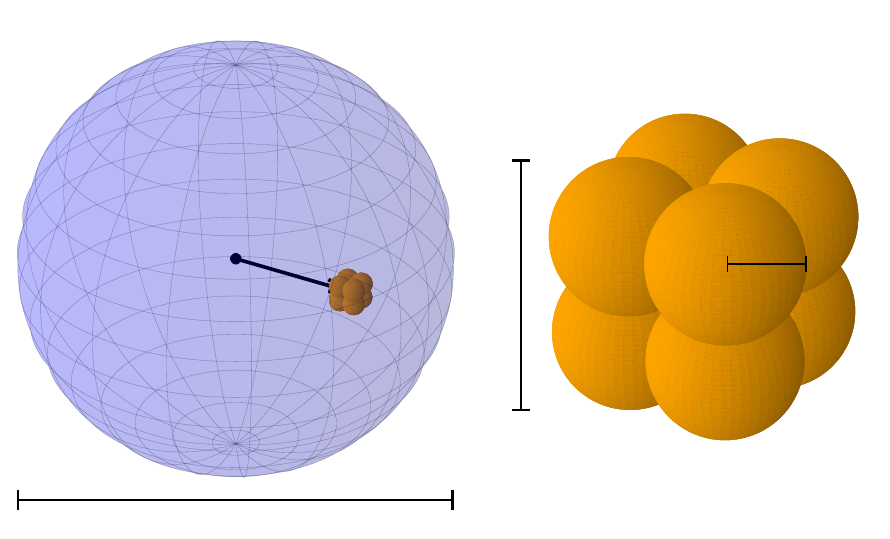}
  \caption{Sketch of the ignition configuration. The blue sphere
   only serves to guide the eye. The enlarged figure shows the
   morphology of the ignition spark.} 
  \label{fig:ini_bub} 
  \centering
  \includegraphics[width=0.99\columnwidth]{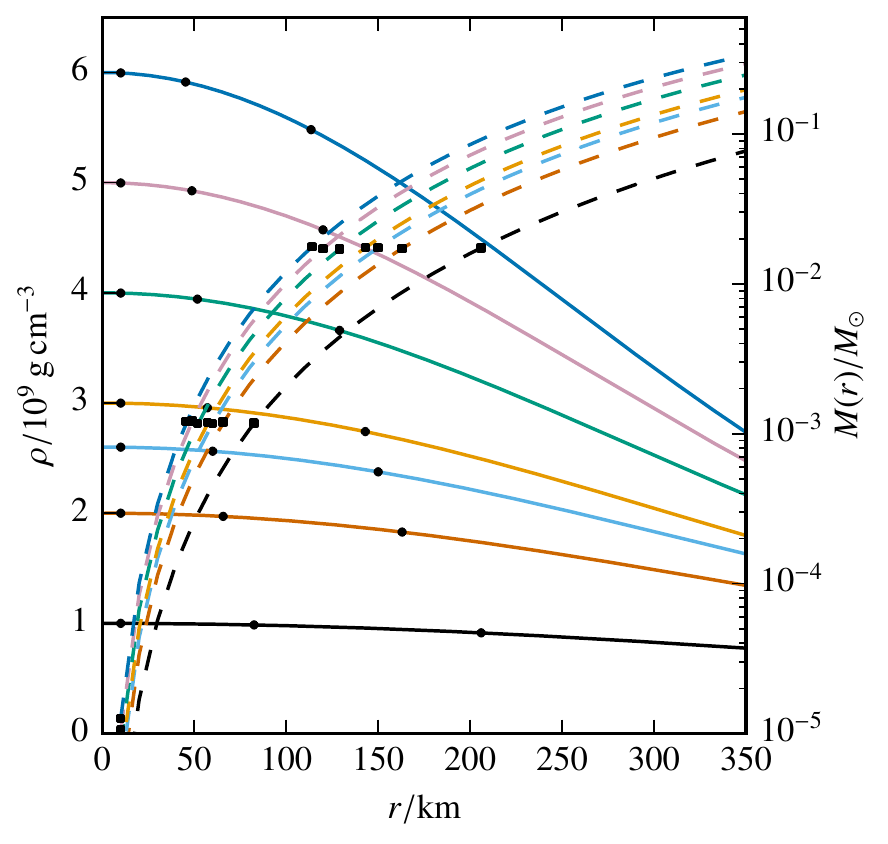}
  \caption{Initial density profiles (solid lines), cumulative masses
    (dashed lines) and ignition radii indicated by scatter points. Solid
    squares depict that models with intermediate to high ignition
    radii are ignited at same mass coordinate.}
  \label{fig:inidens}
\end{figure}

As pointed out in Sect.~\ref{sec:introduction} the most probable
ignition configuration is one single spot.  \citet{nonaka2012a} find
that the ignition is most probable between 40 and $75\,$km off-center
with an upper limit of 100$\,$km. Hence, we restricted this study to
single-spot ignitions. To provide some initial perturbations for
Rayleigh-Taylor instabilities to develop, the spark consists of 8
bubbles of $5\,\si{km}$ radius slightly overlapping (see
Fig.~\ref{fig:ini_bub}). This is the smallest reasonable size for the
initial bubble considering the resolution of $\sim2\,$km per cell. Since
the actual ignition (the thermonuclear runaway) takes place on
very small length scales (cm) the initial bubble should be chosen to be
rather small to capture as much of the flame development as possible.
Its actual shape at the beginning of our simulation is, however, an
assumption. 

\begin{figure*}[tb!]
  \centering
  \resizebox{0.95\textwidth}{!}{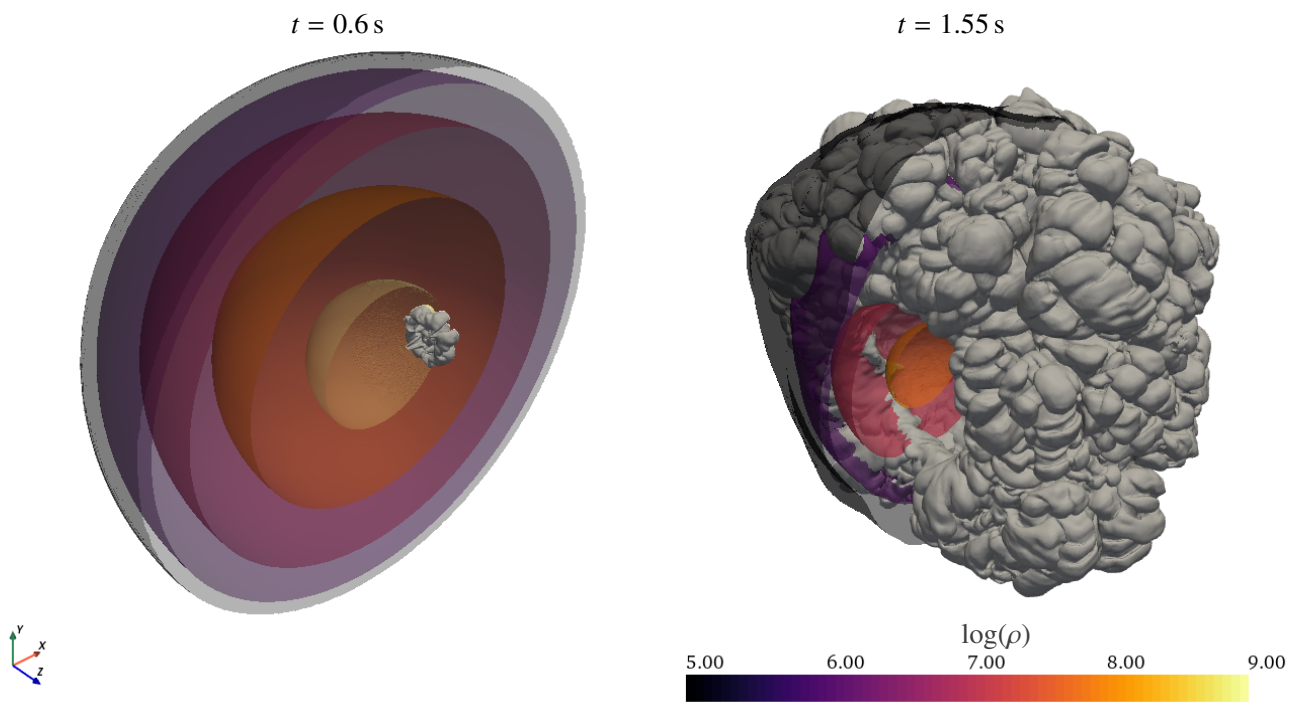}
  \caption{Flame surface (gray) and isosurfaces of the density at
    $\rho=10^{5},\,10^{6},\,10^{7},\,10^{8},\,10^{9}\,\si{g.cm^{-3}}$
    (see colorbar) for Model r60\_d2.6\_Z. The left panel shows the
    rising flame at $t=0.6\,$s while the right panel pictures the flame
    front when it has almost wrapped around the WD. We want to note, that the
    illustration is not to scale. In fact the WD has already expanded
    significantly in the right panel.}
  \label{fig:3d}
\end{figure*}

Our standard initial model is inspired by the results of simulations of
the simmering phase \citep{zingale2009a}. The central density $\rho_c$
is set to $2.6 \times 10^{9}\,\si{g.cm^{-3}}$, the central temperature
to $6 \times 10^{8}\,\si{K}$ with an adiabatic decrease down to $1
\times 10^{8}\,\si{K}$, and constant afterwards. This temperature profile
approximates the conditions during convective carbon burning prior to
explosion and is thus better motivated from a physical point of view
than the assumption of a cold isothermal WD. The actual effect, however,
on the outcome of the simulation is small and should not be
overestimated.  The deflagration is ignited at an off-center radius of
$r_\mathrm{off}=60\,\si{km}$, and, therefore, the model is named
r60\_d2.6\_Z. The model name encodes the most important parameters,
that is, the offset-radius in km (r), the central density in
$10^{9}\,\si{g.cm^{-3}}$ (d), and the metallicity in units of the solar
value $Z_\odot$ (Z).

In this study, we varied $r_\mathrm{off}$ between $10\,$km and $206\,$km,
$\rho_c$ between 1 and $6 \times 10^{9}\,\si{g.cm^{-3}}$ and the
metallicity (only for the standard model) between $1 \times 10^{-4} \,
Z_\odot$ and $2\,Z_\odot$. Since the WD is spherically symmetric, the
location of the ignition spot can be chosen arbitrarily. We always
placed
the spark on the positive $x$-axis with $x=r_\mathrm{off}$ and $y,z =
0$.  A summary of the suite of models and results for the ejecta, that
is,
unbound material with $E_\mathrm{kin} > E_\mathrm{grav}$, can be found
in Table~\ref{tab:ejectasum}. With respect to the $r_\mathrm{off}$
parameter we distinguish three groups in the set of models: (i) models
that were ignited in the inner part of the star, that is, at $10\,$km.
(ii) models in which the ignition is placed at around $60\,$km (standard
model) always at the same mass coordinate, that is, $r_\mathrm{off} =
45-82\,$km. (iii) models where the ignition kernel is located at around
$150\,$km also always at the same mass coordinate, that is, $r_\mathrm{off}
= 114-206\,$km (see also Fig.~\ref{fig:inidens}). 

In addition, we have added two models of rigidly rotating WDs ignited at
$r_\mathrm{off}=60\,\si{km}$ with a central density of $2\times
10^{9}\,\si{g.cm^{-3}}$.  The ignition spark is located perpendicular to
the rotation axis ($z$-axis) for the first model, r60\_d2.0\_Z\_rot1 and
directly on the rotation axis for the second, Model r60\_d2.0\_Z\_rot2.
Due to a more complex setup procedure of the initially rotating WD
compared to the standard model the temperature in the rotating WDs is
held constant at $5\times 10^5\,\si{K}$. The rotation velocity is close
to break-up velocity ($\Omega = 2.73\,\si{rad.s^{-1}}$) increasing the
mass of the WD by about $\sim6\%$ to $1.438\,\mathrm{M}_\odot$. Moreover, we
investigate a scenario with a carbon depleted core. In line with
\citet{lesaffre2006a} and \citet{ohlmann2014a} the inner carbon mass
fraction is reduced to 0.28 and joined smoothly with the outer regions
($X(C)= 0.5$) at a core mass of $\sim1$\msun. This model is labeled
r60\_d2.6\_Z\_c0.28.

% -------------------------------------------------------------------
% ------------------- RESULTS -------------------------------------
% -------------------------------------------------------------------

\section{Hydrodynamic simulations}
\label{sec:results}

The one-sided, single-spot ignition deflagration models of this work
usually proceed as follows: The burnt fuel inside the hot bubble is
lighter than its surroundings, and, thus, is subject to buoyancy. This
is the reason why the flame cannot propagate against the density
gradient and spreads to only one side of the WD. The deflagration itself
is mediated by heat conduction, but the laminar burning speed of the
flame is soon surpassed by the buoyant motion of the rising bubble.
Subsequently, the burning becomes turbulent due to Rayleigh-Taylor and
Kelvin-Helmholtz instabilities further increasing the fuel consumption.
As soon as the turbulent flame reaches the outer parts of the WD, and,
thus, 
lower densities, the burning quenches and the ashes flow around the star
to collide at the far side. This is illustrated in Fig.~\ref{fig:3d}.

\begin{figure}[tp!]
  \centering
  \includegraphics{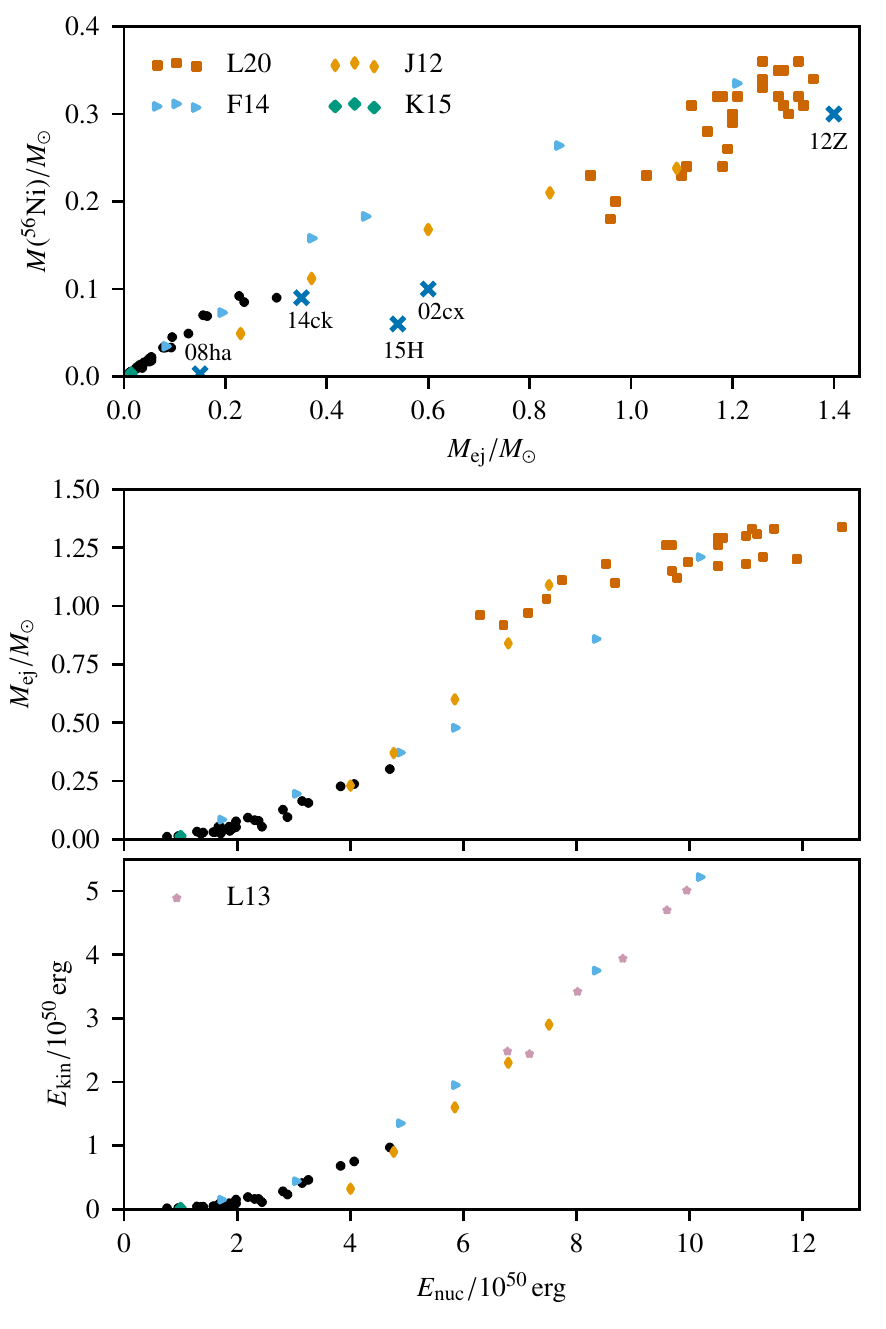}
  \caption{Ejected $^{56}$Ni mass v.s. $M_\mathrm{ej}$ (upper panel),
  $M_\mathrm{ej}$ v.s. $E_\mathrm{nuc}$ (mid panel) and
  $E_\mathrm{kin,ej}$ v.s. $E_\mathrm{nuc}$ (lower panel) for this work
  (black circles) and the works of J12, F14, K15 and
  L20. L12 only provide yields for IGEs, but not for \nifs. To provide
an approximate value for \nifs, we multiplied the IGE masses with a
factor of $0.7$ in agreement with the $M_\mathrm{IGE}$ to \mni ratios of
L14 who use the same initial model. Measurements for \mni and
$M_\mathrm{ej}$ for SN~2002cx, 2008ha, 2012Z and were
taken from \citet{mccully2014b}, and references therein. Values for SN
2015H are from \citet{magee2016a} and for SN~2014ck from
\citet{tomasella2020a}.}
 \label{fig:allrelations}
 \vspace{0.2cm}
  \centering
  \includegraphics{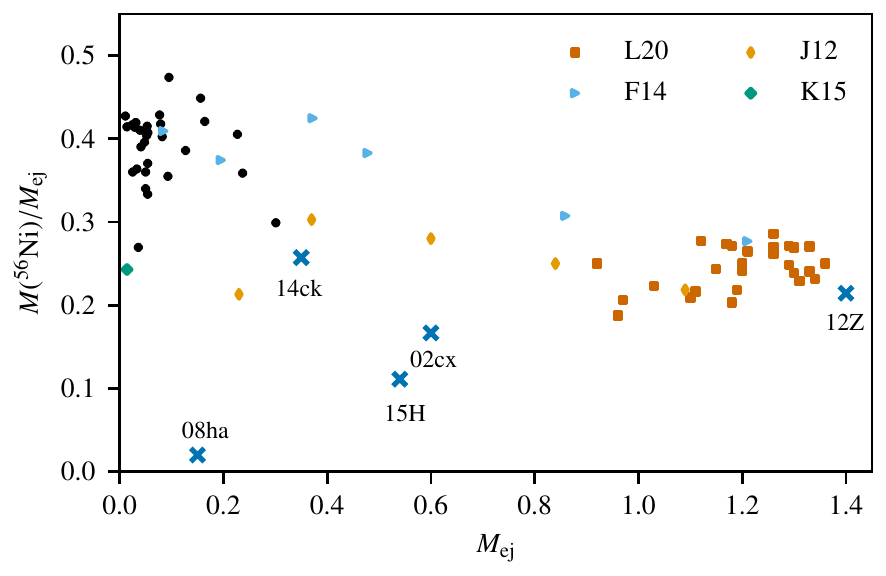}
  \caption{Ejected mass v.s. \mni/$M_\mathrm{ej}$ for this work
(black circles) and the works of J12, F14, K15, and L20.}
  \label{fig:ejvsniige}
\end{figure}

The key outcomes of our parameter study are summarized in
Table~\ref{tab:ejectasum} for ejected material and in
Table~\ref{tab:remnantsum} for the WD remnant. We find \nifs masses from
$0.0058$ to $0.092\,$\msun and ejected masses between 0.014 and
0.301$\,\mathrm{M}_\odot$. It is interesting to note that the lowest as well as
the highest \nifs mass (the mass of \nifs translates to the luminosity
of the event and we therefore refer to bright and faint models when
it comes to high or low \nifs masses, respectively) are found in
high-density models. The brightest model is r10\_d4.0\_Z and the
faintest one is Model r114\_d6.0\_Z. Overall, this suite of models
neither reaches the bright members of the SNe~Iax subclass (SN~2005hk,
SN~2011ay, SN~2012Z, etc.) nor the very faint explosions (SN~2008ha,
SN~2010ae, SN~2019gsc).  However, the \nifs yield of our faintest model
is only ${\sim}\,1.9$ times higher than the estimated value of $3 \times
10^{-3}\,\mathrm{M}_\odot$ for SN~2008ha \citep{foley2009a}. Furthermore, the
brighter models of our sequence (r10\_d4.0\_Z, r10\_d5.0\_Z,
r10\_d6.0\_Z) eject more \nifs than the N3 model of F14 although only
one ignition spark has been used. The \nifs masses in this study and
also in previous works are plotted against ejected mass in
Fig.~\ref{fig:allrelations}. Our models cover the low-energy end of the
pure deflagration model distribution and show slight overlap with the
F14 and J12 studies. Only the hybrid CONe WD model of K15 falls below the
faintest model in the current study. Moreover, the new set of models
extends the almost one-dimensional sequence in the $M(^{56}\mathrm{Ni})
- M_\mathrm{ej}$-plane. Despite varying initial conditions, that is,
central densities and ignition configurations, and different codes
employed, all models follow a linear relation between \mni and
$M_\mathrm{ej}$ to good approximation. However, it has been pointed out
by K13 and K15 that an increase in $M_\mathrm{ej}$ at a constant value
of \mni is highly desirable to obtain broader light curves as observed
in SNe~Iax. The ratio of \mni to $M_\mathrm{ej}$ is shown in
Fig.~\ref{fig:ejvsniige}. These values are comparable to those presented
by F14 and show only little scatter. Therefore, it seems unlikely that
SNe~Iax can be explained by only one scenario taking into account
results of the currently available pure deflagration models. 

Since the new set of single-spot ignited models only covers the fainter
and moderately bright members of the SN~Iax class, we have relaxed the
restriction of one ignition spark. We have recalculated Model N5 (5
ignition kernels) from F14 which has been compared to SN~2005hk by K13.
The N5 ignition configuration has been combined with the initial
condition of our standard model r60\_d2.6\_Z. We find that also with the
updated version of the \textsc{leafs} code brighter SNe~Iax can be
reached. The new version of N5, Model N5\_d2.6\_Z, yields values of
$E_\mathrm{nuc}$, $M(^{56}$Ni$)$, etc. (see Table~\ref{tab:ejectasum})
slightly below those of the F14 version. This is due to the lower
central density of $2.6\times 10^9\,\si{g.cm^{-3}}$ compared to
$2.9\times 10^9\,\si{g.cm^{-3}}$ in the older run. The relations
discussed above, however, do not change compared to the single-spot
ignited models.  Interestingly, Model N5\_d2.6\_Z is the brightest model
in the current study with \mni$=0.136\,\mathrm{M}_\odot$ although Model
r10\_d6.0\_Z releases more nuclear energy. The difference originates
from the more neutron-rich nucleosynthesis at high densities reducing
the amount of \nifs in relation to stable IGEs (see also
Sect.~\ref{subsec:density}).

\subsection{Dependence on central density}
\label{subsec:density}

The influence of the central density on the nuclear energy release is
twofold: (i) Assuming an identical ignition geometry, more mass is
burned initially in the high-density case, and, thus, the energy
production is higher in the early phase of the deflagration.  (ii) Due
to the steeper density gradient the effective gravitational acceleration
is larger at high densities, and, thus, the buoyancy force acting on the
low-density bubble increases. This speeds up the growth of
Rayleigh-Taylor instabilities and also the transition from the laminar
burning regime to turbulent burning (see also \citealp{roepke2006d}).
Moreover, a rather small effect might stem from the fact that the
laminar burning velocity itself is higher for higher densities
\citep{timmes1992a}. This increase of the burning rate at early times
is, however, counteracted by the expansion of the WD. Since the flame is
faster at high densities, the WD expands in a shorter period of time and
the flame reaches the surface significantly earlier than for low
densities. Hence, the burning is quenched earlier and limits the amount
of burnt material. This competition between flame propagation and
expansion of the WD is of particular importance for the energy release
of the models presented in this study.

An increase in the ejected mass of \nifs, that is, the brightness of the
explosion,  is also expected for higher densities. Since more material
is burnt in total and especially at high densities more IGEs, and
therefore also \nifs, will be produced. The synthesis of \nifs, however,
does not increase linearly with the IGEs because of the neutron-rich
environment due to an increasing rate of electron captures at high
densities. If matter burns to NSE the neutron excess of the most
abundant isotope is close to the neutron excess of the burning material,
and, thus, the symmetric isotope \nifs is not favored under these
conditions (see \citealp{woosley1973a}, for instance). 

To isolate the influence of the initial central density, we compare
models ignited at a fixed radius, that is, the models ignited at
$r_\mathrm{off}=10\,\si{km}$ (r10\_dXX\_Z, see
Table~\ref{tab:ejectasum}). We observe a trend expected from theoretical
considerations (see above): The nuclear energy release and the ejected
masses increase from low ($1 \times 10^{9}\,\si{g.cm^{-3}}$) to high ($6
\times 10^{9}\,\si{g.cm^{-3}}$) central densities at ignition. Moreover,
the value of \mni in models r10\_d5.0\_Z and r10\_d6.0\_Z is lower than
in Model r10\_d4.0\_Z although the IGE yields are higher. This is due to
higher electron capture rates at high density, and, therefore, more
neutron-rich environment. This indicates that Model r10\_d4.0\_Z is
close to the brightest explosion we can produce with a single-bubble
ignition (models at higher $r_\mathrm{off}$ are fainter, see
Table~\ref{tab:ejectasum}). The decreasing trend of
\mni/$M_\mathrm{IGE}$ is also present for all other ignition radii (see
Table~\ref{tab:ejectasum}). We find values of \mni/$M_\mathrm{IGE}$ in
the ejecta ranging from 0.88 at low central densities to 0.50 at high
densities.

In summary, the flame evolves faster with increasing central density,
and,
thus, the energy released at early times rises. Therefore, the WD also
starts to expand significantly earlier for higher central densities.
However, the fast burning can always compensate for the expansion in
models with fixed ignition radius ($10\,\si{km}$) which leads to a
monotonic increase of released nuclear energy with increasing central
density.

\subsection{Dependence on ignition radius}
\label{subsec:radius}

For a fixed central density, we observe the expected, decreasing trend
in nuclear energy generation and ejected mass for increasing ignition
radii (see Table~\ref{tab:ejectasum}). This seems rather obvious because
there is less mass available to burn for larger radii since the
deflagration only propagates outward and not against the density
gradient. However, the interplay between the flame velocity and the
expansion of the WD (see Sect.~\ref{subsec:density}) also is a
significant factor here.  We find that for all models except those with
lowest density, that is, $\rho_c=1\times 10^9\,\si{g.cm^{-3}}$, the
total amount of burnt mass decreases for larger ignition radii. The
volume of the flame during the burning phase (until $t\sim2\,\si{s}$) is
always largest for models ignited at large ignition radii reflecting the
faster evolution of the flame due to the higher gravitational
acceleration $g$ (the maximum of $g$ is only reached at
$r>400\,\si{km}$). In contrast, the total mass burned, and with it the
nuclear energy release, is higher for small ignition radii. This
reflects that the average density at which material is burned makes up
for the smaller volume filled by the flame in the first $\sim
2\,\si{s}$.  Only in the case of Model r82\_d1.0\_Z, the increased flame
velocity can overcompensate the expansion of the WD and leads to
slightly higher nuclear energy release than for Model r10\_d1.0\_Z.

In the model sequences of intermediate and high ignition radii, that is,
45 to $82\,\si{km}$ and 114 to $206\,\si{km}$, respectively, we have
eliminated the difference in mass outside the ignition radius by
igniting at the same mass coordinate (see also Fig.~\ref{fig:inidens}).
For the reasons discussed above, the most energetic explosions in the
intermediate sequence are those at low central density (r82\_d1.0\_Z and
r65\_d2.0\_Z). $E_\mathrm{nuc}$ increases again for the highest central
densities ($\rho=5-6 \times 10^{9}\,\si{g.cm^{-3}}$).  This indicates
that the flame speed starts to compensate for the expansion in these
models. In the models with large ignition radii the differences almost
vanish and the trend in $E_\mathrm{nuc}$ becomes monotonically
decreasing. The lowest amount of \nifs ($M(^{56}\mathrm{Ni}) =
0.0058\,\mathrm{M}_\odot$), and, therefore, also the faintest explosion
results from the model at highest central density and largest ignition
radius, that is, Model r114\_d6.0\_Z.  This is still a factor of
${\sim}\, 1.9$ above the \mni mass inferred from observations of
SN~2008ha \citep{foley2009a}, but the trends in this study suggest that
such a low value could easily be reached by increasing the ignition
radius even further.  Moreover, the \mni to $M_\mathrm{IGE}$ ratio does
not drop off very significantly for increasing densities since a larger
part of the inner core is left unburned for a large ignition radius. 

We note that for models ejecting very little mass, that is,
$M_\mathrm{ej} \lesssim 0.08\,\mathrm{M}_\odot$, the data show some
scatter destroying monotonic trends. There is, in general, a decrease in
\mni in the model sequence with large ignition radius from Model
r206\_d1.0\_Z to r114\_d6.0\_Z which is interrupted by an unexpected
increase from Model r163\_d2.0\_Z to r150\_d2.6\_Z, for instance. We are
not able to explain this deviation by any physical properties of the
respective explosion model. Instead, we suspect that the real
differences between these models are too subtle to be captured by our
simulations.

\subsection{Dependence on metallicity}
\label{subsec:metallicity}

The net effect of metallicity is that it introduces a neutron excess
(reduction of $Y_e$) which has several effects on the explosion
dynamics. First, the initial WD becomes more compact and lighter with
increasing metallicity because the degenerate electron pressure
decreases as $Y_e$ decreases. Second, the laminar flame speed is
enhanced by the presence of $^{22}$Ne which is by far the most abundant
species aside from C and O \citep{chamulak2007a}. This is, however, only
important during the very early stages of the deflagration before
turbulence introduced by Rayleigh-Taylor and Kelvin-Helmholtz
instabilities governs the effective flame speed. Third, a neutron-rich
environment leads to a decrease in the production of \nifs in NSE
\citep{timmes2003a} since it favors the synthesis of more tightly bound,
neutron-rich nuclei such as $^{57,58}$Ni. Thus, the energy release is
slightly higher for higher metallicity \citep{townsley2009a}. Finally, a
reduction of $Y_e$ has only negligible influence on the buoyancy force
\citep{townsley2009a}. On average, the effect of metallicity on the
explosion dynamics is rather small. This is supported by Models
r60\_d2.6\_XZ ($X=10^{-4},\,10^{-3},\,10^{-2},\,10^{-1},\,1,\,2$). They
can be seen as identical regarding their values for $E_\mathrm{nuc},
M_\mathrm{ej}, M(^{56}\mathrm{Ni})$ since no real trend is visible in
the data.

A decrease in \mni is expected for low metallicities which might again
help in reducing the \mni to $M_\mathrm{ej}$ ratio. This trend, however,
is too subtle to be observed in this study and the small scatter is more
likely to originate from numerical inaccuracies. The value of $Y_e$ in
the high density regime of \mch explosions, in addition, is dominated by
the effect of electron captures during the explosion whereas in sub-\mch
models the initial metallicity is basically the only parameter
determining the neutron excess.  Nevertheless, the nucleosynthesis
yields are expected to differ for varying metallicity. A detailed
investigation of nucleosynthetic postprocessing data, however, is not
the focus of this work. The nucleosynthesis yields of Model r60\_d2.6\_Z
are included an analyzed in the work of \citet{lach2020a} labeled as
Model R60. The model does produce a value of [Mn/Fe]$=0.11 $, and, thus,
these explosions might play a role in the enrichment of the Universe
with Mn in addition to helium shell detonations
\citep{lach2020a,gronow2021b} and various types of core collapse
supernovae. The role of the latter is not a settled issue yet (compare
the GCE studies of \citealp{prantzos2018a} and \citealp{kobayashi2020a},
for instance). The nucleosynthesis results of our parameter study will
be published on \textsc{HESMA} \citep{kromer2017a} for further use in
GCE studies.

\begin{figure*}[h!]
  \centering
  \includegraphics[width=\textwidth]{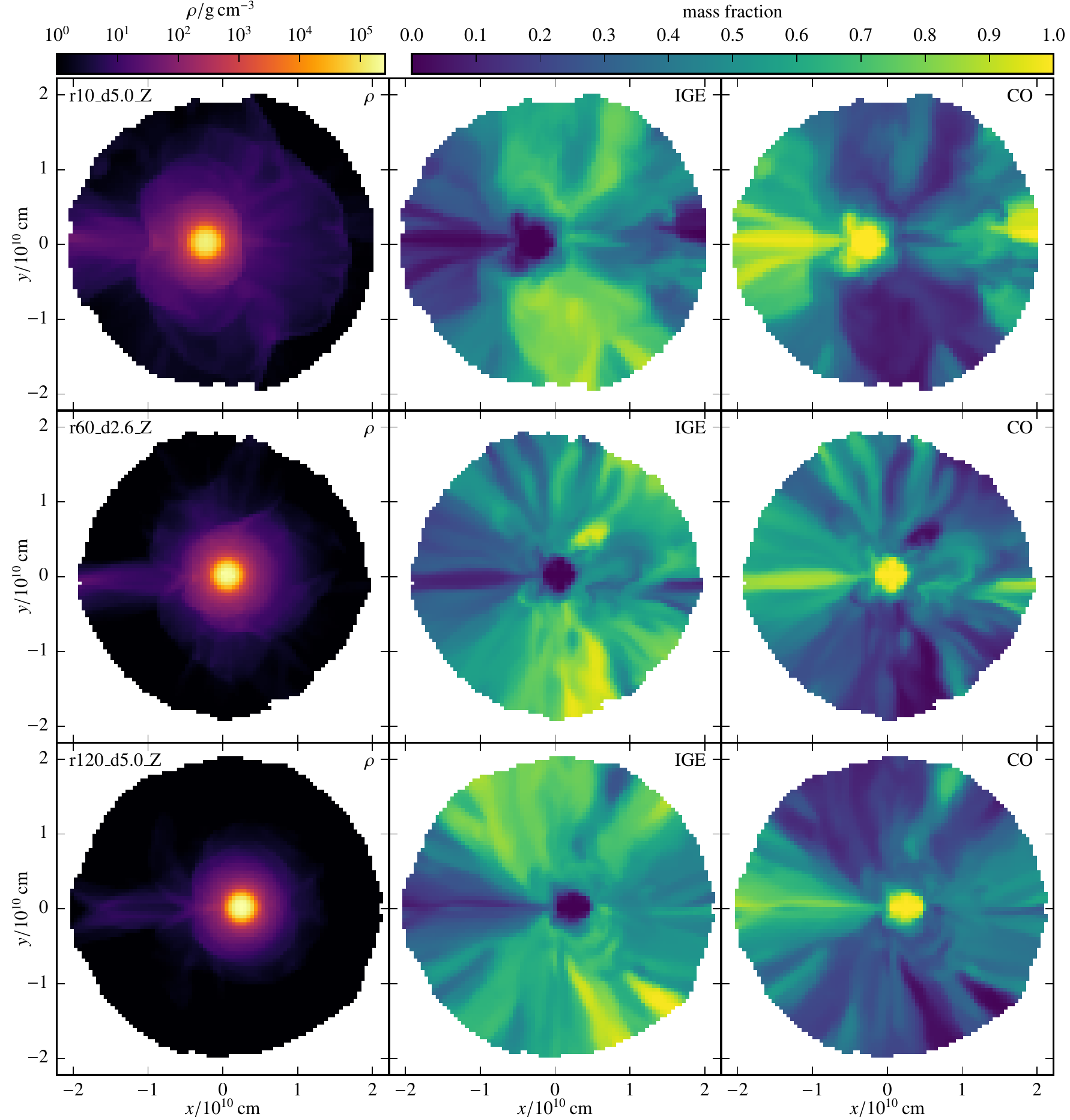}
  \caption{Slices ($x$-$y$-plane) of density, IGE and CO mass fraction in
    the bound remnant for the bright model r10\_5.0\_Z
(upper row) the moderately bright model r60\_d2.6\_Z (middle row) and
the faint model r120\_d5.0\_Z (lower row).}
  \label{fig:remnant}
\end{figure*}

\subsection{Rotating models}
\label{subsec:rot}

In the single degenerate scenario of SNe~Ia the WD is not only expected
to accrete mass from its donor star but also angular momentum (spin-up).
This further stabilizes the WD against gravity and total masses highly
exceeding \mch are possible \citep{yoon2005b}. At some point, the
material available for accretion might be exhausted and the accretion
processes slows down or stops completely and a period of loss of angular
momentum ensues (spin-down). During this spin-down the critical mass
$M_\mathrm{crit}$ of the WD decreases and a deflagration might be
ignited as soon as it is exceeded by the WD mass. A rigidly rotating WD
serves as a lower limit for $M_\mathrm{crit}$.  This spin-up/spin-down
scenario was proposed by \citet{distefano2011a}.

Based on the work of \citet{pfannes2010a}, we expect minor differences
between the rigidly rotating models and the nonrotating model. On the
one hand, more mass is available for burning (${\sim}\,6\%$) in the
rotating case which can lead to a higher nuclear energy release. On the
other hand, the rotating WD progenitor is more tightly bound making it
harder to eject material. Moreover, due to the shallower density
gradient the relative fraction of IGEs compared to the total mass of the
burning products decreases with increasing rotation velocity. These
effects are even more prominent for fast, differentially rotating WDs
(see also \citealp{fink2018a}). In addition, the propagation of the
deflagration front is influenced by rotation. Since buoyancy is stronger
parallel to the rotation axis due to the steeper density gradient and
the flow vertical to the latter is suppressed because the material needs
to gain angular momentum, the flame preferentially rises toward the
poles.  Therefore, the ignition conditions in this study have a large
influence on the explosion compared to centrally distributed, multiple
ignition sparks. 

We compare the rotating models (Model r60\_d2.0\_Z\_rot1 and
r60\_d2.0\_Z\_rot2) to Model r65\_d2.0\_Z. The minor difference in
$r_\mathrm{off}$ is not expected to obscure the differences originating
from rotation. Table~\ref{tab:ejectasum} reveals that the outcome of the
explosion indeed depends on the ignition location. The nonrotating model
is bracketed by the rotating explosions in terms of $E_\mathrm{nuc}$,
$M_\mathrm{ej}$, $M(^{56}$Ni$)$, $M_\mathrm{IGE}$, $M_\mathrm{IME}$ and
$E_\mathrm{kin,ej}$. Model r60\_d2.0\_Z\_rot1 yields higher values than
Model r60\_d2.0\_Z\_rot2 and is therefore expected to be brighter as
well. The reason for this is that Model r60\_d2.0\_Z\_rot2 is ignited on
the rotation axis, and, thus, the flame spreads along this axis very
fast and leads to a fast expansion of the star (see the discussion on
expansion v.s. flame speed in Sect.~\ref{subsec:radius}). In Model
r60\_d2.0\_Z\_rot1 the deflagration is ignited perpendicular to the
rotation axis which prevents burning directly toward the surface by the
rotation and propagates toward north and south pole instead. The
consequence of this is that the expansion is delayed, more material is
burned and the explosion is more vigorous. However, these rigidly
rotating models do not introduce any new characteristics in terms of
their global properties to the model sequence. This indicates that
rigidly rotating WDs as progenitor cannot add any diversity to our
models for SNe~Iax.

\subsection{Carbon-depleted model}
\label{subsec:co}

Another way to modify the progenitor is to vary its C/O ratio. The C
mass fraction in the center of the WD is determined by the initial mass
and metallicity of its zero-age main sequence progenitor and is expected
to lie below $0.5$ \citep{umeda1999a,dominguez2001a}. The value of
$X(\mathrm{C})$ is further reduced during convective carbon burning
(simmering phase) while the outer layers accumulated via accretion and
subsequent shell burning exhibit C/O$\,\sim 1$ \citep{lesaffre2006a}. To
investigate the effects of a reduced C mass fraction, we have calculated
an explosion, Model r60\_d2.6\_Z\_c0.28, set up with a C-depleted core
(see Sect.~\ref{sec:setup}) with a C mass fraction reduced to
$X(\mathrm{C})=0.28$. The total energy released during the explosion
decreases with decreasing $X(\mathrm{C})$ since the nuclear binding
energy of O is higher than of C. This also decreases the buoyancy force
and leads to a slower propagation of the flame. Therefore,
\citet{umeda1999a} claim that higher $X(\mathrm{C})$ leads to brighter
SN events and \citet{khokhlov2000a} and \citet{gamezo2003a} find that
the burning in C depleted material is delayed. However,
\citet{roepke2004c} and \citet{roepke2006b} examine the influence of the
C/O ratio in 3D simulations and detect that its effect on the flame
propagation is more subtle. In detail, the nuclear energy is buffered in
$\alpha$-particles when material is burned to NSE and only released in
later stages of the explosion. This temporary storage of energy is
enhanced for higher $X(\mathrm{C})$ suppressing the buoyancy, and, thus,
the propagation of the flame. Therefore, the C mass fraction does not
affect the total production of IGEs significantly.

We corroborate this result with Model r60\_d2.6\_Z\_c0.28 and find
approximately the same total amount of IGEs and \nifs as in the standard
model r60\_d2.6\_Z (see Table~\ref{tab:ejectasum} and
\ref{tab:remnantsum}) although the nuclear energy released is lower for
the C-depleted model.  This, however, leads to less ejected mass in
Model r60\_d2.6\_Z\_c0.28.  In contrast to the vigorously ignited models
of \citet{roepke2006b}, the C/O ratio has an influence on the ejecta mass
in sparsely ignited models.  Since the relations between
$M_\mathrm{ej}$, $E_\mathrm{nuc}$, $M_\mathrm{IGE}$ and \mni do not show
any deviations from the other models of the sequence (see
Fig.~\ref{fig:allrelations}) the carbon mass fraction is just another
way to vary the brightness of the explosion but does not seem to help
explaining the trends observed among SNe~Iax.

\subsection{Bound remnant}
\label{subsec:remnant}

All models presented in this work are not energetic enough to unbind the
WD completely. They leave behind massive bound remnants.  These objects
consist of a rather dense CO core and a puffed-up envelope of ashes
admixed with CO material. This shell material is settling onto the WD
core. If these stars manage to escape the binary system, for instance
via a natal kick due to the asymmetric ignition, they can potentially be
observed as chemically peculiar, high-velocity WDs. The first candidate
for a SN~Iax postgenitor, LP40-365, was observed by \citet{vennes2017a}
and studied in more detail by \citet{raddi2018a, raddi2018b}. Moreover,
two more possible remnants of thermonuclear supernovae were discussed by
\citet{raddi2019a}. Although there are hints that these stars might be
the product of a failed deflagration their origin is not completely
established yet. Therefore, it is necessary to further investigate the
long-term evolution of the bound remnant beyond 1D models (see
\citealp{shen2017a,zhang2019a}) and also to obtain more detailed
observations of such objects to shed light on their origin. Finally, a
sound understanding of the structure and composition of the envelope is
of vital importance since it may contribute to the light of the actual
SN event \citep{kromer2013a,kromer2015a} and solve the problem of the
fast decreasing light curve in current deflagration models. The 1D study
of \citet{shen2017a} on this problem shows that a post-SN wind driven by
the delayed decay of $^{56}$Ni in the envelope of the remnant
contributes to the late-time bolometric light curve of the postgenitor.
This contribution is in rough agreement with observed late-time light
curves of SNe~Iax. 

\begin{figure*}[hbtp]
  \centering
  \includegraphics[width=1.0\textwidth]{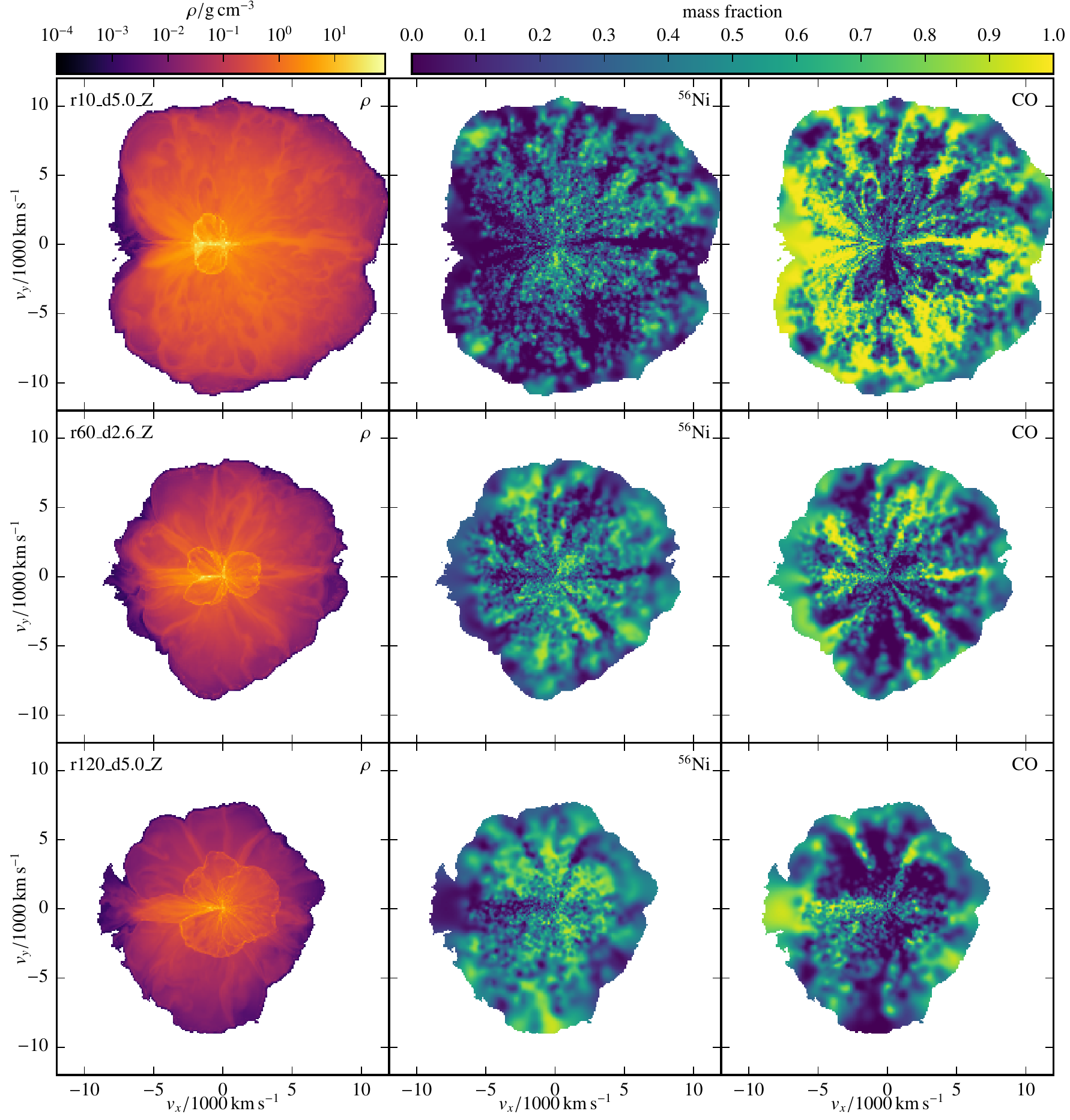}
  \caption{Ejected material mapped to a velocity grid to serve as
  initial model for the RT. Shown are slices ($x$-$y$-plane)
of density, \nifs and CO mass fraction for the bright model r10\_d5.0\_Z
(upper row) the moderately bright model r60\_d2.6\_Z (middle row) and
the faint model r120\_d5.0\_Z (lower row).}
  \label{fig:ejecta}
\end{figure*}

Since the \textsc{leafs} code uses an expanding grid to track the ejecta
the dense core is not very well resolved at the end of the simulations.
Therefore, we can only provide some basic properties of the bound core
directly after the explosion at $t=100\,\si{s}$ (see
Table~\ref{tab:remnantsum}). First, all our models leave behind massive
remnants in the mass range of $1.09$ to $1.38\,\mathrm{M}_\odot$. These
consist of a dense CO core ($\rho \gtrsim 10^5\,\si{g.cm^{-3}}$) with a
diameter of $\sim 4\times 10^9\,\si{cm}$ and a large envelope polluted
with burning products, that is, IGEs and IMEs (see
Fig.~\ref{fig:remnant}).  Unfortunately, we do not have postprocessing
data for the remnant, and, hence cannot provide detailed nucleosynthesis
results. From the hydrodynamic simulation we infer a mass fraction of
$0.78-3.0\,\%$ of IMEs and $4.2-10.1\,\%$ of IGEs including
$3.0-8.7\,\%$ of \nifs in the bound remnant. The amount of IGEs left
inside the remnant is always larger than the amount of ejected IGEs
except for the five most energetic explosion models. Moreover, the
settling material is not well mixed showing alternating plumes of ash
and fuel (see Fig.~\ref{fig:remnant}). The most noticeable feature is
the prominent, cone-shaped region of unburned fuel at the left-hand
side, that is, negative $x$-direction. This is exactly the opposite side
of the ignition spark location and thus was not reached by the
deflagration. 

If we expect these peculiar WDs to be observed as single hyper-velocity
runaway stars they need to escape the binary system via a natal kick,
for example.  While F14 report a maximum kick velocity of only
$36\,\si{km.s^{-1}}$ J12 find velocities of up to $549\,\si{km.s^{-1}}$,
which is sufficient to unbind the WD from its binary system. One
possible explanation for this discrepancy was the monopole gravity
solver employed by F14 which does not capture the full geometry of these
asymmetric explosions. Here, we use an updated version of the
\textsc{LEAFS} code accounting for self-gravity with an FFT gravity
solver (see Sect.~\ref{sec:numerics}). We find kick velocities
$v_\mathrm{kick}$, that is, the center of mass velocity of bound material,
from $6.4$ to $370\,\si{km.s}$. Interestingly, $v_\mathrm{kick}$ does
not simply scale with explosion strength but is highest for low
energetic models, decreases to around zero for intermediate ones and
increases again for the most vigorous explosions (see
Table~\ref{tab:remnantsum}).  Table~\ref{tab:remnantsum} also shows the
kick velocity in $x$-direction $v_\mathrm{x}$. Since the flame is
located off-center (positive $x$-direction in Fig.~\ref{fig:remnant})
and propagates outward, the WD is expected to receive a recoil momentum
in the opposite direction, that is, the negative $x$-direction. This holds
true for the most energetic explosions. The reason why the direction of
the kick changes from negative to positive $x$-direction is hidden in
the dynamical evolution of these kinds of explosions. In general, the
one-sided ignition causes the flame to propagate to one direction only
since a deflagration cannot proceed against the density gradient. In
this case, the flame propagates to the right-hand side (positive
$x$-direction, compare Fig.~\ref{fig:remnant} and \ref{fig:ejecta})
until it reaches the surface of the WD and quenches. It then wraps
around the star, the ashes eventually collide at the opposite side, and
a strong flow is driven toward the inner regions. This flow then
counteracts the initial momentum transferred by the flame propagation.
The more vigorous the explosion the faster is the expansion of the WD
and thus the ashes hardly collide at the far side of the star for our
most energetic models. Therefore, the WD is pushed to the left (negative
$x$-direction).  For decreasing deflagrations strengths the momentum
injected by the initial flame propagation drops and the star expands
more slowly. Thus, the clash of the ashes at the antipode, and,
subsequently, the inward flow of material becomes stronger until both
effects balance each other and lead to almost zero kick velocity
(intermediate models). In the case of our weaker explosions, the kick
caused by the collision of the ashes dominates and leads to a push to
the right-hand side (positive $x$-direction). The very faintest model in
our parameter study, Model r114\_d6.0\_Z, breaks this trend and shows
again a low kick velocity in the negative $x$-direction. Here, the total
amount of mass brought to the surface by the deflagration has become too
low to impact the momentum of the remnant. This shows that the final
value of $v_\mathrm{kick}$ strongly depends on the ignition geometry and
the explosion strength. Finally, the remnant in Model N5\_d2.6\_Z
receives a kick of $264.6\,\si{km.s^{-1}}$ which is significantly higher
than $5.4\,\si{km.s^{-1}}$ reported by F14 employing a monopole gravity
solver. This shows that the new solver has a large impact on the
dynamics of these asymmetric explosion simulations. 

\begin{figure*}[t!]
  \centering
  \includegraphics[width=1.0\textwidth]{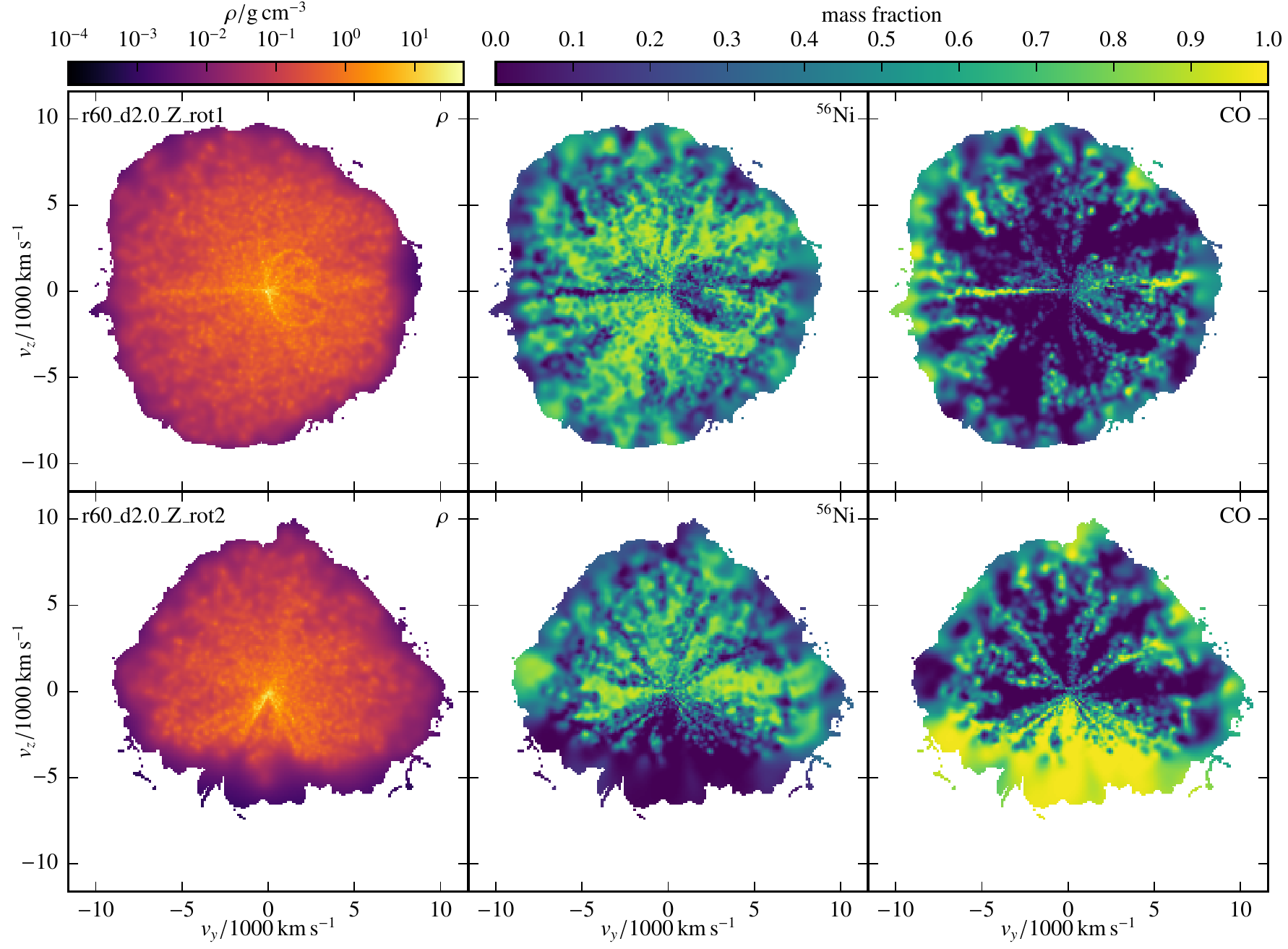}
  \caption{Same as Fig.~\ref{fig:ejecta} ($y$-$z$-plane), but for the
rotating Models r60\_2.0\_Z\_rot1 (upper row) and r60\_d2.0\_Z\_rot2
(lower row).}
  \label{fig:ejecta_rot}
\end{figure*}

We emphasize that there is a large scatter in the values of
$v_\mathrm{kick}$. The very coarse treatment of the inner parts of the
WD at late times is most probably the reason for this, and, especially
the differences in $v_\mathrm{kick}$ for the models at different
metallicity (r60\_d2.6\_Xz) lack an explanation since their energy
release does not differ significantly. This suggests that the values can
vary on the order of $\sim 100\,\si{km.s^{-1}}$ and can only serve as a
rough estimate. However, the trend in $v_\mathrm{x}$ explained above is
not affected. 

\subsection{Ejecta}
\label{subsec:ejecta}

\begin{figure}[htbp]
  \centering
  \includegraphics{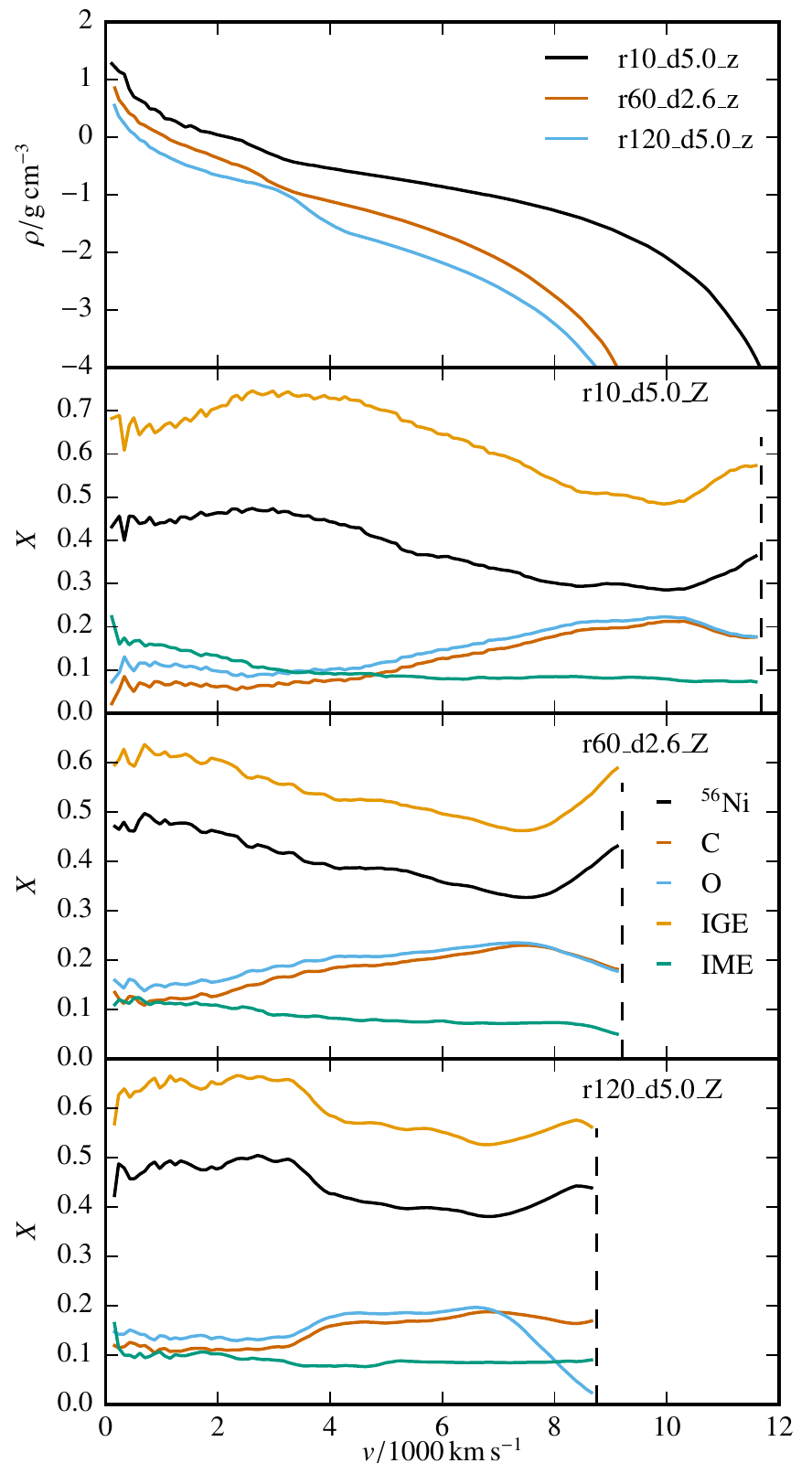}
  \caption{1D averaged density profile of the ejecta for Models
  r10\_5.0\_Z, r60\_2.6\_Z, r120\_5.0\_Z (upper panel). The lower three panels
show 1D IGE, \nifs, IME, C, and O profiles for the respective models. The
dashed lines indicate a cutoff at densities below
$10^{-4}\,\si{g.cm^{-3}}$.}
  \label{fig:1Dvel}
\end{figure}

We have compiled slices through the ejecta of three representative
models in Fig.~\ref{fig:ejecta} displaying the density, \nifs mass
fraction, and CO mass fraction, respectively. These models include
r10\_d5.0\_Z (bright event, upper row), r60\_d2.6\_Z (intermediate
brightness, middle row) and r120\_d5.0\_Z (faint event, lower row).
Despite the asymmetric ignition, the ejecta in general appear spherically
symmetric. Only for the bright model r10\_d5.0\_Z the material is skewed
slightly to the side of the initial ignition (right-hand side in
Fig.~\ref{fig:ejecta}). This is due to the fact that the fast expansion
prevents the ashes from wrapping around the core completely (see also
Sect.~\ref{subsec:remnant}). Moreover, it is also most apparent in
r10\_d5.0\_Z that the opposite side of the ignition kernel is deficient
in \nifs (also burning products in general) which may introduce some
viewing angle dependency in the synthetic spectra and light curves (see
Sect.~\ref{subsec:viewingangle}). This characteristic was also reported
by J12. Apart from that, the ejecta are well-mixed (see also
Fig.~\ref{fig:1Dvel}) showing a star-shaped pattern created by the
Rayleigh-Taylor plumes. The rotating models, in contrast, show a more
asymmetric structure (see Fig.~\ref{fig:ejecta_rot}). The ejecta in
Model r60\_d2.0\_Z\_rot1, for instance, are slightly shifted to the
negative $y$-axis (left-hand side in Fig.~\ref{fig:ejecta_rot}).  This
is due to the angular momentum barrier (see Sect.~\ref{subsec:rot})
hindering the flame from sweeping around the WD as easily as in the
nonrotating case. Model r60\_d2.0\_rot2, ignited on the rotation axis,
is even more extreme concerning the asymmetry of the ejecta. The flame
burns toward the north pole very quickly but is prevented from
propagating around the core almost completely making the south pole
deficient of burning products. Furthermore, the ejecta only reach
velocities of ${\sim}\, 6000\,\si{km.s^{-1}}$ toward the south pole and
${\sim}\, 10000\,\si{km.s^{-1}}$ at the north pole. This large-scale
anisotropy might introduce significant viewing angle dependencies. 

To analyze the chemical composition of the ejecta we have compiled 1D
average velocity profiles of the three reference models in
Fig.~\ref{fig:1Dvel}. They show that the ejecta are mixed in the sense
that IGEs, IMEs and unburned CO fuel are present at all velocities.
However, we observe a weak decreasing trend for IGEs (including \nifs)
and IMEs toward high velocities and an increase in C and O. At the very
edge of the ejecta, IGEs and \nifs even begin to rise again. However,
these trends are not as strong as predicted by \citet{barna2018a} and
the increase of IGEs at very high velocities is in strong conflict with
their work. Their abundance tomography study yields similar mass
fractions for IMEs, IGEs and O in the inner regions but they find that C
is virtually absent in the inner region and O dominates at high
velocities. This comparison is, however, not too revealing since
\citet{barna2018a} investigate the brightest members of the SN~Iax
subclass (SNe~2011ay, 2012Z, 2005hk, 2002cx) and our models represent
faint and moderate members of the class. In a recent study,
\citet{barna2021a} focus on the moderately bright SN~Iax SN~2019muj
(comparable to r10\_d1.0\_Z, r82\_d1.0\_Z, r65\_d2.0\_Z, r45\_d6.0\_Z in
terms of $M(^{56}$Ni$)$) and find no significant stratification except
for carbon.  Although their results are uncertain above ${\sim}\,
6500\,\si{km.s^{-1}}$ due to a sharp drop-off in density, they conclude
that C is virtually absent below this value and steeply increases above
in agreement with their earlier work \citep{barna2018a} and in strong
contrast to the models presented here. This conclusion, however, is not
really strong since earlier spectra are needed to accurately model the
outer layers of the ejecta. Apart from the case of C, a qualitative
comparison of Fig.~\ref{fig:1Dvel} to Fig.~11 in \citet{barna2021a}
shows that our abundance profiles are concordant with their abundance
tomography models showing a shallow decline of \nifs and IMEs and an
increase in the O mass fraction.

%------------------------------------------------------------------------
%-------------------------- Observables --------------------------------
%------------------------------------------------------------------------

\section{Synthetic observables}
\label{sec:rt}

To determine how the parameters varied in this study impact the
synthetic observables, we carried out time-dependent 3D Monte-Carlo RT
simulations using the \textsc{artis} code \citep{sim2007b,kromer2009a}.
Tab.~\ref{tab:lightcurve_data_bessel} lists the models for which RT
simulations were carried out along with bolometric band, BVRI Bessel
band, and ugriz Sloan band light curve parameters.  We selected models
that cover the range of $^{56}$Ni masses produced in the model sequence
to explore its full diversity and investigate the impact of the varied
parameters.  For each RT simulation, $3 \times 10^{7}$ energy packets
were tracked through the ejecta for 150 logarithmically spaced time
steps between 0.3 and 35 days post explosion. We use the atomic data set
described by \citet{gall2012a}. A gray approximation is used in cells
that are optically thick (cf. \citealp{kromer2009a}) and local thermal
equilibrium (LTE) is assumed for times earlier than 0.4 days post
explosion. Line of sight dependent light curves are calculated for 100
equal solid angle bins.

After the choice of initial conditions we make in the models (e.g.,
central density, ignition radius, metallicity, rigid rotation), we have a
fully self-consistent modeling pipeline. This consists of the
hydrodynamic explosion simulations, nucleosynthesis postprocessing
step, and, finally, the RT simulations producing synthetic observables.
This means we are comparing the predictions of our simulations, given a
choice of initial parameters, to measured data. We are not providing any
further input parameters (e.g., temperature, luminosity etc.) in order to
fit the data and the comparisons we make should be interpreted within
this context. Our self consistent pipeline also means it is important to
take into account any assumptions made throughout our simulations. Of
particular note for these models is the approximate non-LTE treatment of
the ionization and excitation conditions in the plasma we adopt for our
\textsc{artis} RT simulations (see \citealp{kromer2009a} for more
details).  The non-LTE treatment of the ionization and excitation
conditions in the plasma is an important ingredient in the modeling of
SNe~Ia (\citealp{dessart2014a}). No direct comparisons have been made
between a \mch pure deflagration model simulated using approximate
non-LTE and full non-LTE treatment in the regime that matches SNe~Ia
models. It is therefore uncertain how adopting a full non-LTE treatment
for our models would change our results.  However, previous works
employing a full non-LTE treatment of the plasma conditions in models
for SNe~Ia (e.g., \citealp{blondin2013a}, \citealp{dessart2014a},
\citealp{shen2021a}) indicate that it will have noticeable effects on
the synthetic observables produced on scales  relevant for detailed
comparisons to data.  We are currently working on follow-up work in
which we will re-simulate a subset of the models from the sequence
presented here using the updated version of the \textsc{artis} code
developed by \cite{shingles2020a} which implements a full non-LTE
treatment of the excitation and ionization conditions in the plasma.
The results presented here will help prioritize models for further
study.     

\subsection{Angle-averaged light curves}
\label{subsec:angleav_lc}

\begin{figure}[h]
  \centering
  \includegraphics[width=1.\linewidth,trim={2.1cm 2.1cm 2.1cm 2.1cm},clip]{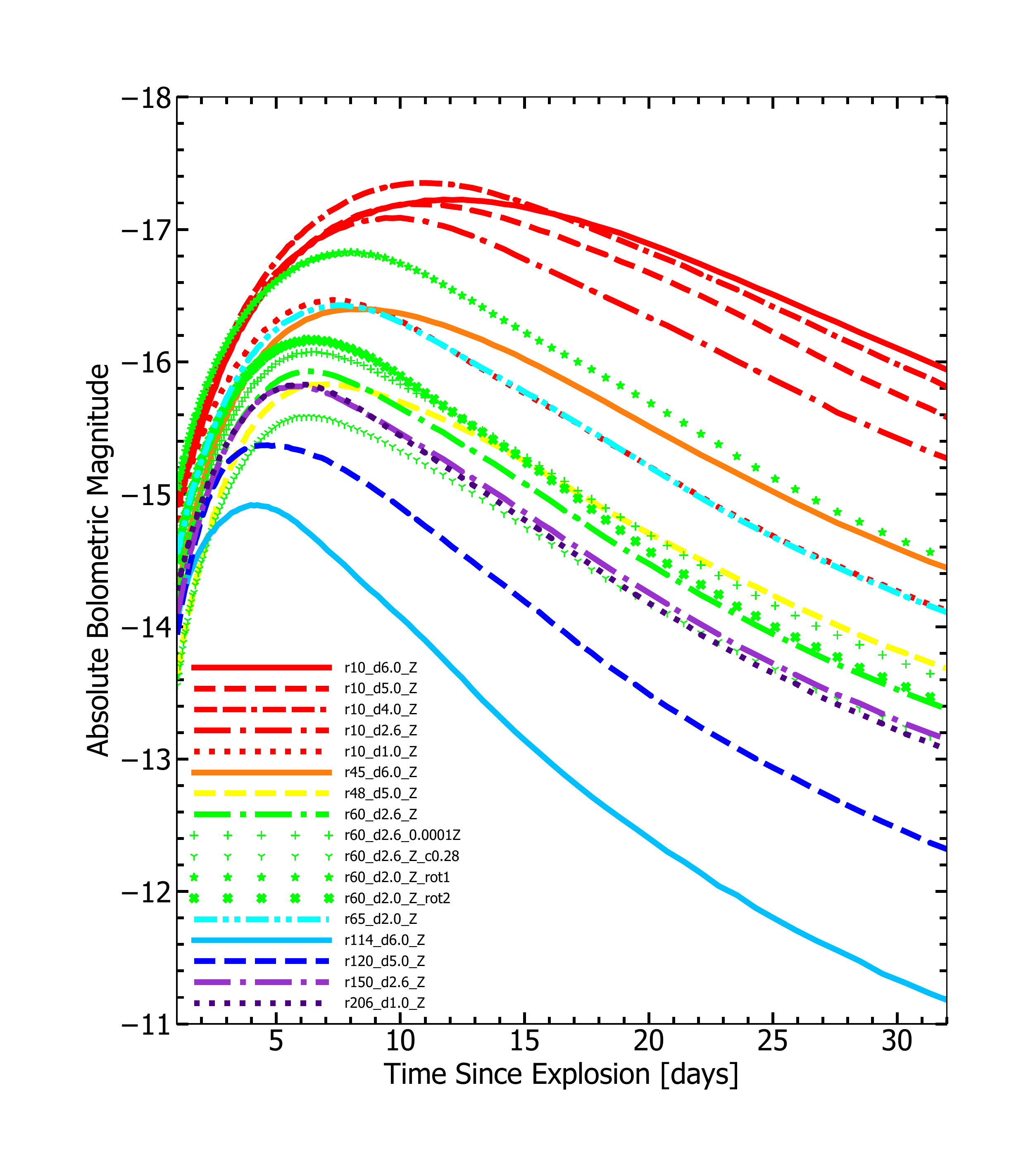}
  \caption{Angle averaged bolometric light curves for a selection of our
  models.}
  \label{fig:bol_lightcurves}
\end{figure}
 
\begin{figure*}[htbt]
  \centering
  \includegraphics[width=.98\linewidth,trim={0cm 0cm 1.5cm 1.5cm},clip]{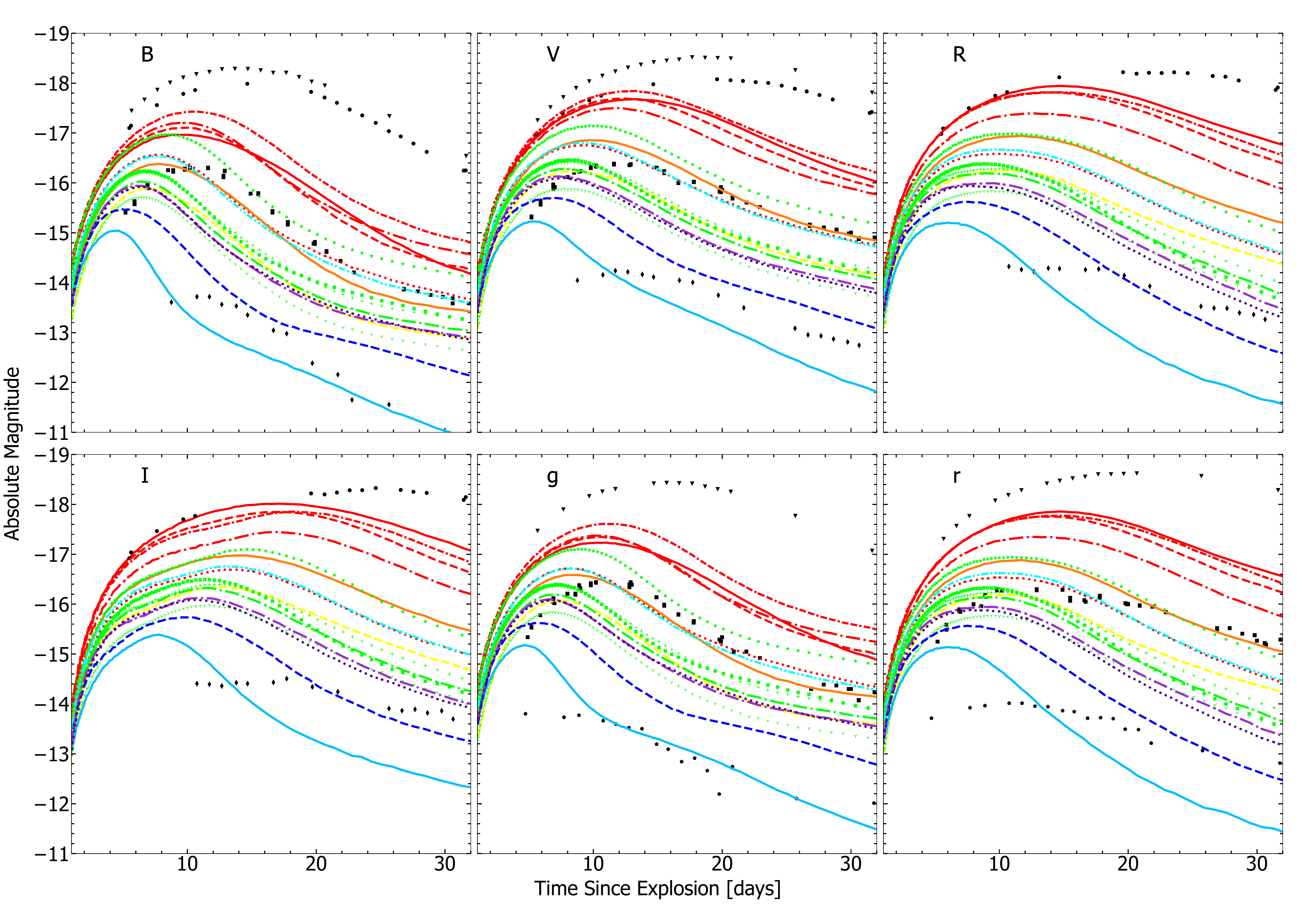}
  \includegraphics[width=.98\linewidth]{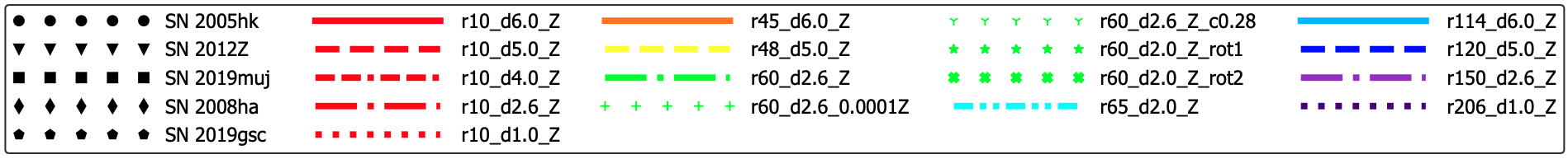}
  \caption{Angle averaged BVRIgr band light curves for selected models.
    Observed light curves of the SNe~Iax SN~2005hk
    (\citealp{phillips2007a}), SN~2012Z (\citealp{stritzinger2015a}), SN
    2019muj (\citealp{barna2021a}), SN~2008ha (\citealp{foley2009a}) and
    SN~2019gsc (\citealp{srivastav2020a} and \citealp{tomasella2020a})
    are included for comparison.}
  \label{fig:band_lightcurves}
\end{figure*}
 
Our bolometric light curves  are shown in
Fig.~\ref{fig:bol_lightcurves}. From this, it can be seen that the model
light curves show a clear relationship between their bolometric
evolution time scales and peak bolometric brightness, that is, the brighter
models are slower in rise and decline. This is in agreement with the
trend observed in the deflagration study of F14 driven by $^{56}$Ni
synthesized in the explosion. In addition to this, the trends discussed
in Sect.~\ref{sec:results} for the choice of initial conditions and
varied parameters, that is, ignition radius, central density, and
metallicity, are confirmed by the results of the RT simulations: in
general for a fixed central density the smaller the ignition radius of
the model the brighter and broader its bolometric light curve will be.
Additionally, for a fixed ignition radius the higher the central density
the brighter and broader the model light curve.  These trends, however,
are not uniform for the whole model sequence and break down for the
models with the highest central densities and also for those models
which have both high central density and large ignition radius (see
Sect.~\ref{sec:results} for a more detailed discussion). 

Fig.~\ref{fig:band_lightcurves} shows the BVRIgr band light curves for
the selected models. Example SNe~Iax are included for comparison.  We
emphasize that while we show observed SNe~Iax for comparison here, the
primary aim of this section is to comment on the overall light curve
properties of the model sequence. In
Sect.~\ref{subsec:best_agreement_model} we show a more detailed
comparison to an observed SNe~Iax.  The estimated explosion epochs used
for the observed SNe~Iax 2005hk, 2012Z, 2019muj, 2008ha, and 2019gsc in
Fig.~\ref{fig:band_lightcurves} are taken from previous studies,
specifically \cite{phillips2007a}, \cite{yamanaka2015a},
\cite{barna2021a}, \cite{valenti2009a}, and \cite{srivastav2020a},
respectively. We note that this approach of comparing all light curves
relative to literature values for explosion date may make the agreement
of the model light curves with observed SNe~Iax light curves seem
relatively poor. However, if we allow the freedom to shift the explosion
epochs of the observed SNe~Iax later by even $\sim 2$ days, (which is
reasonable on the scale of the uncertainties in the estimated explosion
epochs), significantly better agreement can be achieved between model
and observed light curves, particularly for intermediate luminosity
SNe~Iax such as SN~2019muj (see Sect.~\ref{subsec:best_agreement_model}
and Fig.~\ref{fig:r48_d5.0_Z_19muj_band_lc}).

From Fig.~\ref{fig:band_lightcurves} it can be seen that the light
curves show a similar relationship between their evolution timescales
and peak magnitude in all bands as was observed for the bolometric light
curves.  Some models do show small variations in their band colors.
However, these differences are relatively small and have little impact
on the overall trends observed for the suite of models. Furthermore, any
differences between models are too small to significantly impact the
agreement between the models and observed SNe~Iax. 

The models show reasonably good agreement between their rise time scales
and the data in both red and blue bands, although the red bands do
appear to match the rise to peak of observed SNe~Iax slightly better.
However, the decline post peak of the data is matched much better by the
models in the blue bands. The models are significantly too fast in the
red bands to compare favorably with the light curve evolution of the
data post peak (see also Fig.~\ref{fig:B_band_dm15_viewing_angles} and
Fig.~\ref{fig:rband_risetime_dm15_peakmag}). This is similar to the
trend observed previously in the deflagration study of F14. As noted
above, when we move from brighter to fainter models in the sequence the
models become faster in both rise and decline. This results in an
evolution of the light curves of the fainter models that is
significantly too fast to produce good agreement with the faintest
members of the SN~Iax class such as SN~2008ha and SN~2019gsc. 

The choice of initial conditions leads to a variation in the $^{56}$Ni
mass synthesized in the models and this is the characteristic which
overwhelmingly controls the light curve properties of the models in
terms of their absolute bolometric and band magnitudes and evolution
time scales, regardless of what initial conditions are chosen. However,
one parameter which leads to some variation in the light curve evolution
is the $^{56}$Ni mass to ejecta mass ratio. The model with the lowest
value of \mni/$M_\mathrm{ej}$, Model r48\_d5.0\_Z (see
Tab.~\ref{tab:ejectasum}, Tab.~\ref{tab:lightcurve_data_bessel}, yellow
dashed line in Fig.~\ref{fig:bol_lightcurves} and
Fig.~\ref{fig:band_lightcurves}), is slightly slower in both rise and
decline in the red bands than models with similar peak absolute band
magnitudes (e.g., Model r60\_d2.6\_Z, green dash-dot line).  While this
does improve agreement between the model light curve evolution compared
to the data in red bands (see r-band in Fig.~\ref{fig:band_lightcurves}
and Fig.~\ref{fig:rband_risetime_dm15_peakmag}) the red light curves for
this model are not slowed down sufficiently to account for the slower
light curve evolution of real SNe~Iax.
 
\subsection{Angle-averaged spectra}
\label{subsec:angleav_sp}

\begin{figure}[tp!]
  \centering
  \includegraphics[width=.98\linewidth,trim={0.05 1.02cm 0 0},clip]{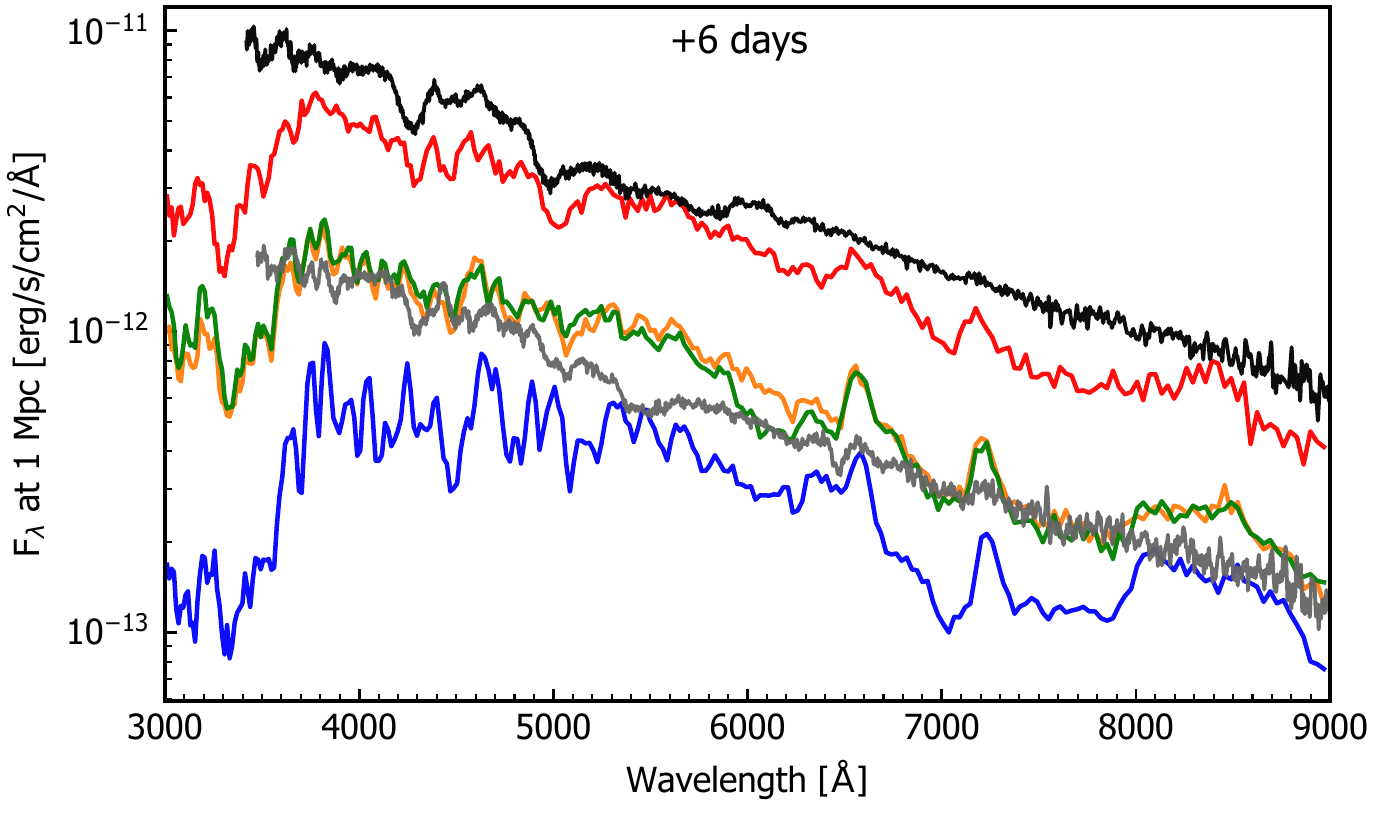}
  \includegraphics[width=.98\linewidth,trim={0.05 1.02cm 0 0},clip]{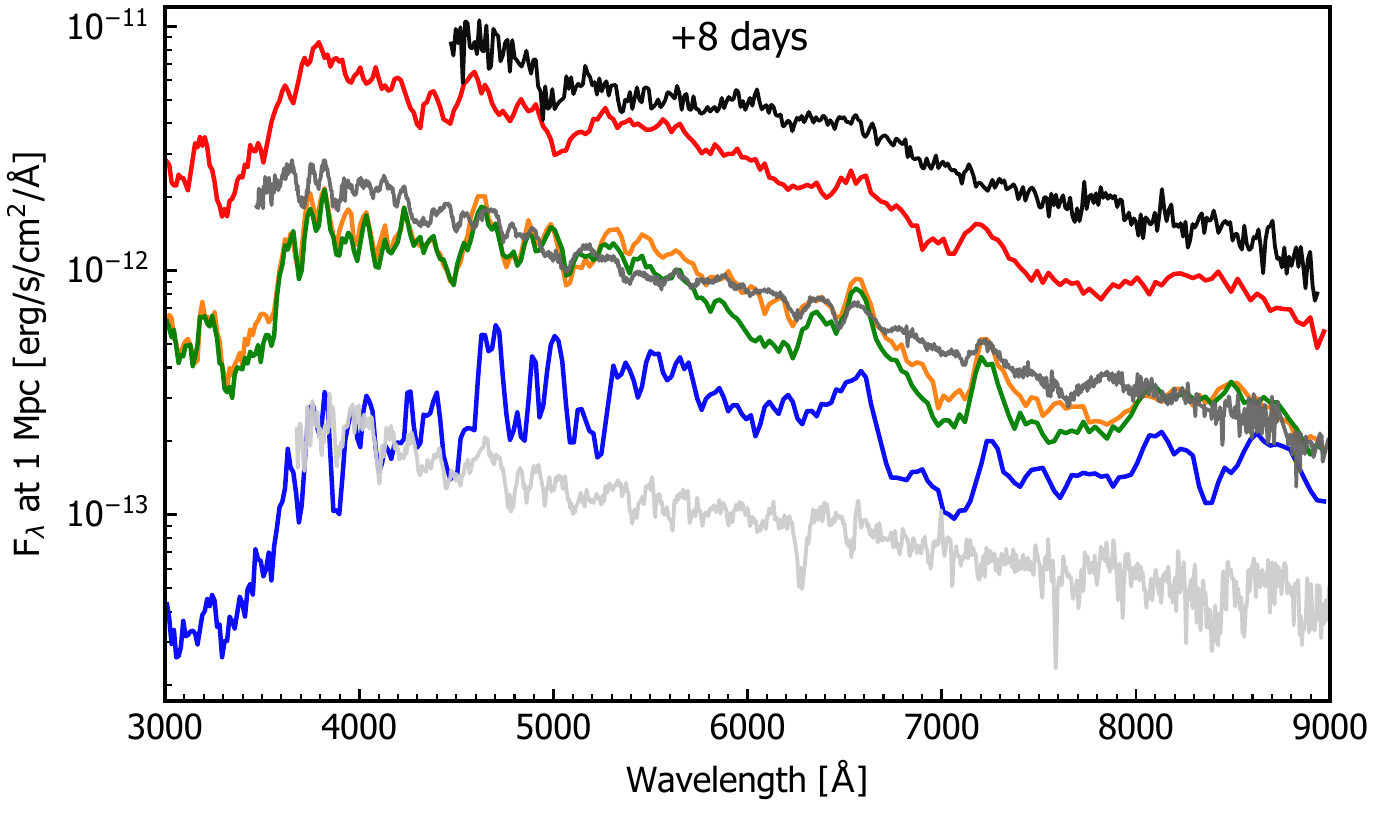}
  \includegraphics[width=.98\linewidth,trim={0.05 1.02cm 0 0},clip]{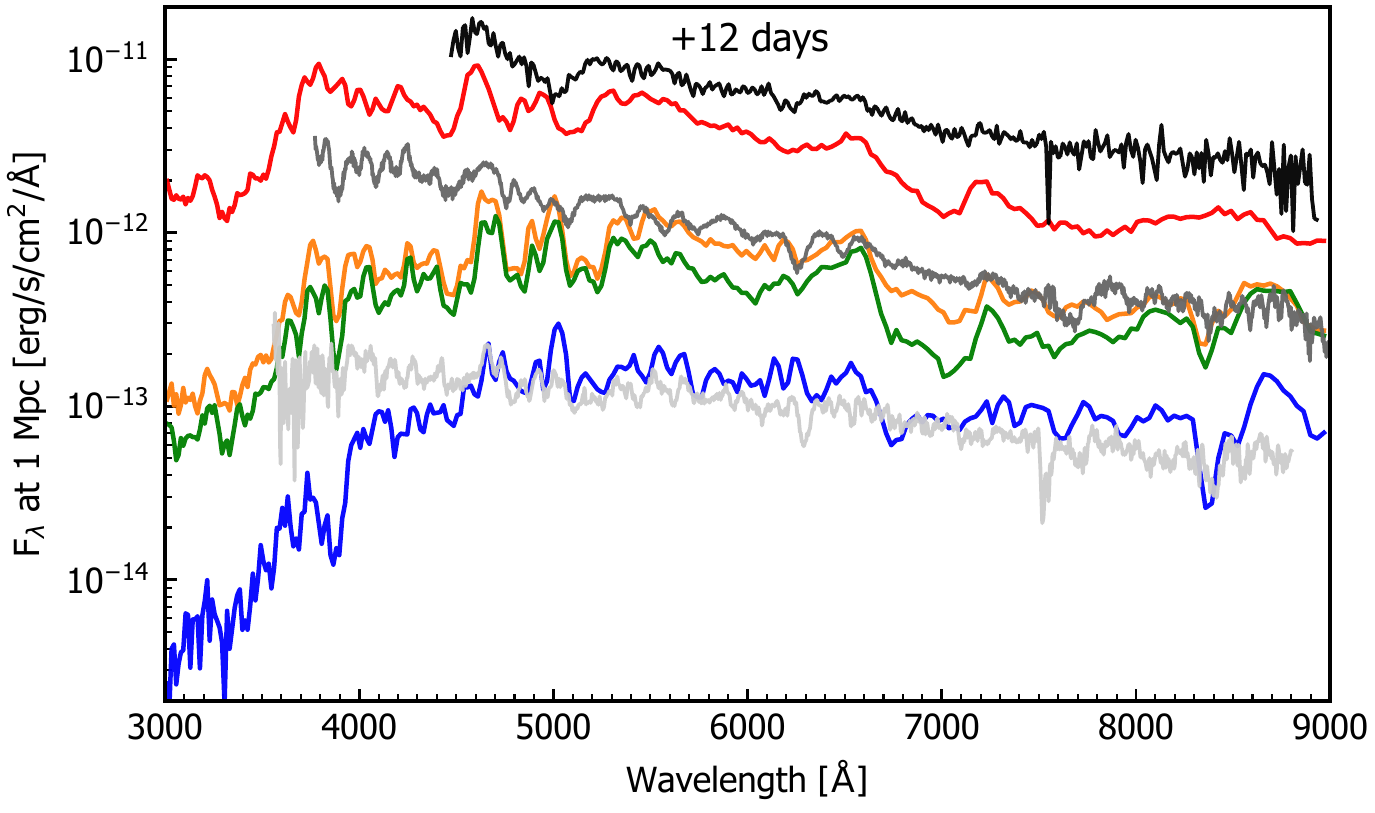}
  \includegraphics[width=.98\linewidth,trim={0.05 0.12cm 0 0},clip]{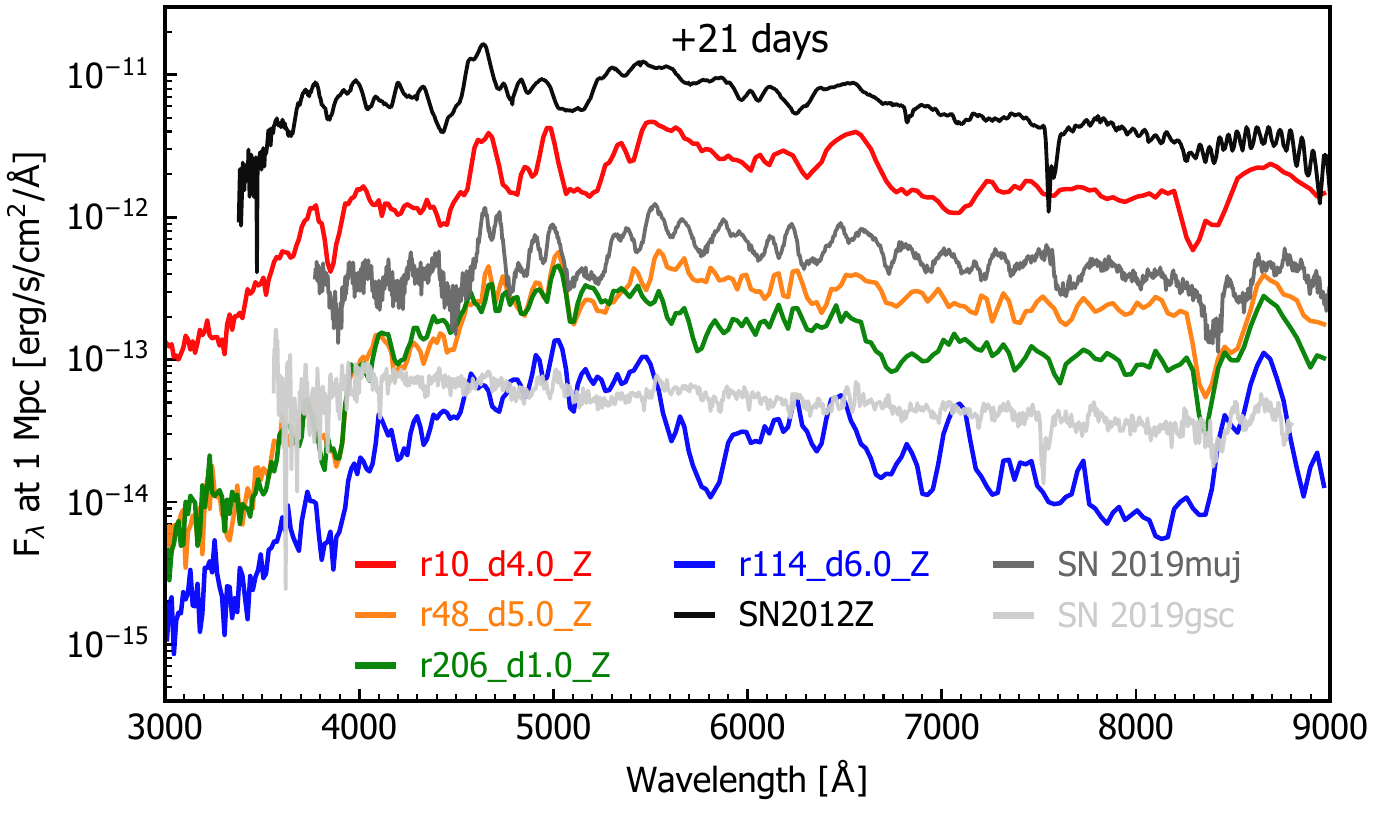}
  \caption{Spectra over a variety of epochs for selected models which
    range from the bright to faint end of the model sequence alongside
    three observed SNe~Iax: SN~2012Z (\citealp{stritzinger2015a}), one
    of the brighter members of the SN~Iax class, SN~2019muj
    (\citealp{barna2021a}) which is an intermediate brightness SNe~Iax,
    and SN~2019gsc (\citealp{srivastav2020a}) which is one of the
    faintest SNe~Iax (see Fig.~\ref{fig:bol_band_viewing_angles} and
    Fig.~\ref{fig:rband_risetime_dm15_peakmag}). The observed spectra
    shown here were obtained from WISeREP\protect\footnotemark
    (\citealp*{yaron2012a}). All times are relative to explosion and in
    all cases real flux is plotted on a logarithmic scale. There is no
    data for SN~2019gsc 6 days after explosion and in the lowest panel
    the spectra were taken 23 days after explosion for SN~2012Z and
  SN~2019gsc. All observed spectra are flux calibrated to match the
photometry and corrected for distance, red shift and reddening.}
  \label{fig:spectra}
\end{figure}

Fig.~\ref{fig:spectra} shows spectroscopic comparisons between a
selection of  models ranging from faintest (r114\_d6.0\_Z, blue) to
brightest (r10\_d4.0\_Z, red) and three observed SNe~Iax which span the
diversity in brightness of the SNe~Iax class. A variety of epochs from
pre-peak to post-peak are chosen such that comparisons between the
overall spectroscopic evolution of the model sequence and data can be
made. As in Sect.~\ref{subsec:angleav_lc} we note, that the purpose of
this section is not to make detailed comparisons between the spectra of
individual observed SNe~Iax and the best agreeing model (see instead
Sect.~\ref{subsec:best_agreement_model} for such a detailed comparison).
Therefore, again for consistency, all times in Fig.~\ref{fig:spectra}
are relative to explosion with the same estimates for explosion epochs
used for the data as referenced in Sect.~\ref{subsec:angleav_lc}.
 
\footnotetext{https://wiserep.weizmann.ac.il}
From Fig.~\ref{fig:spectra} we see that the bright and intermediate
luminosity models show best spectroscopic agreement with the data in
terms of their spectroscopic features for later epochs. The brightest
model (r10\_d4.0\_Z, red) shows good agreement with one of the brighter
SNe~Iax, that is, SN~2012Z, for the latest epoch shown in terms of its
spectroscopic features. In addition, the two intermediate luminosity
models (r48\_d5.0\_Z, orange and r206\_d1.0\_Z, green) agree well with
the intermediate luminosity SN~Iax, that is, SN~2019muj, in terms of their
spectroscopic features for all but the earliest epoch shown.  However,
while the bright and intermediate luminosity models match the slope and
overall flux profile of observed SNe~Iax spectra well at earlier epochs
(at later times the flux agreement is poorer due to the model light
curves evolving faster than the observed light curves), the synthetic
spectra show too many distinct spectral features compared to the data at
earlier times (6, 8, and 12 days after explosion for SN~2012Z and 6 days
after explosion for SN~2019muj).   

The faintest model (r114\_d6.0\_Z, blue) shows poor spectroscopic
agreement with the faint SN~Iax, SN~2019gsc.  This model has spectra
that have significantly too many spectral features compared to the data
at all epochs.  Therefore, the fainter models provide a much poorer
match to the spectroscopic evolution of the data than the bright and
intermediate luminosity models.  From Fig.~\ref{fig:spectra} it is also
clear that Model r114\_d6.0\_Z is noticeably spectroscopically different
to all the other models.  This model has significantly more absorption
in the blue region of the spectrum and more emission in the red
wavelengths compared to the rest of the models which are more typical of
the models in the sequence. This difference observed for Model
r114\_d6.0\_Z  is due to greater absorption by singly ionized IGEs such
as iron, cobalt, nickel, and chromium in the blue region of the spectrum
and subsequent re-emission in the red part of the spectrum. Since it is
the faintest in the suite of models this characteristic can be
attributed to this model having cooler ejecta which leads to a greater
fraction of IGEs being in a lower ionization state.  

Overall, we can see that all our models are spectroscopically very
similar across all epochs, although the faintest model shows more
noticeable differences. Therefore, the spectroscopic differences between
models are much smaller than differences between models and data,
although spectroscopic agreement between the models and data is
reasonably good for bright and intermediate luminosity models
particularly at later epochs. However, the poorer spectroscopic
agreement at earlier times between models and data, even for the bright
and moderate luminosity models shows there is disagreement in the early
phase spectroscopic evolution of the models compared to data. This
suggests that there are systematic differences in the spectroscopic
evolution of the models and data which these simulations do not account
for. We also note, that while the brightest model in our sequence does
not quite reach the flux level of bright SNe~Iax such as SN~2012Z,
previous work by K13 and F14 has already shown that pure deflagration
models are able to produce reasonable agreement with bright observed
SNe~Iax (although there are still some spectroscopic systematic
differences). To confirm these findings still hold true when using the
updated version of the \textsc{leafs} code we re-simulated the N5def
model from K13 and F14 and simulated the RT using \textsc{artis}.  The
results obtained are consistent with those reported by K13 and F14.

\subsection{Viewing-angle dependencies}
\label{subsec:viewingangle}

\begin{figure*}[htbt]
  \centering
  \includegraphics[width=0.48\linewidth,trim={0cm 0cm 0cm 0cm},clip]{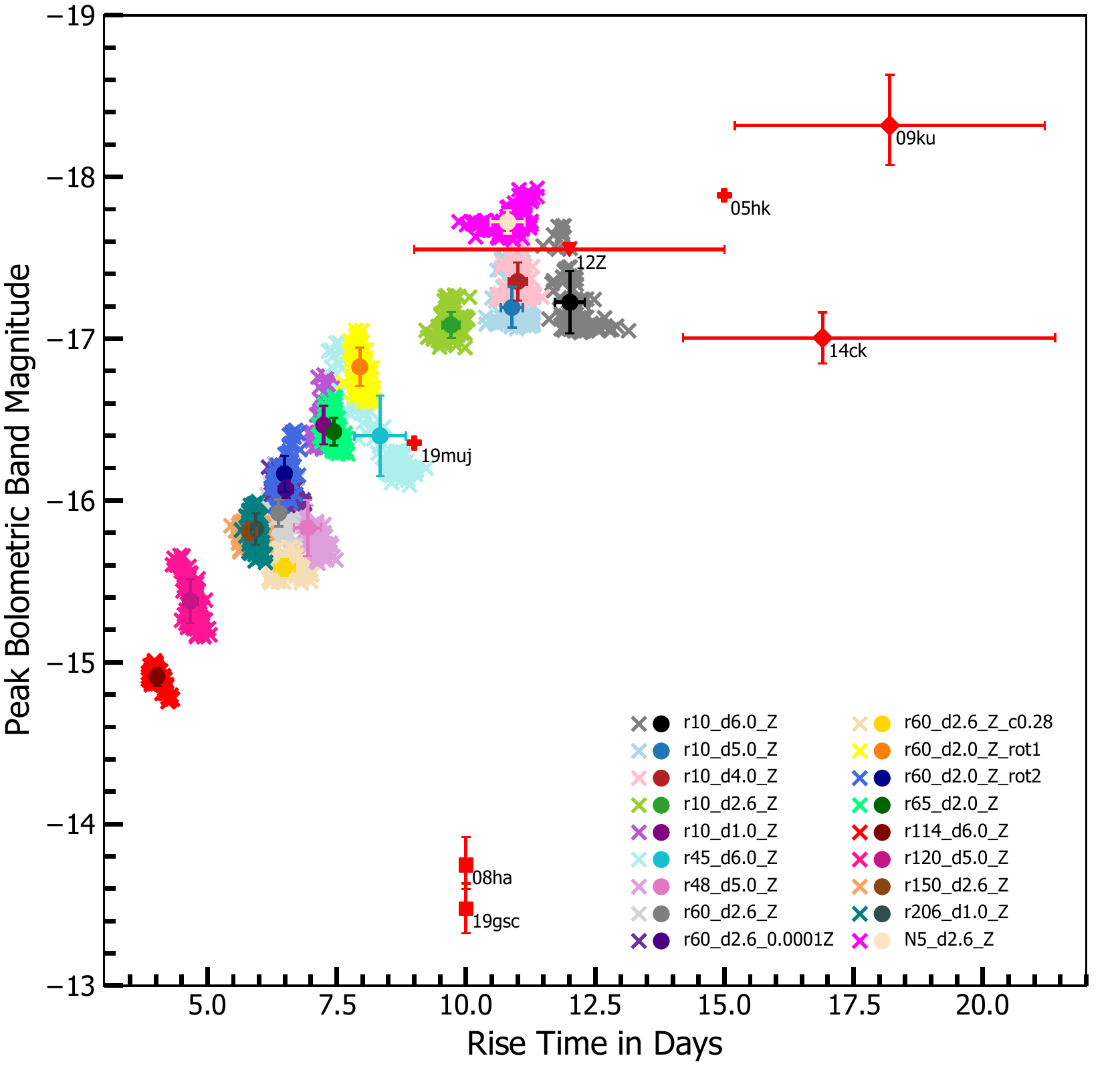}
  \includegraphics[width=0.48\linewidth,trim={0cm 0cm 0cm 0cm},clip]{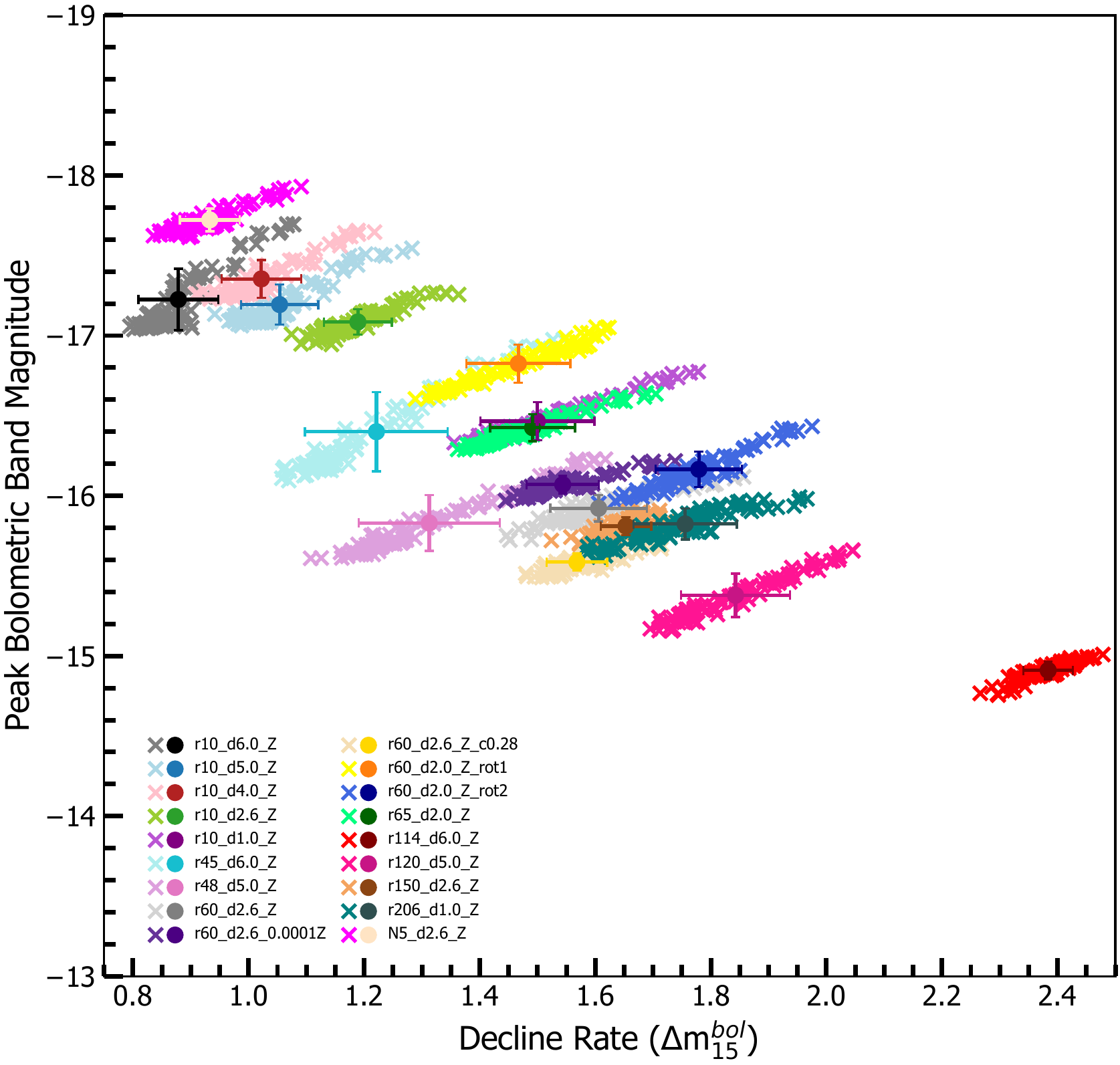}
  \caption{Viewing angle dependencies of the bolometric light curve
    properties for selected models. The solid circles represent the
    angle-averaged values and the crosses represent values from 100
    different viewing angles for each model. The error bars depict the
    standard deviation of the viewing angle distribution for each model.
    In the left panel the peak bolometric-band magnitude is plotted
    against rise time. Alongside the models, estimates of peak
    bolometric magnitude and rise times for the observed SNe~Iax
    SN~2019gsc (\citealp{srivastav2020a}, taking the values calculated
    for their full blackbody bolometric light curve), SN~2008ha
    (\citealp{foley2009a}), SN~2019muj (\citealp{barna2021a}), SN~2014ck
    (\citealp{tomasella2016a}), SN~2012Z (\citealp{yamanaka2015a}),
    SN~2005hk (\citealp{phillips2007a}), and SN~2009ku
    (\citealp{narayan2011a}) are included (represented by red markers
    and labeled by name). The different marker styles representing the
    SNe~Iax of diamond, square, triangle, and plus correspond to SNe~Iax
    where errors were quoted for both rise time and peak bolometric
    magnitude, only for peak bolometric magnitude, only for rise time
    and no errors quoted respectively. Where uncertainties
    were not quoted the uncertainties are still significant,
    particularly in rise times that can have uncertainties as much as
    several days. The right hand panel shows peak bolometric-band
    magnitude v.s decline rate in terms of ${\Delta}m_{15}^{bol}$.}
  \label{fig:bol_band_viewing_angles}
\end{figure*}

Fig.~\ref{fig:bol_band_viewing_angles} illustrates the viewing angle
dependencies of the bolometric light curve properties (peak absolute
bolometric magnitude, bolometric rise time to peak and bolometric
decline post peak) for the selection of models for which RT simulations
were carried out. In addition, Fig.~\ref{fig:B_band_dm15_viewing_angles}
shows the viewing angle dependencies of B-band peak absolute magnitude
and rise time for the selected models and includes observed SNe~Iax for
comparison. As can be seen from Fig.~\ref{fig:bol_band_viewing_angles}
and Fig.~\ref{fig:B_band_dm15_viewing_angles} the models occupy a
one-dimensional sequence where the brighter the model the slower its
rise to peak and decline post peak (the models of K13 and F14 also lie
on this sequence). As discussed in Sect.~\ref{subsec:angleav_lc} this
trend is driven by the $^{56}$Ni mass synthesized in the models with
secondary parameters having little impact on this overall trend.
However, there can be variations of up to a magnitude at peak in both
the bolometric and band light curves for certain models due to viewing
angle effects, as well as significant variation in light curve evolution
in both rise and decline. In particular, from
Fig.~\ref{fig:bol_band_viewing_angles} (right panel) we can see that
there is a clear correlation between the peak bolometric brightness of
the model viewing angles and their rate of decline: the brighter viewing
angles have faster declining light curves post peak. An explanation for
this is that the brighter viewing angles correspond to those in which
there is a higher concentration of $^{56}$Ni near to the surface of the
ejecta.  Therefore the radiation due to the decay of this $^{56}$Ni will
have to travel through less material to escape the ejecta leading to a
faster evolution post peak.

\subsubsection{Viewing-angle dependencies of nonrotating models}
\label{subsec:non-rotating_viewingangle}

The nonrotating model spectra do not show particularly strong viewing
angle dependencies. The most noticeable effect is the velocity shifts of
spectral lines: spectral features exhibit different blue shifts
depending on the direction they are viewed from.  The velocity shifts of
spectral features can differ by up to ${\sim}\,1000\,\si{km.s^{-1}}$. For
all models the faint viewing angles have spectral features which are
more blue shifted than for the bright viewing angles. Towards the blue
wavelengths (less than $\sim 6000\,$\r{A}) the bright models show less
strong absorption than the faint models which is in better agreement
with what is observed in the data (see Fig.~\ref{fig:spectra}). However,
this effect is still too small to significantly improve the agreement of
any of the individual models with the spectra of observed SNe~Iax.
Conversely, the red wavelengths show only very small differences in
their flux when observed from different viewing angles. Variations in
flux between different viewing angles can however lead to improved
agreement with the overall flux of the spectra of an observed SNe~Iax
for selected viewing angles.

The nonrotating models also show some viewing angle dependencies in
their light curves. Viewing-angles closer to the ignition spark (which
is always on the positive $x$-axis for the nonrotating models) tend to
be fainter, whereas those viewing angles further from the ignition spark
tend to be the brightest. This is the case for all models for which RT
simulations were carried out except the two faintest models in the
sequence (r114\_d6.0\_Z and r120\_d5.0\_Z) which, by contrast, have
brighter viewing angles closer to the ignition spark and fainter viewing
angles further from the ignition spark.  These two models have a
noticeably different structure from the other models in the sequence
(see r120\_d5.0\_Z compared to the other models shown in
Fig.~\ref{fig:ejecta} that have structures more representative of the
rest of the models in the sequence).

\begin{figure}[htbt]
  \centering
  \includegraphics[width=1.0\linewidth,trim={0cm 0cm 0cm 0cm},clip]{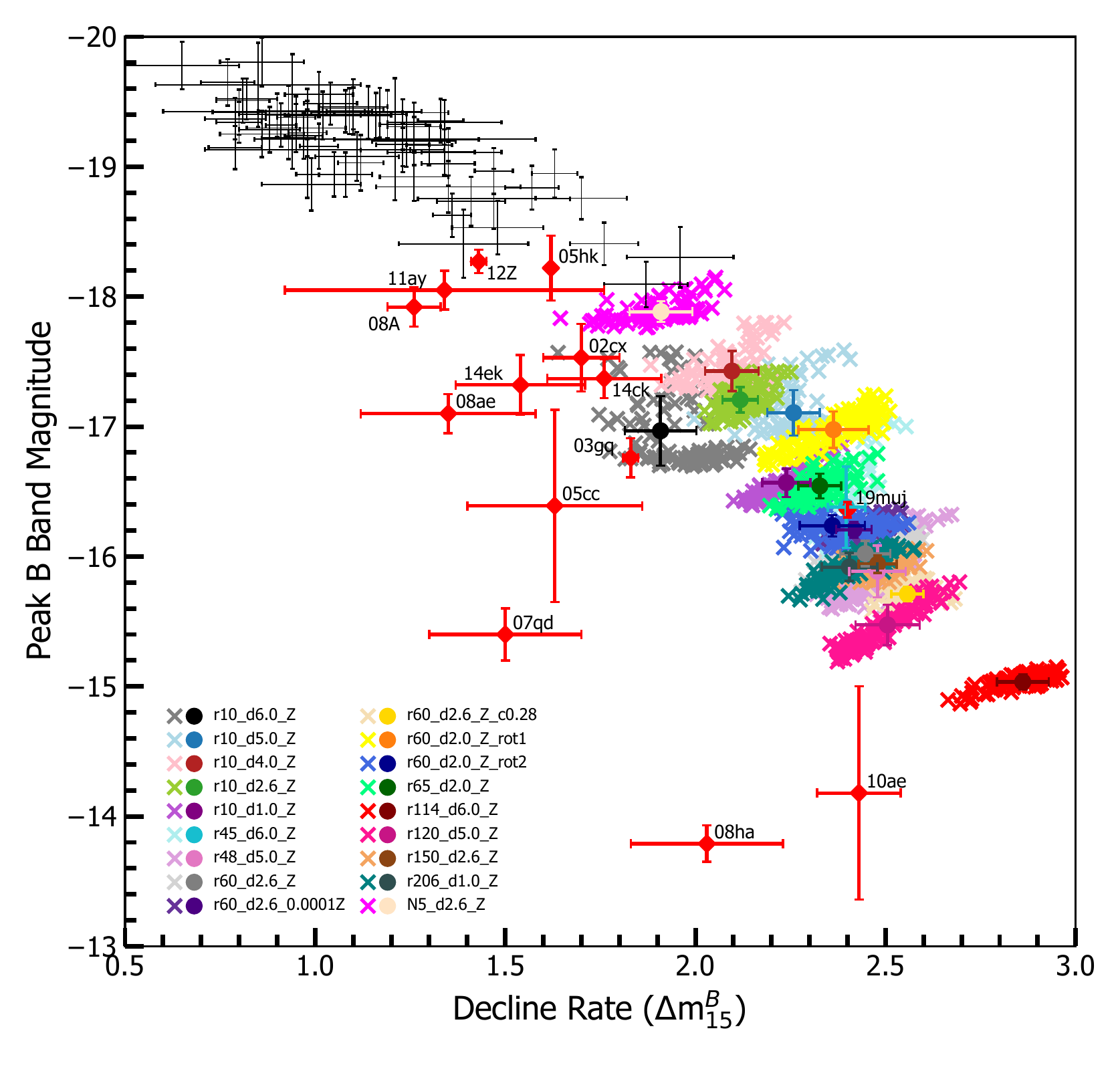}
  \caption{B-band magnitude v.s ${\Delta}m_{15}^{B}$ plotted in the same
  way as in Fig.~\ref{fig:bol_band_viewing_angles}. Also included is
  data compiled by \cite{taubenberger2017a}: normal SNe~Ia are
  shown as black crosses and SNe~Iax (as in
  Fig.~\ref{fig:bol_band_viewing_angles}) are labeled by name and
  represented by red diamonds. Additional SNe~Iax have also been added:
  SN~2014ck, SN~2019muj (see Fig.~\ref{fig:bol_band_viewing_angles} for
  references), and SN~2014ek (\citealp{li2018a}).  B-band is chosen as
viewing angle effects are easier to identify in the blue bands.}
  \label{fig:B_band_dm15_viewing_angles}
\end{figure}

In general, it is clear that viewing angle effects are modest but
noticeable even for well mixed pure deflagration models. These viewing
angle effects, while unable to explain all the differences between our
models and observed SNe~Iax, are relevant on the scale of comparisons
between observed SNe~Iax and the models. This can clearly be seen from
Fig.~\ref{fig:B_band_dm15_viewing_angles} where the variation of
${\Delta}m_{15}^B$ observed for individual models is comparable to the
differences in ${\Delta}m_{15}^B$ between observed SNe~Iax, particularly
for the brighter SNe~Iax. In addition, the spread of ${\Delta}m_{15}^B$
values for the bright models due to viewing angle effects puts certain
model viewing angles in better agreement with observed SNe~Iax than the
angle-averaged values for the models.  Finally, we note that neither the
models with different metallicity nor the carbon-depleted model show any
noticeable differences in their viewing angle dependencies compared to
the other models in the sequence.  

\subsubsection{Viewing-angle dependencies of rigidly rotating models }
\label{subsec:rigidrotationsRT}

As discussed in Sect.~\ref{subsec:rot} the rigidly rotating models are
sensitive to whether the model is ignited perpendicular to the rotation
axis (r60\_d2.0\_Z\_rot1) or along the rotation axis
(r60\_d2.0\_Z\_rot2).  The different choice of the ignition location in
the rotational models leads to a variation in their brightnesses and
produces a noticeably different ejecta structure for each model (see
Fig.~\ref{fig:ejecta_rot}). The rigidly rotating models exhibit most
significant differences with the nonrotating models in their viewing
angle dependent spectra. Like the nonrotating models the rigidly rotating
models show differences in their spectra depending on what line of sight
they are viewed from, with the most noticeable differences in flux being
seen  in the blue wavelengths. Additionally, as was the case for the
nonrotating models, the rigidly rotating models show spectral features
which are more blue shifted for faint viewing angles and less blue
shifted for bright viewing angles. However, the rigidly-rotating models
show angle dependent differences between the blue shifts of their
spectral features of over $2000\,\si{km.s^{-1}}$; double the difference
in velocity shifts observed for the spectral features of the nonrotating
models.  The rigidly rotating models therefore exhibit significantly
more asymmetry in their ejecta velocities compared to the nonrotating
models.  These models do not, however, show larger variations in their
viewing angle dependent light curves compared to the nonrotating models.  

\subsection{Best agreeing model comparisons with SN~2019muj}
\label{subsec:best_agreement_model}

\begin{figure*}[htbt]
  \centering
  \includegraphics[width=.98\linewidth,trim={0cm 0cm 1.5cm 1.5cm},clip]{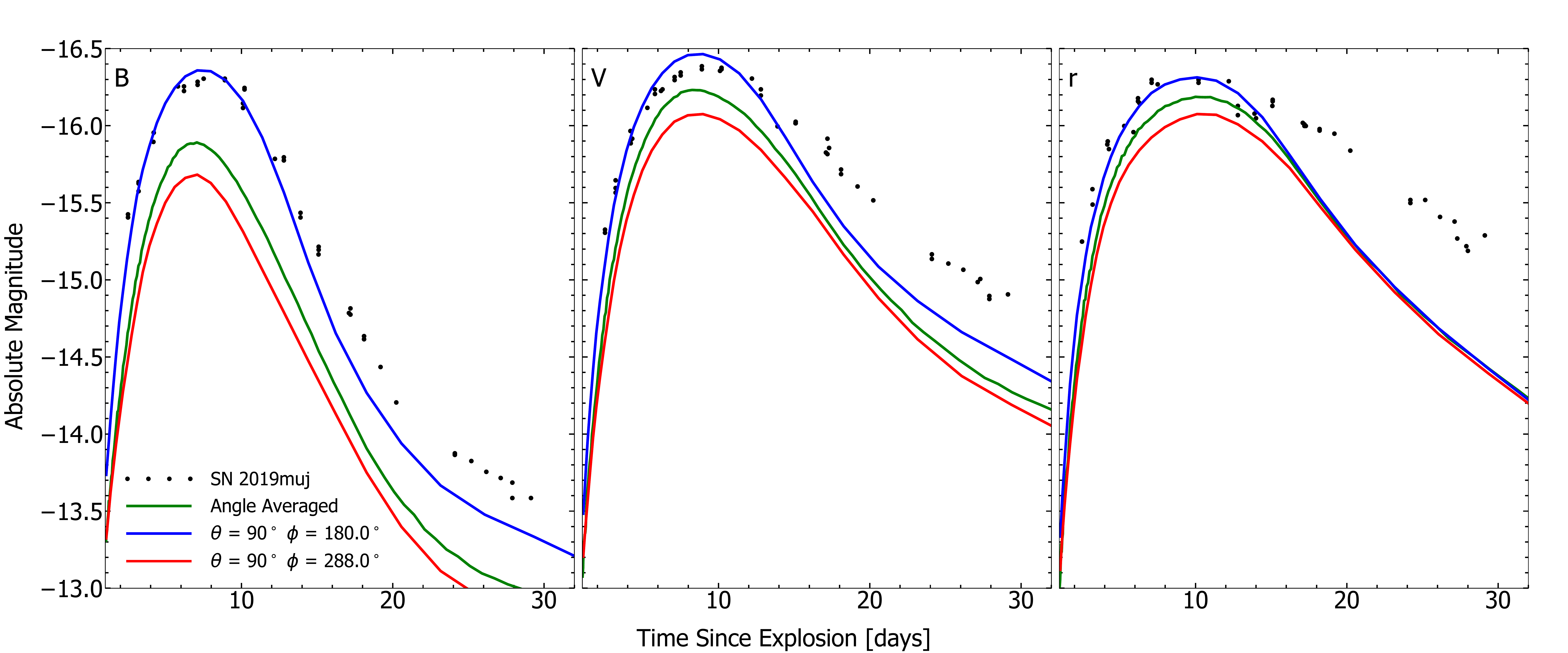}
  \caption{BVr band light curves for Model r48\_d5.0\_Z 
  (angle-averaged as well as a faint and bright viewing angle)
  compared with SN~2019muj \citep{barna2021a}. Model r48\_d5.0\_Z 
  produces the best agreement in terms of its light curve and spectra compared to 
  SN~2019muj. The model light curves are plotted relative to the explosion 
  time predicted by our simulations, whereas
  the band light curves of SN~2019muj have all been shifted 2 days earlier 
  than the explosion epoch estimated by \cite{barna2021a} such that SN~2019muj 
  matches the brightest viewing angle shown (blue) for the model at B peak.
  The viewing angle looking along the axis that the ignition spark
  is placed is ${\theta = 90^{\circ}~\phi = 0^{\circ}}$ meaning the fainter 
  viewing angle shown (red) is looking almost along this axis whereas 
  the brighter viewing angle is looking along the axis directly opposite.}
  \label{fig:r48_d5.0_Z_19muj_band_lc}
\end{figure*}

In this section, we compare the intermediate luminosity SN~Iax,
SN~2019muj (\citealp{barna2021a}), with the model from our sequence in
best agreement, to allow a more detailed comparisons between the light
curves and spectra of one of our models and an observed SNe~Iax. We have
chosen SN~2019muj as it provides a good brightness match to the
intermediate luminosity models in our sequence and also has good quality
spectra to compare to (also including earlier epochs).
Fig.~\ref{fig:r48_d5.0_Z_19muj_band_lc} shows the BVr light curves of
SN~2019muj and the best agreeing model, r48\_d5.0\_Z. The angle-averaged
as well as light curves for both a faint and bright viewing angle for
this model are included.  The model light curves are plotted relative to
explosion time while we have shifted the band light curves of SN~2019muj
such that their explosion epoch occurs 2 days later (MJD 58700) compared
to the estimate from \cite{barna2021a} based on early light curve
fitting, spectral fitting and bolometric light curve fitting (MJD
58697.5 - 58698.1).  This means the B band peak of SN~2019muj matches
the brightest viewing angle shown for Model r48\_d5.0\_Z (blue).
\cite{barna2021a} do not include uncertainties in their estimates of the
explosion epoch, however the first \textsc{atlas} detection of
SN~2019muj is on MJD 58702.5 with only a marginal 1.8$\sigma$ detection
before this on MJD 58700.5. We therefore argue that shifting the
explosion epoch 2 days later than what was estimated by
\cite{barna2021a} is justified within the uncertainty of the explosion
epoch of SN~2019muj.  We have flux calibrated the spectra of SN~2019muj
to match the photometry and corrected the light curves and spectra of
SN~2019muj for redshift, reddening, and distance (taking the values
estimated by \citealp{barna2021a} of z = 0.007035, E(B - V) = 0.02, d =
34.1 Mpc and using $R_v$ = 3.1).  As discussed in
Sect.~\ref{subsec:angleav_lc}, this model has the lowest value of
\mni/$M_\mathrm{ej}$ which results in it being slightly slower to rise
and decline in the red bands compared to models with similar peak
absolute band magnitudes.  This leads to slightly better agreement with
SN~2019muj. However, the improvement in agreement for the Model
r48\_d5.0\_Z compared to other models of similar brightness (e.g., Models
r10\_d1.0\_Z, r60\_d2.0\_Z\_rot2, and r45\_d6.0\_Z) is small and the same
conclusions are reached if we compare to one of these models instead.

Fig.~\ref{fig:r48_d5.0_Z_19muj_band_lc} demonstrates the importance of
taking into account the viewing angle dependencies of our models.  The
brightest viewing angle shown (blue) has band light curves in noticeably
better agreement with those of SN~2019muj than both the fainter viewing
angle (red) and angle-averaged band light curves (green).  Additionally,
these brightest viewing angle band light curves are noticeably brighter
than both the angle-averaged and faint viewing angle band light curves.
This is especially true in B band where it is brighter than the angle
averaged light curve by almost half a magnitude and more than this
compared to the faint viewing angle shown. From
Fig.~\ref{fig:r48_d5.0_Z_19muj_band_lc} we see the brightest viewing
angle shown for the r48\_d5.0\_Z model matches the rise to peak of
SN~2019muj very well in all bands. Moreover, this viewing angle matches
the decline of SN~2019muj well in all bands until approximately 5 days
after B peak.  Therefore, up until this time, this viewing angle
provides a very good match to the colors of SN~2019muj.  After this
time, the decline of SN~2019muj is matched much better by the model in
the B band than the V and particularly r band where the model light
curves decline too quickly compared to those of SN~2019muj.  As
discussed in Sect.~\ref{subsec:angleav_lc} (see also
Sect.~\ref{subsec:comparisons_to_observations}) this is a systematic
discrepancy which effects all models in the sequence. 

\begin{figure*}[htbt]
  \centering
  \includegraphics[width=.90\linewidth,trim={0.05 1.00cm 0 0},clip]{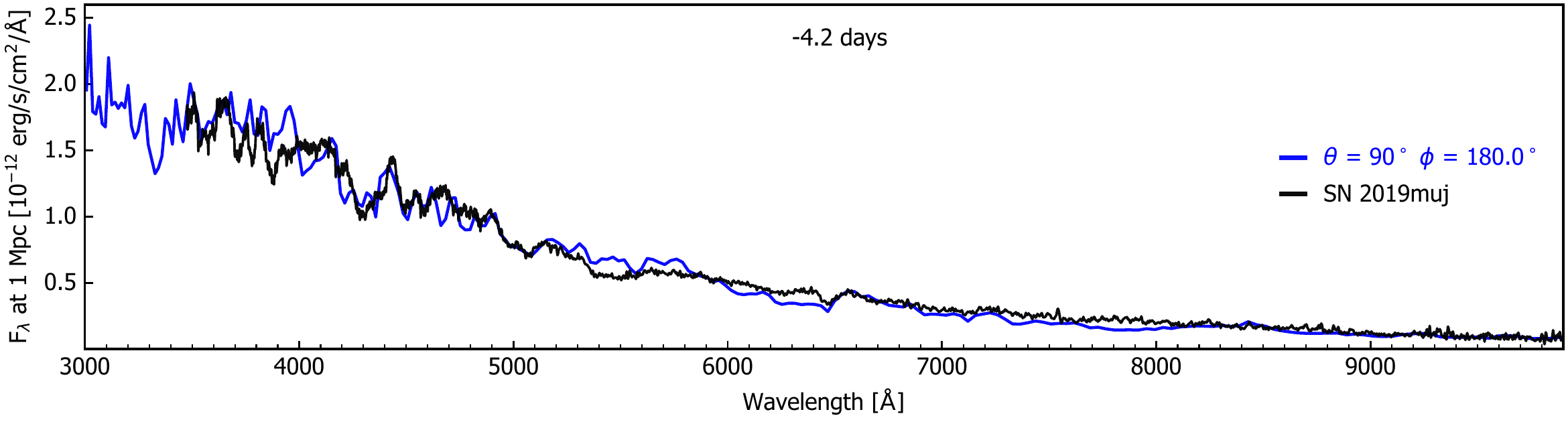}
  \includegraphics[width=.90\linewidth,trim={0.05 1.01cm 0 0},clip]{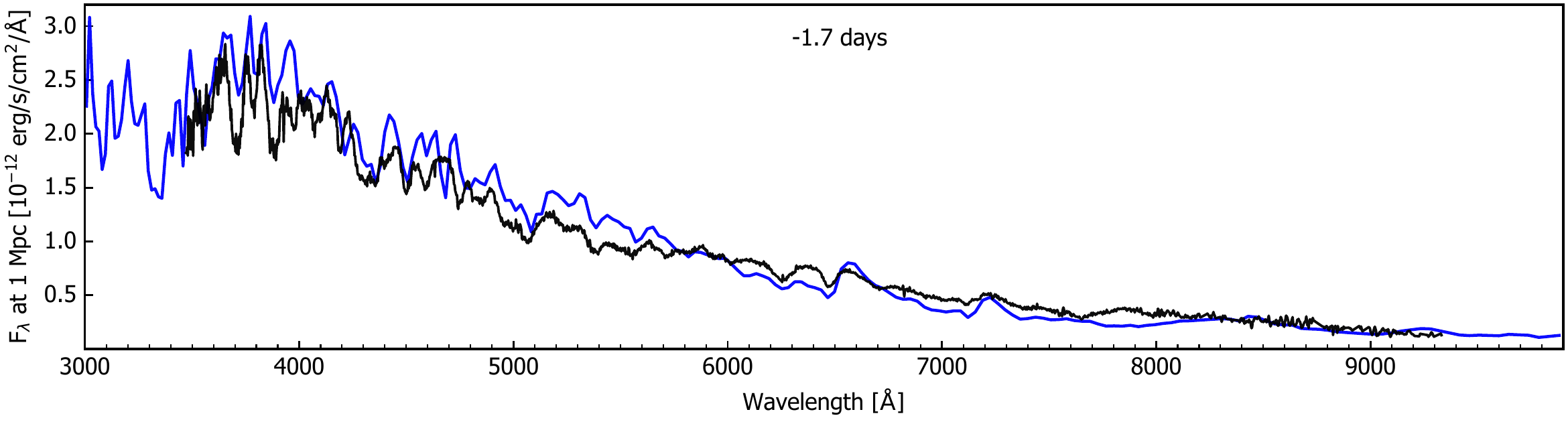}
  \includegraphics[width=.90\linewidth,trim={0.05 1.01cm 0 0},clip]{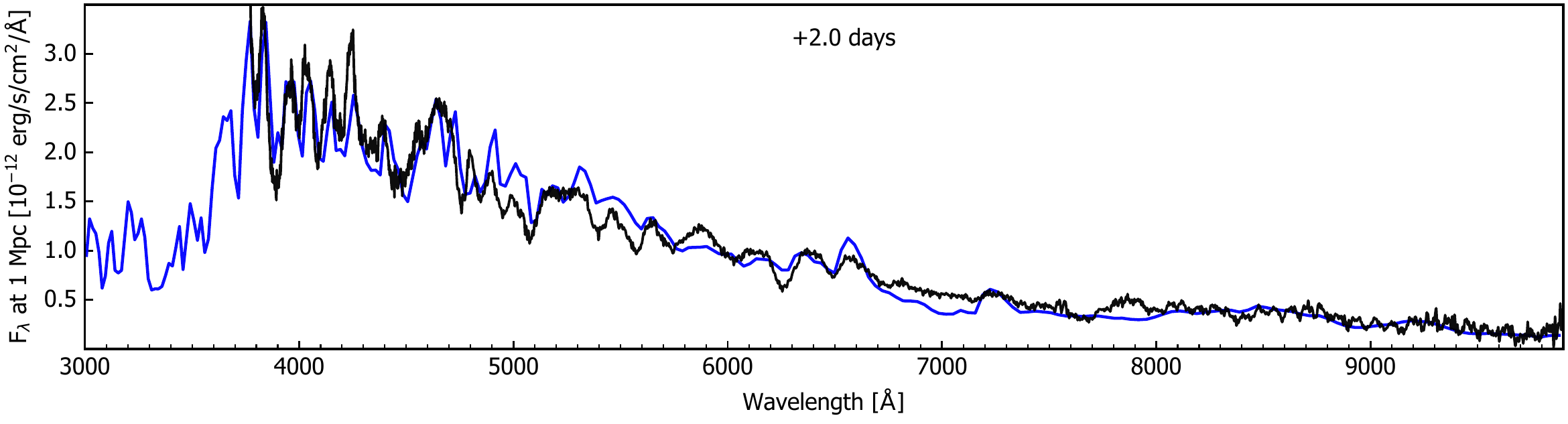}
  \includegraphics[width=.90\linewidth,trim={0.05 0.08cm 0 0},clip]{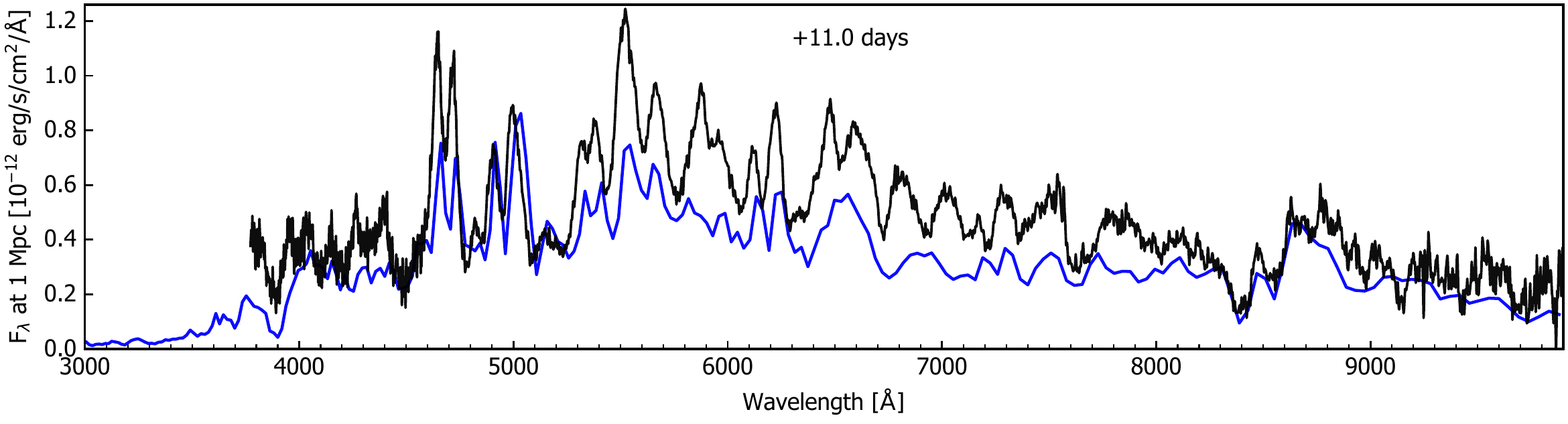}
  \caption{Spectroscopic comparisons in absolute flux between SN~2019muj and 
  the viewing angle for Model r48\_d5.0\_Z
  (blue in Fig.~\ref{fig:r48_d5.0_Z_19muj_band_lc}). 
  Times are relative to B band maximum.}
  \label{fig:r48_d5.0_Z_19muj_spectra}
\end{figure*}

Fig.~\ref{fig:r48_d5.0_Z_19muj_spectra} shows spectroscopic comparisons
over a variety of epochs between SN~2019muj (black) and the viewing
angle for Model r48\_d5.0\_Z in best agreement with SN~2019muj (blue in
Fig.~\ref{fig:r48_d5.0_Z_19muj_band_lc}). From
Fig.~\ref{fig:r48_d5.0_Z_19muj_spectra} we see that the model produces
very good agreement in terms of the overall flux and spectral shape
compared to SN~2019muj at all epochs apart from the latest epoch shown.
For this latest epoch the model is too faint as we move to red
wavelengths because the model declines faster than the data in the red
bands at later epochs.  The model is also able to successfully reproduce
a significant proportion of the spectral features observed for SN
2019muj both in terms of their strength and location, although for the
earliest epoch shown (4.2  days before B peak) the model does show
slightly too many distinct spectral features compared to SN~2019muj (see
also discussion in Sect.~\ref{subsec:angleav_sp}). We note that our
synthetic spectra produce C features (at 4268, 4746, 6580, and
7234$\,${\AA}) that are broadly consistent with observations of
SN~2019muj (\citealp{barna2021a}).  In particular, unlike the uniform
composition model tested by \cite{barna2021a}, we do not find the carbon
features are significantly too strong at epochs around peak.  Overall,
from Fig.~\ref{fig:r48_d5.0_Z_19muj_band_lc} and
Fig.~\ref{fig:r48_d5.0_Z_19muj_spectra} it is clear that our models are
able to reproduce many of the observed characteristics of the light
curves, and, in particular, spectra of intermediate SNe~Iax such as SN
2019muj very well, although some systematic differences remain.

\subsection{Overall model sequence comparisons to observations}
\label{subsec:comparisons_to_observations}

\begin{figure*}[htbt]
  \centering
  \includegraphics[width=0.48\linewidth, trim={1cm 1.5cm 1cm 1.5cm},clip]{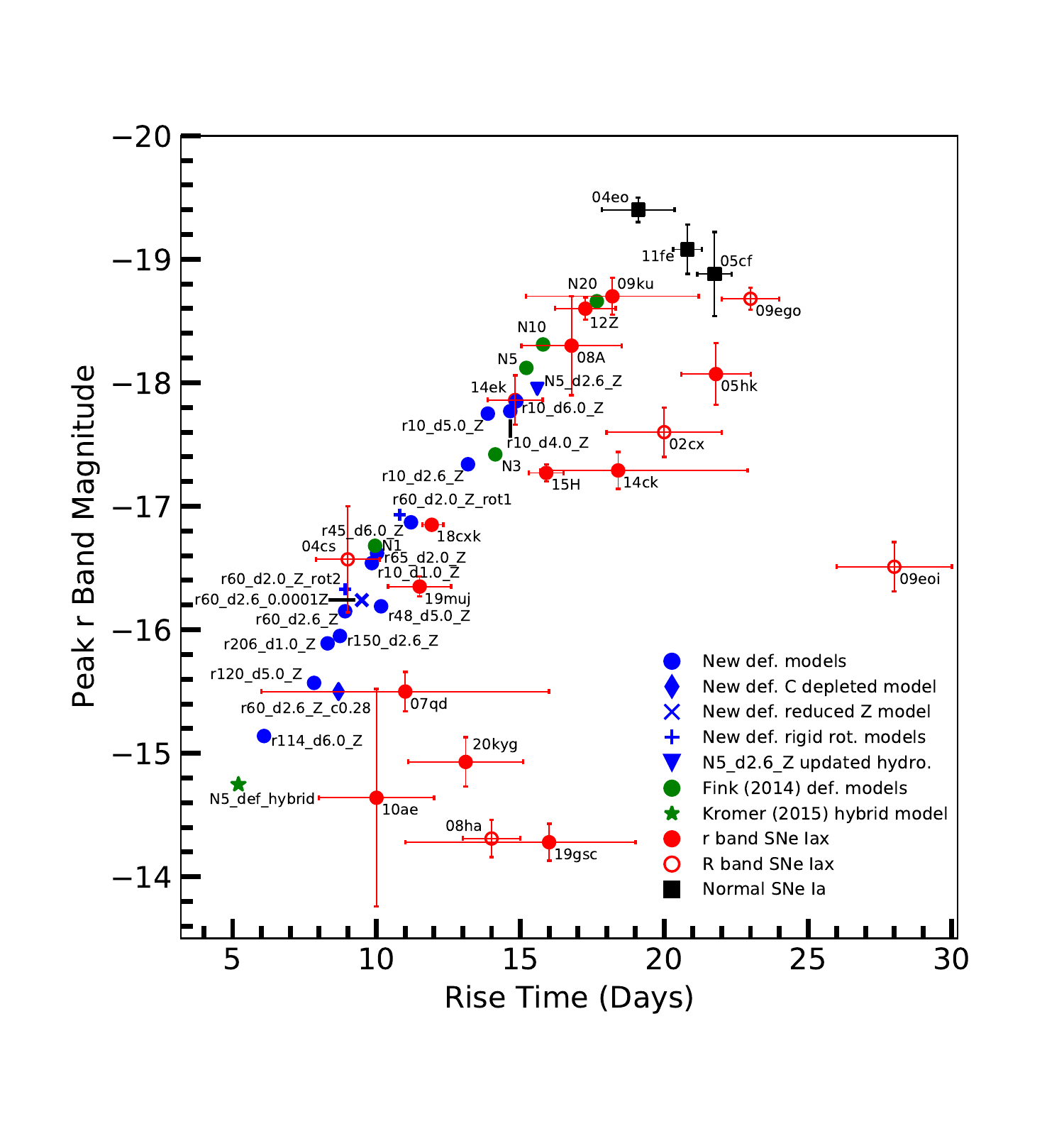}
  \includegraphics[width=0.48\linewidth, trim={1cm 1.5cm 1cm 1.5cm},clip]{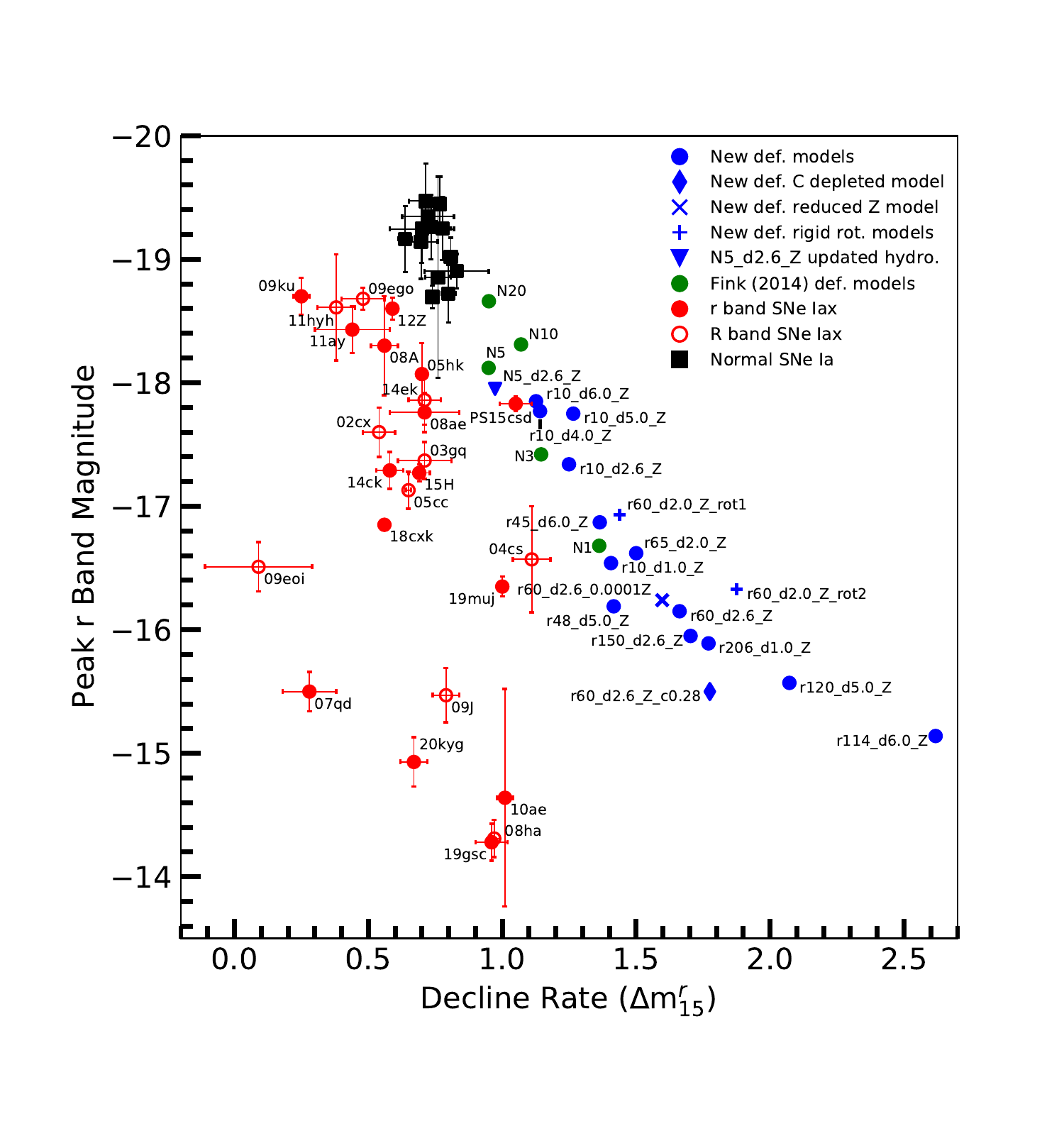}  
  \caption{Peak r-band magnitudes v.s. rise time (left) and peak r-band
    magnitudes v.s.  decline rate in terms of ${\Delta}m_{15}$ (right)
    for the models of F14 (green), our new sequence of models (blue),
    real SNe~Iax  (red) and normal SNe~Ia (black). The filled red
    circles represent measurements in the r-band while the unfilled red
    circles represent measurements in the R-band. The observations
    include those compiled by \citet{magee2016a} as well as SN~2014ck
    (\citealp{tomasella2016a}), SN~2014ek (\citealp{li2018a}), SN
    2018cxk (\citealp{yao2019a}), SN~2019muj (\citealp{barna2021a}),
    SN~2019gsc (\citealp{tomasella2020a}, \citealp{srivastav2020a}) and 
    SN~2020kyg (Srivastav et al. 2021, in prep).
    The blue diamond, cross and plus markers correspond to models which
    are variations on the standard r60\_d2.6\_Z model.  The blue diamond
    represents r60\_d2.6\_Z\_c0.28 (carbon depleted model), the cross
    depicts r60\_d2.6\_0.0001Z (reduced Z model) and the blue plus signs
    represent r60\_d2.0\_Z\_rot1 and r60\_d2.0\_Z\_rot2 (rigidly
    rotating models).  The blue triangle represents the N5\_d2.6\_Z model from
    the F14 deflagration study simulated again using the updated version
    of the LEAFS code to quantify the difference caused by using the
    newer version of the code. The green star represents the hybrid CONe
  WD model produced by K15.}
  \label{fig:rband_risetime_dm15_peakmag}  
\end{figure*}

Fig.~\ref{fig:rband_risetime_dm15_peakmag} shows peak absolute r-band
magnitude with rise time to r-peak (left panel) and decline rate post
r-peak (right panel) for our models, models from the F14 and K15 studies
as well as observed SNe~Iax, and normal SNe~Ia for comparison.  From
Fig.~\ref{fig:rband_risetime_dm15_peakmag} we can see that our models
map a significant portion of the wide variety of brightnesses covered by
the SNe~Iax class. They are, however, unable to reach the luminosities
of the very brightest as well as the faintest SNe~Iax. The new models
extend the 1D sequence previously observed by F14 to more than a
magnitude fainter at r-band peak and the new models appear to connect
the F14 suite of models to the hybrid CONe models produced in the K15
study as an attempt to reach the faintest members of the Iax class (in
particular SN~2008ha). 

It is interesting to note that single spark models from our model
sequence are very similar to models from the F14 sequence of similar
peak brightness both in terms of their light curve evolution (see
Fig.~\ref{fig:bol_band_viewing_angles},
Fig.~\ref{fig:B_band_dm15_viewing_angles}, and
Fig.~\ref{fig:rband_risetime_dm15_peakmag}   where the N5\_d2.6\_Z model
from the F14 sequence, re-simulated using the newest version of the
\textsc{leafs} code, is included) and spectra. However, as expected, the
single spark models do show greater viewing angle dependencies than the
multi-spark models (see Fig.~\ref{fig:bol_band_viewing_angles} and
Fig.~\ref{fig:B_band_dm15_viewing_angles}).  This suggests that while
using a multi-spark ignition is a physically improbable scenario there
is relatively little signature of how the models were ignited in the
synthetic observables they produce. This also provides some confirmation
of the validity of the multispot ignition approach taken by J12, L14 and
F14 to vary the strength of the deflagration in different models. 

The 1D sequence followed by the models of previous works (see
Sect.~\ref{sec:previous}) and also our new model sequence is driven by
the $^{56}$Ni mass synthesized in the explosion. Although the $^{56}$Ni
mass to $M_\mathrm{ej}$ ratio may have a small impact (see
Sect.~\ref{subsec:angleav_lc}), no secondary parameters appear to have
any dramatic impact on the overall behavior of the model sequence.
Additionally, further changes to initial setups (such as models with
different metallicity, the carbon-depleted model and rigidly rotating
models) do not lead to any major break away from the general behavior of
the model sequence with these different setups only being a further way
in which the $^{56}$Ni synthesized in the explosions can be varied.    

The sequence occupied by the models means they do not produce good
agreement in their light curve evolution time scales, especially in
their decline in the red bands (see
Fig.~\ref{fig:rband_risetime_dm15_peakmag}) where the majority of the
flux is observed. The agreement also becomes increasingly poor as we
move to the faint models in our sequence because the models become
faster in both rise and decline times for decreasing brightness. Faint
SNe~Iax, on the other hand, do not appear to evolve significantly faster
than the brighter members of the SN~Iax class (see
Fig.~\ref{fig:B_band_dm15_viewing_angles},
Fig.~\ref{fig:rband_risetime_dm15_peakmag}). In addition, the model
spectra show some systematic differences with observed SNe~Iax,
especially at early times and as we move to fainter models. 

\section{Conclusions}
\label{sec:conclusion}

In this work we presented an extensive parameter study of 3D explosion
simulations of deflagrations in \mch CO WDs. The aim of this research is
to gain insights into the explosion mechanism of SNe~Iax, a subluminous
subclass of SNe~Ia, since the pure deflagration scenario yields results
in broad agreement with observations
\citep{long2014a,fink2014a,kromer2013a,kromer2015a,jordan2012b}.
However, none of the studies carried out to date captures the full range
in brightness and variations in decline rates and rise times, especially
for the faint objects among this subclass. Moreover, multispot ignition
was used as a tool to vary the explosion strength although the ignition
in a single spark seems to be more realistic
\citep{zingale2009a,zingale2011a,nonaka2012a}. Therefore, we restricted
the models to single-spot ignition and varied the location of the
ignition spark from $10$ to $206\,\si{km}$ off from the center of the
WD. We also employed central densities from $1$ to $6 \times
10^{9}\,\si{g.cm^{-3}}$ and metallicities between $1 \times
10^{-4}\,Z_\odot$ and $2\,Z_\odot$.  Finally, two rigidly rotating
models and a progenitor with a carbon depleted core were added to the
sequence to widen the parameter space and search for additional,
physically motivated characteristics of a WD breaking the 1D trends
found in previous studies. 

We find that the sequence of models covers a large range in bolometric
brightness ranging from $-14.91\,$mag to $-17.35\,$mag  although the
explosion is not controlled via the number of ignition bubbles. This
demonstrates that single-spark ignition models can account for a wide
range of luminosity as required to match the SN~Iax observations. The
faintest model is still about one order of magnitude brighter than the
faint SN~2008ha and SN~2019gsc. However, a further reduction in
brightness, that is,  \nifs mass, is rather easy to achieve by further
increasing the ignition radius. On the other hand, it is hard to reach
the brightest members of the SN~Iax class with the restriction to a
single spark ignition. Furthermore, we validate that, regardless of
ignition, deflagrations do produce well mixed ejecta apart from some
shallow abundance gradients. In addition, we report kick velocities of
the bound explosion remnant of up to $369.8\,\si{km.s^{-1}}$ largely
exceeding those reported by \citet{fink2014a}. The kick velocity is,
however, not a simple function of the deflagration strength. The
direction of the natal kick changes by $180^\circ$ for decreasing energy
release leading to a minimum in the absolute value of the kick velocity.

Significant viewing angle effects in rise and decline times as well as
variations of up to $\sim 1\,$mag in peak brightness further enhance the
variations in observational properties. Additionally, both nonrotating
and rigidly-rotating models show spectroscopic differences depending on
the line of sight with variations in velocity shifts of
$\sim1000\,\si{km.s^{-1}}$ and $\sim2000\,\si{km.s^{-1}}$ for the
nonrotating and rotating models, respectively.  Spectra, rise times and
decline rates for the bright models of our sequence are in rough
agreement with observations as already found by \cite{fink2014a} and
\cite{kromer2013a}. The intermediate luminosity models in our sequence
also show reasonably good agreement.  However, although the model
observational properties show a significant diversity introduced by the
wide exploration of the parameter space of the initial conditions and
viewing angle effects, there are systematic differences between models
and data which become increasingly apparent when moving to lower
luminosities. For instance, the light curve decline is too fast in the
red bands for all models, but as we move to the faint models this worsens
and the overall light curve evolution also becomes too fast in all
bands. The synthetic spectra of the faint models also show worse
agreement across all epochs.

Overall, our findings suggest the pure deflagration scenario remains an
appealing explanation for bright and intermediate luminosity SNe~Iax.
However, some systematic differences remain which need to be addressed.
The work presented here, therefore leaves a few open questions.  First,
it needs to be investigated whether the decline of the model light
curves can be slowed down within the framework of the \mch deflagration
model. It seems that strong coupling between the ejected mass of \nifs
and the total ejected mass cannot be overcome easily by varying the
parameters of our simulations.  The most uncertain assumptions are the
ignition conditions which need to be further investigated. Also the RT
calculations introduce inaccuracies which need to be quantified and the
effects of the new non-LTE version of \textsc{artis}
\citep{shingles2020a} will be tested in future studies, which may help
explain some of the systematic differences (particularly in spectral
evolution) between models and observed SNe~Iax. Another explanation for
these systematic differences may be the properties of the models
themselves: perhaps a more stratified ejecta structure as suggested by
\citet{stritzinger2015a}, \citet{barna2017a}, \citet{barna2018a},
and \citet{barna2021a} may help match the spectral evolution of SNe~Iax
better. Moreover, we note our simulations do not take into account the
possible impacts of the burning products in the bound remnant on the
light curves and spectra. While this impact is very uncertain it has
been suggested that the contribution of the burned material in the bound
core may help to explain the long term evolution of SNe~Iax light curves
(\citealp{kromer2013a}, \citealp{foley2014b}, \citealp{shen2017a}) and
their peculiar late time spectra \citep{foley2016a}. In addition, the
effects due to the bound remnant could be increasingly relevant for the
faint explosions, and, depending on the structure of the bound remnant,
could contribute to SNe~Iax light curves and spectra at earlier times
\citep{kromer2015a}.  A greater contribution of the remnant may be
expected for fainter explosions as in fainter models the $^{56}$Ni mass
in the bound remnant is predicted to be significantly higher than in the
ejected material (e.g., Model r114\_d6.0\_Z has bound remnant $^{56}$Ni
mass which is 5 times higher than the $^{56}$Ni mass in the ejected
material). The bound remnant, therefore, may help explain some of the
systematic differences between our models (particularly the fainter
ones) and observed SNe~Iax. However, as the impact, if any, of the
burned material on top of the bound remnant on observed SNe~Iax light
curves and spectra is very uncertain future studies which better
quantify the possible contribution of the bound remnant are key to
understanding if this really can help explain some of the systematic
differences between our models and observed SNe~Iax.

The fact that the models can match some of the bright and intermediate
luminosity events, but the observational trend does not naturally explain
the faint explosions, may suggest that a different scenario might be at
work. In particular, \citet{valenti2009a} argue that the most probable
explanation for the very faint SN~Iax supernova, SN~2008ha, is that it
was produced in the low-energy core-collapse explosion of a
hydrogen-deficient massive star. However, the fainter models in our
sequence as well as the hybrid CONe deflagration model produced by
\citet{kromer2015a} are able to reach luminosities approaching the low
luminosities of the faintest SNe~Iax such as SN~2008ha.  In addition,
while \citet{valenti2009a} argue there is a striking resemblance between
SN~2008ha and the under-luminous type IIP SN~2005cs
\citep{li2006a,pastorello2009a} the comparisons they make between the
early time spectra of SN~2008ha and those of SN~2002cx and SN~2005hk
with their line velocities shifted by $-3000\,\si{km.s^{-1}}$ show
similarly good agreement.  Moreover, speculations about ONe WD - neutron
star mergers as a candidate for faint SNe Iax have been put forward by
\citet{bobrick2021a}. In summary, it seems likely that SNe~Iax can be
modeled by a combination of deflagrations in \mch WDs and other
scenarios that may be needed to account for the fainter members of this
class of transients.

\begin{acknowledgements} 
This work was supported by the Deutsche Forschungsgemeinschaft (DFG,
German Research Foundation) -- Project-ID 138713538 -- SFB 881 (``The
Milky Way System'', subproject A10), by the ChETEC COST Action
(CA16117), and by the National Science Foundation under Grant No.
OISE-1927130 (IReNA).  FL and FKR acknowledge support by the Klaus
Tschira Foundation. FPC acknowledges an STFC studentship and SAS
acknowledges funding from STFC Grant Ref: ST/P000312/1. 

NumPy and SciPy \citep{oliphant2007a}, IPython \citep{perez2007a}, and
Matplotlib \citep{hunter2007a} were used for data processing and
plotting. The authors gratefully acknowledge the Gauss Centre for
Supercomputing e.V.  (www.gauss-centre.eu) for funding this project by
providing computing time on the GCS Supercomputer JUWELS
\citep{juwels2019} at J\"{u}lich Supercomputing Centre (JSC). Part of
this work was performed using the Cambridge Service for Data Driven
Discovery (CSD3), part of which is operated by the University of
Cambridge Research Computing on behalf of the STFC DiRAC HPC Facility
(www.dirac.ac.uk). The DiRAC component of CSD3 was funded by BEIS
capital funding via STFC capital grants ST/P002307/1 and ST/R002452/1
and STFC operations grant ST/R00689X/1. DiRAC is part of the National
e-Infrastructure.  We thank James Gillanders for assisting with the flux
calibrations of the observed spectra.
\end{acknowledgements}

\bibliography{astrofritz}
\bibliographystyle{aa}

\begin{appendix}
\section{Simulation summary tables}
\label{sec:app}
\input{tables.tex}

\end{appendix}

\end{document}

%% file: journals.tex
\def\apjl{ApJ}%
\def\apjs{ApJS}%
\def\ao{Appl.~Opt.}%
\def\aap{A\&A}%

%% file: ignition_config.pdf_tex
%% Creator: Inkscape inkscape 0.91, www.inkscape.org
%% Accompanies image file 'ignition_config.pdf' (pdf, eps, ps)
\begingroup%
  \makeatletter%
  \providecommand\color[2][]{%
    \errmessage{(Inkscape) Color is used for the text in Inkscape, but the package 'color.sty' is not loaded}%
    \renewcommand\color[2][]{}%
  }%
  \providecommand\transparent[1]{%
    \errmessage{(Inkscape) Transparency is used (non-zero) for the text in Inkscape, but the package 'transparent.sty' is not loaded}%
    \renewcommand\transparent[1]{}%
  }%
  \providecommand\rotatebox[2]{#2}%
  \ifx\svgwidth\undefined%
    \setlength{\unitlength}{256.07501221bp}%
    \ifx\svgscale\undefined%
      \relax%
    \else%
      \setlength{\unitlength}{\unitlength * \real{\svgscale}}%
    \fi%
  \else%
    \setlength{\unitlength}{\svgwidth}%
  \fi%
  \global\let\svgwidth\undefined%
  \global\let\svgscale\undefined%
  \makeatother%
  \begin{picture}(1,0.60529139)%
    \put(0,0){\includegraphics[width=\unitlength,page=1]{ignition_config.pdf}}%
    \put(0.21358592,0.00776416){\color[rgb]{0,0,0}\makebox(0,0)[lb]{\smash{$200\,\mathrm{km}$}}}%
    \put(0.29053981,0.31553262){\color[rgb]{0,0,0}\makebox(0,0)[lb]{\smash{$r_\mathrm{off}$}}}%
    \put(1.17934179,0.2780436){\color[rgb]{0,0,0}\makebox(0,0)[lb]{\smash{}}}%
    \put(0.5752151,0.23689135){\color[rgb]{0,0,0}\rotatebox{90}{\makebox(0,0)[lb]{\smash{$17\,\mathrm{km}$}}}}%
    \put(0.82632028,0.2741385){\color[rgb]{0,0,0}\makebox(0,0)[lb]{\smash{$5\,\mathrm{km}$}}}%
  \end{picture}%
\endgroup%

%% file: 3Dflame.pdf_tex
%% Creator: Inkscape inkscape 0.91, www.inkscape.org
%% Accompanies image file '3Dflame.pdf' (pdf, eps, ps)
\begingroup%
  \makeatletter%
  \providecommand\color[2][]{%
    \errmessage{(Inkscape) Color is used for the text in Inkscape, but the package 'color.sty' is not loaded}%
    \renewcommand\color[2][]{}%
  }%
  \providecommand\transparent[1]{%
    \errmessage{(Inkscape) Transparency is used (non-zero) for the text in Inkscape, but the package 'transparent.sty' is not loaded}%
    \renewcommand\transparent[1]{}%
  }%
  \providecommand\rotatebox[2]{#2}%
  \ifx\svgwidth\undefined%
    \setlength{\unitlength}{467.1574707bp}%
    \ifx\svgscale\undefined%
      \relax%
    \else%
      \setlength{\unitlength}{\unitlength * \real{\svgscale}}%
    \fi%
  \else%
    \setlength{\unitlength}{\svgwidth}%
  \fi%
  \global\let\svgwidth\undefined%
  \global\let\svgscale\undefined%
  \makeatother%
  \begin{picture}(1,0.53322995)%
    \put(0,0){\includegraphics[width=\unitlength,page=1]{3Dflame.pdf}}%
    \put(0.74205568,0.04693164){\color[rgb]{0.24313725,0.24313725,0.24313725}\makebox(0,0)[lb]{\smash{$\log(\rho)$}}}%
    \put(0.23730158,0.50934061){\color[rgb]{0,0,0}\makebox(0,0)[lb]{\smash{$t=0.6\,\mathrm{s}$}}}%
    \put(0.6938013,0.50909603){\color[rgb]{0,0,0}\makebox(0,0)[lb]{\smash{$t=1.55\,\mathrm{s}$}}}%
  \end{picture}%
\endgroup%

%% file: tables.tex
\begin{table*}[b!]
  \centering
  \caption{Summary of the main properties of the ejected material and
  the initial conditions.}
 \begin{tabularx}{\textwidth}{c @{\extracolsep{\fill}} cccccccccc}
    \toprule
    model & $\rho_c$                   & $r_\mathrm{ign}$ & Z          &
    $E_\mathrm{nuc}$ & $M_\mathrm{ej}$ & $M(^{56}$Ni) & $M_\mathrm{IGE}$
    & $M_\mathrm{IME}$ & $M(^{56}$Ni$)/M_\mathrm{IGE}$ & $E_\mathrm{kin,ej}$  \\ 
     & \small $(10^{9}\,\si{g.cm^{-3}})$ & \small(km)             &\small$(Z_\odot$) &
    \small$(10^{50}\si{erg})$     & \small(\msun)  & \small(\msun) & \small(\msun) & \small(\msun) &
    \small& $(10^{50}\si{erg})$ \\
    \midrule
    r10\_d1.0\_Z & 1.0 &  10   & 1 & 1.98 & 0.077 & 0.033 & 0.039 & 0.0072 & 0.84 & 0.15 \\     
    r10\_d2.0\_Z & 2.0 &  10   & 1 & 2.81 & 0.127 & 0.049 & 0.066 & 0.011  & 0.74 & 0.28 \\
    r10\_d2.6\_Z & 2.6 &  10   & 1 & 3.15 & 0.164 & 0.069 & 0.094 & 0.015  & 0.74 & 0.41\\
    r10\_d3.0\_Z & 3.0 &  10   & 1 & 3.26 & 0.174 & 0.070 & 0.100 & 0.014  & 0.70 & 0.43 \\
    r10\_d4.0\_Z & 4.0 &  10   & 1 & 3.83 & 0.227 & 0.092 & 0.144 & 0.020 & 0.64 & 0.68 \\
    r10\_d5.0\_Z & 5.0 &  10   & 1 & 4.07 & 0.237 & 0.085 & 0.150 & 0.022 & 0.58 & 0.75 \\
    r10\_d6.0\_Z & 6.0 &  10   & 1 & 4.70 & 0.301 & 0.090 & 0.178 & 0.027 & 0.50 & 0.97 \\
    r82\_d1.0\_Z & 1.0 &  82   & 1 & 2.31 & 0.082 & 0.033 & 0.038 & 0.0071 & 0.86 & 0.16 \\
    r65\_d2.0\_Z & 2.0 &  65   & 1 & 2.38 & 0.079 & 0.033 & 0.041 & 0.0074 & 0.78 & 0.16 \\
    r60\_d2.6\_0.0001Z & 2.6 &  60   & 1e-4 & 1.95 & 0.053 & 0.022 & 0.028 & 0.0045 & 0.79 & 0.089 \\
    r60\_d2.6\_0.001Z & 2.6 &  60   & 1e-3 & 1.98 & 0.052 & 0.021 & 0.026 & 0.0044 & 0.80 & 0.088 \\
    r60\_d2.6\_0.01Z & 2.6 &  60   & 1e-2 & 1.93 & 0.048 & 0.019 & 0.024 & 0.0045 & 0.79 & 0.084 \\
    r60\_d2.6\_0.1Z & 2.6 &  60   & 1e-1 & 1.92 & 0.049 & 0.020 & 0.026 & 0.0042 & 0.79 & 0.79 \\
    r60\_d2.6\_Z & 2.6 &  60   & 1 & 1.93 & 0.050 & 0.018 & 0.025 & 0.0045 & 0.75 & 0.082 \\
    r60\_d2.6\_2Z & 2.6 &  60   & 2 & 1.97 & 0.054 & 0.020 & 0.028 & 0.0052 & 0.71 & 0.096 \\
    r60\_d2.6\_Z\_co0.28 & 2.6 &  60   & 1 & 1.87 & 0.036 & 0.012 & 0.018 & 0.0036 & 0.68 & 0.039 \\
    r57\_d3.0\_Z & 3.0 &  57   & 1 & 1.86 & 0.054 & 0.022 & 0.030 & 0.0053 & 0.74 & 0.093 \\
    r51\_d4.0\_Z & 4.0 &  51   & 1 & 1.29 & 0.033 & 0.012 & 0.019 & 0.0033 & 0.67 & 0.042 \\
    r48\_d5.0\_Z & 5.0 &  48   & 1 & 1.67 & 0.054 & 0.018 & 0.030 & 0.0047 & 0.59 & 0.072 \\
    r45\_d6.0\_Z & 6.0 &  45   & 1 & 2.19 & 0.093 & 0.033 & 0.056 & 0.0086 & 0.58 & 0.19 \\
    r206\_d1.0\_Z & 1.0 &  206  & 1 & 1.67 & 0.041 & 0.016 & 0.018 & 0.0037 & 0.88 & 0.064 \\ 
    r163\_d2.0\_Z & 2.0 &  163  & 1 & 1.40 & 0.029 & 0.012 & 0.015 & 0.0025 & 0.80 & 0.038 \\
    r150\_d2.6\_Z & 2.6 &  150  & 1 & 1.75 & 0.039 & 0.016 & 0.021 & 0.0036 & 0.78 & 0.061 \\
    r143\_d3.0\_Z & 3.0 &  143  & 1 & 1.61 & 0.031 & 0.013 & 0.017 & 0.0028 & 0.77 & 0.041\\ 
    r129\_d4.0\_Z & 4.0 &  129  & 1 & 1.58 & 0.031 & 0.013 & 0.017 & 0.0029 & 0.75 & 0.047 \\
    r120\_d5.0\_Z & 5.0 &  120  & 1 & 1.36 & 0.024 & 0.010 & 0.014 & 0.0023 & 0.74 & 0.034 \\
    r114\_d6.0\_Z & 6.0 &  114  & 1 & 0.96 & 0.014 & 0.0058 & 0.0081 & 0.0012 & 0.72 & 0.018 \\
    r60\_d2.0\_Z\_rot1 & 2.0 &  60   &  1 & 2.89 & 0.095 & 0.045 & 0.058 & 0.0076 & 0.77 & 0.23 \\  
    r60\_d2.0\_Z\_rot2 & 2.0 &  60   &  1 & 2.44 & 0.054 & 0.022 & 0.028 & 0.0038 &  0.78 & 0.11 \\
    N5\_d2.6\_Z & 3.6 &  N5  &  1 & 4.30 & 0.294 & 0.136 & 0.178 & 0.0352 &  0.76 & 0.983 \\
    \bottomrule
  \end{tabularx}
  \label{tab:ejectasum}
\end{table*}
\begin{table*}[t!]
  \centering
  \caption{Main properties of the bound remnant and the respective
  initial conditions. } \begin{tabularx}{\textwidth}{c @{\extracolsep{\fill}} ccccccccccc}
    \toprule
    model & $\rho_c$                   & $r_\mathrm{ign}$ & Z          &
    $M_\mathrm{bound}$ & $M(^{56}$Ni) & $M_\mathrm{IGE}$
    & $M_\mathrm{IME}$ & $\rho_\mathrm{max}$ & $v_\mathrm{kick}$ &
    $v_x$  \\ 
    & \small $(10^{9}\,\si{g.cm^{-3}})$ & \small (km)             &\small $(Z_\odot$) &
    \small (\msun)  & \small (\msun) & \small (\msun) & \small (\msun) &
    \small $(10^{5}\si{g.cm^{-3}})$ & \small $(\si{km.s^{-1}})$
    & \small $(\si{km.s^{-1}})$ \\
    \midrule
    r10\_d1.0\_Z & 1.0 &  10   & 1 & 1.27 & 0.053 & 0.059 & 0.027 & 2.19 & 139.3 & -139.3 \\     
    r10\_d2.0\_Z & 2.0 &  10   & 1 & 1.24 & 0.073 & 0.087 & 0.023 & 2.17 & 76.6 & -76.3 \\
    r10\_d2.6\_Z & 2.6 &  10   & 1 & 1.21 & 0.064 & 0.078 & 0.022 & 1.49 & 109.8 & -109.4\\
    r10\_d3.0\_Z & 3.0 &  10   & 1 & 1.21 & 0.064 & 0.080 & 0.021 & 1.30 & 182.6 & -182.0 \\
    r10\_d4.0\_Z & 4.0 &  10   & 1 & 1.16 & 0.054 & 0.067 & 0.020 & 1.27 & 157.7 & -157.7 \\
    r10\_d5.0\_Z & 5.0 &  10   & 1 & 1.16 & 0.057 & 0.082 & 0.016 & 1.43 & 250.2 & -250.0 \\
    r10\_d6.0\_Z & 6.0 &  10   & 1 & 1.09 & 0.058 & 0.091 & 0.014 & 1.12 & 241.9 & -241.8 \\
    r82\_d1.0\_Z & 1.0 &  82   & 1 & 1.27 & 0.070 & 0.078 & 0.030 & 2.48 & 22.9 & -22.3\\
    r65\_d2.0\_Z & 2.0 &  65   & 1 & 1.29 & 0.073 & 0.085 & 0.024 & 1.92 & 7.5 & -6.9 \\
    r60\_d2.6\_0.0001Z& 2.6 &  60   & 1e-4 & 1.33 & 0.065 & 0.075 & 0.020 & 2.33 & 53.3 & 53.2\\
    r60\_d2.6\_0.001Z& 2.6 &  60   & 1e-3 & 1.33 & 0.068 & 0.078 & 0.020 & 2.62 & 20.2 & 16.7\\
    r60\_d2.6\_0.01Z& 2.6 &  60   & 1e-2 & 1.33 & 0.066 & 0.077 & 0.021 & 2.61 & 119.9 & 119.7 \\
    r60\_d2.6\_0.1Z& 2.6 &  60   & 1e-1 & 1.33 & 0.065 & 0.076 & 0.020 & 2.22 & 46.1 & 45.9 \\
    r60\_d2.6\_Z & 2.6 &  60   & 1 & 1.33 & 0.064 & 0.077 & 0.020 & 2.34 & 16.8 & 16.5\\
    r60\_d2.6\_2Z& 2.6 &  60   & 2 & 1.33 & 0.066 & 0.076 & 0.020 & 2.62 & 8.6 & -5.4 \\
    r60\_d2.6\_Z\_co0.28 & 2.6 &  60   & 1 & 1.34 & 0.069 & 0.088 & 0.021 & 3.28 & 271.6 & -271.4 \\
    r57\_d3.0\_Z & 3.0 &  57   & 1 & 1.33 & 0.056 & 0.069 & 0.018 & 2.25 & 135.1 & 134.7 \\
    r51\_d4.0\_Z & 4.0 &  51   & 1 & 1.36 & 0.033 & 0.051 & 0.011 & 2.90 & 365.6 & 364.0\\
    r48\_d5.0\_Z & 5.0 &  48   & 1 & 1.34 & 0.038 & 0.064 & 0.010 & 2.02 & 38.1 & 38.0\\
    r45\_d6.0\_Z & 6.0 &  44   & 1 & 1.30 & 0.037 & 0.070 & 0.0078 & 1.98 & 10.7 & 6.2\\
    r206\_d1.0\_Z& 1.0 &  206  & 1 & 1.31 & 0.057 & 0.062 & 0.027 & 2.76 & 6.4 & -2.1 \\ 
    r163\_d2.0\_Z& 2.0 &  163  & 1 & 1.34 & 0.048 & 0.056 & 0.018 & 3.16 & 54.8 & 53.1\\
    r150\_d2.6\_Z& 2.6 &  150  & 1 & 1.34 & 0.061 & 0.071 & 0.019 & 2.64 & 86.5 & 86.2 \\
    r143\_d3.0\_Z& 3.0 &  143  & 1 & 1.35 & 0.058 & 0.069 & 0.017 & 2.70 & 130.8 & 130.6&\\ 
    r129\_d4.0\_Z& 4.0 &  129  & 1 & 1.36 & 0.056 & 0.068 & 0.015 & 2.21 & 369.8 & 369.6 \\
    r120\_d5.0\_Z& 5.0 &  120  & 1 & 1.37 & 0.047 & 0.058 & 0.014 & 2.54 & 225.4 & 225.2 \\
    r114\_d6.0\_Z& 6.0 &  114  & 1 & 1.38 & 0.030 & 0.042 & 0.011 & 3.63 & 43.2 & -42.6 \\
    r60\_d2.0\_Z\_rot1 & 2.0 &  60   &  1 & 1.34 & 0.084 & 0.096 & 0.027 & 1.84 & 232.4 & - \\
    r60\_d2.0\_Z\_rot2 & 2.0 &  60   &  1 & 1.38 & 0.087 & 0.101 & 0.026 & 2.15 & 17.6 & - \\
    N5\_d2.6\_Z & 2.6 & N5 &  1 & 1.08 & 0.050 & 0.058 & 0.018 & 1.11 & 264.6 & -137.0 \\
    \bottomrule
  \end{tabularx}
  \label{tab:remnantsum}
\end{table*}
\begin{sidewaystable*}[t!]
%\begin{table*}
  \centering
  \caption{Angle averaged light curve properties for bolometric band and BVRI Bessel
  bands (top) as well as ugriz Sloan bands (bottom) for the selection of models
  for which RT simulations were carried out. }
  \begin{tabularx}{\textwidth}{c @{\extracolsep{\fill}} cccccccccccccccc}
    \toprule
    model & $t_{rise}^{bol}$  & $M_{peak}^{bol}$  & ${\Delta}m_{15}^{bol}$ & 
    $t_{rise}^B$  & $M_{peak}^B$  & ${\Delta}m_{15}^B$ &  
    $t_{rise}^V$  & $M_{peak}^V$  & ${\Delta}m_{15}^V$ &  
    $t_{rise}^R$  & $M_{peak}^R$  & ${\Delta}m_{15}^R$ &  
    $t_{rise}^I$  & $M_{peak}^I$  & ${\Delta}m_{15}^I$ \\ 
    
    \midrule    
    r\_10\_d1.0\_Z & 7.27 & -16.47 & 1.5 & 7.69 & -16.57 & 2.24 & 9.07 & -16.76 & 1.49 & 9.46 & -16.59 & 1.29 & 13.99 & -16.68 & 1.49 \\
    r\_10\_d2.6\_Z & 9.7 & -17.08 & 1.19 & 9.61 & -17.21 & 2.12 & 11.56 & -17.49 & 1.37 & 12.88 & -17.38 & 1.13 & 17.06 & -17.45 & 1.25 \\
    r\_10\_d4.0\_Z & 11.02 & -17.35 & 1.03 & 10.42 & -17.43 & 2.09 & 12.85 & -17.83 & 1.31 & 14.62 & -17.81 & 1.05 & 18.35 & -17.86 & 0.98 \\
    r\_10\_d5.0\_Z & 10.93 & -17.19 & 1.06 & 9.7 & -17.11 & 2.26 & 12.25 & -17.69 & 1.41 & 13.96 & -17.81 & 1.17 & 17.03 & -17.86 & 1.22 \\
    r\_10\_d6.0\_Z & 12.07 & -17.23 & 0.89 & 9.55 & -16.97 & 1.92 & 13.15 & -17.67 & 1.32 & 15.04 & -17.94 & 1.02 & 16.82 & -18.02 & 0.93 \\
    r\_65\_d2.0\_Z & 7.45 & -16.43 & 1.49 & 7.63 & -16.54 & 2.33 & 9.1 & -16.8 & 1.57 & 9.67 & -16.67 & 1.37 & 13.84 & -16.73 & 1.52 \\
    r\_60\_d2.6\_0.0001Z & 6.52 & -16.07 & 1.55 & 6.79 & -16.2 & 2.42 & 8.2 & -16.42 & 1.62 & 9.31 & -16.29 & 1.48 & 12.31 & -16.38 & 1.53 \\
    r\_60\_d2.6\_Z & 6.39 & -15.92 & 1.61 & 6.48 & -16.02 & 2.45 & 7.84 & -16.28 & 1.66 & 8.83 & -16.2 & 1.57 & 11.44 & -16.3 & 1.59 \\
    r\_60\_d2.6\_Z\_c0.28 & 6.48 & -15.58 & 1.57 & 6.58 & -15.71 & 2.56 & 7.96 & -15.89 & 1.64 & 9.19 & -15.84 & 1.46 & 11.53 & -15.96 & 1.48 \\
    r\_48\_d5.0\_Z & 6.94 & -15.83 & 1.32 & 6.94 & -15.88 & 2.49 & 8.53 & -16.23 & 1.57 & 10.27 & -16.25 & 1.32 & 12.73 & -16.34 & 1.33 \\
    r\_45\_d6.0\_Z & 8.35 & -16.4 & 1.22 & 7.9 & -16.38 & 2.41 & 9.76 & -16.86 & 1.53 & 11.26 & -16.93 & 1.26 & 14.08 & -16.97 & 1.28 \\
    r\_206\_d1.0\_Z & 5.91 & -15.82 & 1.76 & 6.3 & -15.92 & 2.41 & 7.51 & -16.1 & 1.68 & 8.08 & -15.95 & 1.66 & 10.54 & -16.07 & 1.61 \\
    r\_150\_d2.6\_Z & 5.79 & -15.81 & 1.65 & 6.09 & -15.94 & 2.48 & 7.45 & -16.13 & 1.67 & 8.65 & -16.0 & 1.61 & 11.11 & -16.11 & 1.62 \\
    r\_120\_d5.0\_Z & 4.68 & -15.38 & 1.84 & 5.1 & -15.47 & 2.5 & 6.76 & -15.7 & 1.81 & 7.81 & -15.62 & 2.01 & 9.55 & -15.74 & 1.91 \\
    r\_114\_d6.0\_Z & 4.02 & -14.91 & 2.38 & 4.17 & -15.04 & 2.84 & 5.31 & -15.21 & 2.13 & 6.06 & -15.2 & 2.55 & 7.72 & -15.36 & 2.38 \\
    r\_60\_d2.0\_Z\_rot1 & 7.96 & -16.83 & 1.47 & 8.41 & -16.98 & 2.37 & 9.73 & -17.15 & 1.52 & 10.36 & -16.98 & 1.3 & 15.1 & -17.09 & 1.57 \\
    r\_60\_d2.0\_Z\_rot2 & 6.52 & -16.16 & 1.79 & 6.76 & -16.24 & 2.36 & 8.11 & -16.46 & 1.69 & 8.71 & -16.38 & 1.79 & 11.47 & -16.48 & 1.83 \\
    N5\_d2.6\_Z & 10.81 & -17.72 & 0.93 & 10.75 & -17.89 & 1.9 & 13.45 & -18.16 & 1.19 & 15.37 & -17.98 & 0.87 & 20.06 & -18.06 & 0.72 \\
    \bottomrule
    
        \toprule
    model & $t_{rise}^{u}$  & $M_{peak}^{u}$  & ${\Delta}m_{15}^{u}$ & 
    $t_{rise}^g$  & $M_{peak}^g$  & ${\Delta}m_{15}^g$ &
    $t_{rise}^r$  & $M_{peak}^r$  & ${\Delta}m_{15}^r$ &
    $t_{rise}^i$  & $M_{peak}^i$  & ${\Delta}m_{15}^i$ &
    $t_{rise}^z$  & $M_{peak}^z$  & ${\Delta}m_{15}^z$ &    \\ 

    \midrule    
    r10\_d1.0\_Z & 6.42 & -16.23 & 4.11 & 8.20 & -16.72 & 1.83 & 9.85 & -16.54 & 1.41 & 9.04 & -15.96 & 0.87 & 14.65 & -16.45 & 1.30 \\
    r10\_d2.6\_Z & 8.14 & -16.78 & 3.69 & 10.21 & -17.38 & 1.75 & 13.18 & -17.34 & 1.25 & 16.88 & -16.75 & 1.12 & 17.57 & -17.22 & 1.16 \\
    r10\_d4.0\_Z & 9.16 & -16.9 & 3.47 & 11.14 & -17.61 & 1.73 & 14.65 & -17.77 & 1.14 & 18.38 & -17.21 & 0.85 & 18.98 & -17.58 & 0.93 \\
    r10\_d5.0\_Z & 8.56 & -16.54 & 3.65 & 10.45 & -17.35 & 1.88 & 13.87 & -17.75 & 1.26 & 16.97 & -17.28 & 1.08 & 17.63 & -17.56 & 1.12 \\
    r10\_d6.0\_Z & 8.02 & -16.4 & 2.78 & 10.42 & -17.23 & 1.67 & 14.86 & -17.85 & 1.12 & 16.73 & -17.52 & 0.85 & 17.57 & -17.68 & 0.85 \\
    r65\_d2.0\_Z & 6.42 & -16.11 & 4.27 & 8.14 & -16.72 & 1.92 & 10.00 & -16.63 & 1.5 & 9.34 & -16.02 & 0.91 & 14.32 & -16.49 & 1.34 \\
    r60\_d2.6\_0.0001Z & 5.64 & -15.8 & 4.47 & 7.24 & -16.36 & 1.99 & 9.49 & -16.24 & 1.60 & 11.08 & -15.68 & 1.30 & 12.85 & -16.12 & 1.32 \\
    r60\_d2.6\_Z & 5.49 & -15.61 & 4.56 & 6.94 & -16.19 & 2.01 & 8.92 & -16.15 & 1.66 & 10.54 & -15.62 & 1.43 & 12.01 & -16.00 & 1.36 \\
    r60\_d2.6\_Z\_c0.28 & 5.49 & -15.34 & 4.63 & 7.00 & -15.84 & 2.08 & 9.13 & -15.78 & 1.54 & 11.11 & -15.33 & 1.38 & 12.04 & -15.62 & 1.27 \\
    r48\_d5.0\_Z & 5.82 & -15.44 & 4.39 & 7.39 & -16.07 & 2.04 & 10.15 & -16.19 & 1.42 & 12.88 & -15.75 & 1.27 & 13.03 & -15.99 & 1.18 \\
    r45\_d6.0\_Z & 6.64 & -15.88 & 4.02 & 8.47 & -16.59 & 2.01 & 11.20 & -16.87 & 1.36 & 13.99 & -16.40 & 1.18 & 14.68 & -16.65 & 1.18 \\
    r206\_d1.0\_Z & 5.19 & -15.57 & 4.45 & 6.73 & -16.08 & 1.97 & 8.32 & -15.89 & 1.77 & 8.32 & -15.38 & 1.36 & 11.47 & -15.79 & 1.35 \\
    r150\_d2.6\_Z & 4.98 & -15.59 & 4.66 & 6.55 & -16.09 & 2.03 & 8.74 & -15.95 & 1.7 & 10.21 & -15.41 & 1.46 & 11.68 & -15.82 & 1.38 \\
    r120\_d5.0\_Z & 3.78 & -15.22 & 4.84 & 5.67 & -15.63 & 2.05 & 7.84 & -15.57 & 2.07 & 9.01 & -15.06 & 1.92 & 10.03 & -15.42 & 1.58 \\
    r114\_d6.0\_Z & 3.06 & -14.74 & 4.97 & 4.47 & -15.16 & 2.32 & 6.09 & -15.14 & 2.62 & 7.48 & -14.7 & 2.62 & 8.08 & -14.98 & 1.90 \\
    r60\_d2.0\_Z\_rot1 & 6.91 & -16.66 & 4.18 & 8.89 & -17.11 & 1.93 & 10.66 & -16.94 & 1.41 & 15.01 & -16.35 & 1.48 & 15.70 & -16.85 & 1.40 \\
    r60\_d2.0\_Z\_rot2 & 5.46 & -15.9 & 4.18 & 7.21 & -16.39 & 1.93 & 8.92 & -16.33 & 1.88 & 9.37 & -15.79 & 1.59 & 12.22 & -16.20 & 1.55 \\
    N5\_d2.6\_Z & 8.83 & -17.46 & 3.01 & 11.62 & -18.05 & 1.6 & 15.58 & -17.95 & 0.97 & 20.63 & -17.33 & 0.57 & 20.84 & -17.86 & 0.71 \\
    \bottomrule
  \end{tabularx}

  \label{tab:lightcurve_data_bessel}
%\end{table*}
\end{sidewaystable*}

%% file: Iax.bbl
\begin{thebibliography}{157}
\expandafter\ifx\csname natexlab\endcsname\relax\def\natexlab#1{#1}\fi

\bibitem[{{Asplund} {et~al.}(2009){Asplund}, {Grevesse}, {Sauval}, \&
  {Scott}}]{asplund2009a}
{Asplund}, M., {Grevesse}, N., {Sauval}, A.~J., \& {Scott}, P. 2009, \araa, 47,
  481

\bibitem[{{Barna} {et~al.}(2021){Barna}, {Szalai}, {Jha}, {Camacho-Neves},
  {Kwok}, {Foley}, {Kilpatrick}, {Coulter}, {Dimitriadis}, {Rest},
  {Rojas-Bravo}, {Siebert}, {Brown}, {Burke}, {Padilla Gonzalez}, {Hiramatsu},
  {Howell}, {McCully}, {Pellegrino}, {Dobson}, {Smartt}, {Swift}, {Stacey},
  {Rahman}, {Sand}, {Andrews}, {Wyatt}, {Hsiao}, {Anderson}, {Chen}, {Della
  Valle}, {Galbany}, {Gromadzki}, {Inserra}, {Lyman}, {Magee}, {Maguire},
  {M{\"u}ller-Bravo}, {Nicholl}, {Srivastav}, \& {Williams}}]{barna2021a}
{Barna}, B., {Szalai}, T., {Jha}, S.~W., {et~al.} 2021, \mnras, 501, 1078

\bibitem[{Barna {et~al.}(2018)Barna, Szalai, Kerzendorf, Kromer, Sim, Magee, \&
  Leibundgut}]{barna2018a}
Barna, B., Szalai, T., Kerzendorf, W.~E., {et~al.} 2018, Monthly Notices of the
  Royal Astronomical Society, 480, 3609

\bibitem[{{Barna} {et~al.}(2017){Barna}, {Szalai}, {Kromer}, {Kerzendorf},
  {Vink{\'o}}, {Silverman}, {Marion}, \& {Wheeler}}]{barna2017a}
{Barna}, B., {Szalai}, T., {Kromer}, M., {et~al.} 2017, \mnras, 471, 4865

\bibitem[{Bauer {et~al.}(2019)Bauer, White, \& Bildsten}]{bauer2019a}
Bauer, E.~B., White, C.~J., \& Bildsten, L. 2019, The Astrophysical Journal,
  887, 68

\bibitem[{{Blinnikov} \& {Khokhlov}(1986)}]{blinnikov1986a}
{Blinnikov}, S.~I. \& {Khokhlov}, A.~M. 1986, Soviet Astronomy Letters, 12, 131

\bibitem[{{Blondin} {et~al.}(2013){Blondin}, {Dessart}, {Hillier}, \&
  {Khokhlov}}]{blondin2013a}
{Blondin}, S., {Dessart}, L., {Hillier}, D.~J., \& {Khokhlov}, A.~M. 2013,
  \mnras, 429, 2127

\bibitem[{Bobrick {et~al.}(2021)Bobrick, Zenati, Perets, Davies, \&
  Church}]{bobrick2021a}
Bobrick, A., Zenati, Y., Perets, H.~B., Davies, M.~B., \& Church, R. 2021,
  arXiv preprint arXiv:2104.03415

\bibitem[{{Brachwitz} {et~al.}(2000){Brachwitz}, {Dean}, {Hix}, {Iwamoto},
  {Langanke}, {Mart{\'{\i}}nez-Pinedo}, {Nomoto}, {Strayer}, {Thielemann}, \&
  {Umeda}}]{brachwitz2000a}
{Brachwitz}, F., {Dean}, D.~J., {Hix}, W.~R., {et~al.} 2000, \apj, 536, 934

\bibitem[{{Branch} {et~al.}(2004){Branch}, {Baron}, {Thomas}, {Kasen}, {Li}, \&
  {Filippenko}}]{branch2004a}
{Branch}, D., {Baron}, E., {Thomas}, R.~C., {et~al.} 2004, \pasp, 116, 903

\bibitem[{{Branch} {et~al.}(2006){Branch}, {Dang}, {Hall}, {Ketchum},
  {Melakayil}, {Parrent}, {Troxel}, {Casebeer}, {Jeffery}, \&
  {Baron}}]{branch2006a}
{Branch}, D., {Dang}, L.~C., {Hall}, N., {et~al.} 2006, \pasp, 118, 560

\bibitem[{{Branch} {et~al.}(1993){Branch}, {Fisher}, \& {Nugent}}]{branch1993a}
{Branch}, D., {Fisher}, A., \& {Nugent}, P. 1993, \aj, 106, 2383

\bibitem[{Bravo(2019)}]{bravo2019a}
Bravo, E. 2019, Astronomy \& Astrophysics, 624, A139

\bibitem[{{Bravo} {et~al.}(2016){Bravo}, {Gil-Pons}, {Guti{\'e}rrez}, \&
  {Doherty}}]{bravo2016a}
{Bravo}, E., {Gil-Pons}, P., {Guti{\'e}rrez}, J.~L., \& {Doherty}, C.~L. 2016,
  \aap, 589, A38

\bibitem[{{Bulla} {et~al.}(2015){Bulla}, {Sim}, \& {Kromer}}]{bulla2015a}
{Bulla}, M., {Sim}, S.~A., \& {Kromer}, M. 2015, \mnras, 450, 967

\bibitem[{Catalan {et~al.}(2008)Catalan, Ribas, Isern, \&
  Garc{\'\i}a-Berro}]{catalan2008a}
Catalan, S., Ribas, I., Isern, J., \& Garc{\'\i}a-Berro, E. 2008, Astronomy \&
  Astrophysics, 477, 901

\bibitem[{{Chamulak} {et~al.}(2007){Chamulak}, {Brown}, \&
  {Timmes}}]{chamulak2007a}
{Chamulak}, D.~A., {Brown}, E.~F., \& {Timmes}, F.~X. 2007, \apjl, 655, L93

\bibitem[{{Chamulak} {et~al.}(2008){Chamulak}, {Brown}, {Timmes}, \&
  {Dupczak}}]{chamulak2008a}
{Chamulak}, D.~A., {Brown}, E.~F., {Timmes}, F.~X., \& {Dupczak}, K. 2008,
  \apj, 677, 160

\bibitem[{{Chornock} {et~al.}(2006){Chornock}, {Filippenko}, {Branch}, {Foley},
  {Jha}, \& {Li}}]{chornock2006a}
{Chornock}, R., {Filippenko}, A.~V., {Branch}, D., {et~al.} 2006, \pasp, 118,
  722

\bibitem[{{Colella} \& {Woodward}(1984)}]{colella1984a}
{Colella}, P. \& {Woodward}, P.~R. 1984, Journal of Computational Physics, 54,
  174

\bibitem[{{Denissenkov} {et~al.}(2013){Denissenkov}, {Herwig}, {Truran}, \&
  {Paxton}}]{denissenkov2013c}
{Denissenkov}, P.~A., {Herwig}, F., {Truran}, J.~W., \& {Paxton}, B. 2013,
  \apj, 772, 37

\bibitem[{{Denissenkov} {et~al.}(2015){Denissenkov}, {Truran}, {Herwig},
  {Jones}, {Paxton}, {Nomoto}, {Suzuki}, \& {Toki}}]{denissenkov2015a}
{Denissenkov}, P.~A., {Truran}, J.~W., {Herwig}, F., {et~al.} 2015, \mnras,
  447, 2696

\bibitem[{{Dessart} {et~al.}(2014){Dessart}, {Hillier}, {Blondin}, \&
  {Khokhlov}}]{dessart2014a}
{Dessart}, L., {Hillier}, D.~J., {Blondin}, S., \& {Khokhlov}, A. 2014, \mnras,
  441, 3249

\bibitem[{{Di Stefano} {et~al.}(2011){Di Stefano}, {Voss}, \&
  {Claeys}}]{distefano2011a}
{Di Stefano}, R., {Voss}, R., \& {Claeys}, J.~S.~W. 2011, \apjl, 738, L1

\bibitem[{{Dom{\'{\i}}nguez} {et~al.}(2001){Dom{\'{\i}}nguez}, {H{\"o}flich},
  \& {Straniero}}]{dominguez2001a}
{Dom{\'{\i}}nguez}, I., {H{\"o}flich}, P., \& {Straniero}, O. 2001, \apj, 557,
  279

\bibitem[{Fern{\'a}ndez \& Metzger(2013)}]{fernandez2013a}
Fern{\'a}ndez, R. \& Metzger, B.~D. 2013, The Astrophysical Journal, 763, 108

\bibitem[{Fink {et~al.}(2018)Fink, Kromer, Hillebrandt, R{\"o}pke, Pakmor,
  Seitenzahl, \& Sim}]{fink2018a}
Fink, M., Kromer, M., Hillebrandt, W., {et~al.} 2018, Astronomy \&
  Astrophysics, 618, A124

\bibitem[{{Fink} {et~al.}(2014){Fink}, {Kromer}, {Seitenzahl},
  {Ciaraldi-Schoolmann}, {R{\"o}pke}, {Sim}, {Pakmor}, {Ruiter}, \&
  {Hillebrandt}}]{fink2014a}
{Fink}, M., {Kromer}, M., {Seitenzahl}, I.~R., {et~al.} 2014, \mnras, 438, 1762

\bibitem[{{Foley} {et~al.}(2013){Foley}, {Challis}, {Chornock},
  {Ganeshalingam}, {Li}, {Marion}, {Morrell}, {Pignata}, {Stritzinger},
  {Silverman}, {Wang}, {Anderson}, {Filippenko}, {Freedman}, {Hamuy}, {Jha},
  {Kirshner}, {McCully}, {Persson}, {Phillips}, {Reichart}, \&
  {Soderberg}}]{foley2013b}
{Foley}, R.~J., {Challis}, P.~J., {Chornock}, R., {et~al.} 2013, \apj, 767, 57

\bibitem[{{Foley} {et~al.}(2009){Foley}, {Chornock}, {Filippenko},
  {Ganeshalingam}, {Kirshner}, {Li}, {Cenko}, {Challis}, {Friedman}, {Modjaz},
  {Silverman}, \& {Wood-Vasey}}]{foley2009a}
{Foley}, R.~J., {Chornock}, R., {Filippenko}, A.~V., {et~al.} 2009, \aj, 138,
  376

\bibitem[{{Foley} {et~al.}(2016){Foley}, {Jha}, {Pan}, {Zheng}, {Bildsten},
  {Filippenko}, \& {Kasen}}]{foley2016a}
{Foley}, R.~J., {Jha}, S.~W., {Pan}, Y.-C., {et~al.} 2016, \mnras, 461, 433

\bibitem[{Foley \& Kirshner(2013)}]{foley2013a}
Foley, R.~J. \& Kirshner, R.~P. 2013, The Astrophysical Journal Letters, 769,
  L1

\bibitem[{{Foley} {et~al.}(2014){Foley}, {McCully}, {Jha}, {Bildsten}, {Fong},
  {Narayan}, {Rest}, \& {Stritzinger}}]{foley2014b}
{Foley}, R.~J., {McCully}, C., {Jha}, S.~W., {et~al.} 2014, \apj, 792, 29

\bibitem[{{Fryxell} {et~al.}(1989){Fryxell}, {M{\"u}ller}, \&
  {Arnett}}]{fryxell1989a}
{Fryxell}, B.~A., {M{\"u}ller}, E., \& {Arnett}, W.~D. 1989, Hydro\-dynamics
  and nuclear burning, MPA Green Report 449, Max-Planck-Institut f\"ur
  Astrophysik, Garching

\bibitem[{{Gall} {et~al.}(2012){Gall}, {Taubenberger}, {Kromer}, {Sim},
  {Benetti}, {Blanc}, {Elias-Rosa}, \& {Hillebrandt}}]{gall2012a}
{Gall}, E.~E.~E., {Taubenberger}, S., {Kromer}, M., {et~al.} 2012, \mnras, 427,
  994

\bibitem[{{Gamezo} {et~al.}(2005){Gamezo}, {Khokhlov}, \& {Oran}}]{gamezo2005a}
{Gamezo}, V.~N., {Khokhlov}, A.~M., \& {Oran}, E.~S. 2005, \apj, 623, 337

\bibitem[{{Gamezo} {et~al.}(2003){Gamezo}, {Khokhlov}, {Oran}, {Chtchelkanova},
  \& {Rosenberg}}]{gamezo2003a}
{Gamezo}, V.~N., {Khokhlov}, A.~M., {Oran}, E.~S., {Chtchelkanova}, A.~Y., \&
  {Rosenberg}, R.~O. 2003, Science, 299, 77

\bibitem[{Gronow {et~al.}(2021)Gronow, Cote, Lach, Seitenzahl, Collins, Sim, \&
  Roepke}]{gronow2021b}
Gronow, S., Cote, B., Lach, F., {et~al.} 2021, Metallicity-dependent
  nucleosynthetic yields of Type Ia supernovae originating from double
  detonations of sub-M$_{\text{Ch}}$ white dwarfs

\bibitem[{{Hamuy} {et~al.}(1996){Hamuy}, {Phillips}, {Suntzeff}, {Schommer},
  {Maza}, \& {Aviles}}]{hamuy1996a}
{Hamuy}, M., {Phillips}, M.~M., {Suntzeff}, N.~B., {et~al.} 1996, \aj, 112,
  2391

\bibitem[{{Hillebrandt} {et~al.}(2013){Hillebrandt}, {Kromer}, {R{\"o}pke}, \&
  {Ruiter}}]{hillebrandt2013a}
{Hillebrandt}, W., {Kromer}, M., {R{\"o}pke}, F.~K., \& {Ruiter}, A.~J. 2013,
  Frontiers of Physics, 8, 116

\bibitem[{{Hix} \& {Thielemann}(1999)}]{hix1999a}
{Hix}, W.~R. \& {Thielemann}, F. 1999, \apj, 511, 862

\bibitem[{{H{\"o}flich} \& {Khokhlov}(1996)}]{hoeflich1996a}
{H{\"o}flich}, P. \& {Khokhlov}, A. 1996, \apj, 457, 500

\bibitem[{Hunter(2007)}]{hunter2007a}
Hunter, J.~D. 2007, Computing in Science \& Engineering, 9, 90

\bibitem[{{Iben} \& {Tutukov}(1984)}]{iben1984a}
{Iben}, Jr., I. \& {Tutukov}, A.~V. 1984, \apjs, 54, 335

\bibitem[{{Iwamoto} {et~al.}(1999){Iwamoto}, {Brachwitz}, {Nomoto},
  {Kishimoto}, {Umeda}, {Hix}, \& {Thielemann}}]{iwamoto1999a}
{Iwamoto}, K., {Brachwitz}, F., {Nomoto}, K., {et~al.} 1999, \apjs, 125, 439

\bibitem[{{Jha} {et~al.}(2006){Jha}, {Branch}, {Chornock}, {Foley}, {Li},
  {Swift}, {Casebeer}, \& {Filippenko}}]{jha2006b}
{Jha}, S., {Branch}, D., {Chornock}, R., {et~al.} 2006, \aj, 132, 189

\bibitem[{Jha {et~al.}(2007)Jha, Riess, \& Kirshner}]{jha2007a}
Jha, S., Riess, A.~G., \& Kirshner, R.~P. 2007, The Astrophysical Journal, 659,
  122

\bibitem[{Jha(2017)}]{jha2017type}
Jha, S.~W. 2017, Handbook of Supernovae, 375

\bibitem[{{Jordan} {et~al.}(2012){Jordan}, {Perets}, {Fisher}, \& {van
  Rossum}}]{jordan2012b}
{Jordan}, IV, G.~C., {Perets}, H.~B., {Fisher}, R.~T., \& {van Rossum}, D.~R.
  2012, \apjl, 761, L23

\bibitem[{{J\"{u}lich Supercomputing Centre}(2019)}]{juwels2019}
{J\"{u}lich Supercomputing Centre}. 2019, Journal of large-scale research
  facilities, 5

\bibitem[{{Kasen}(2006)}]{kasen2006b}
{Kasen}, D. 2006, \apj, 649, 939

\bibitem[{Kashyap {et~al.}(2018)Kashyap, Haque, Lor{\'e}n-Aguilar,
  Garc{\'\i}a-Berro, \& Fisher}]{kashyap2018a}
Kashyap, R., Haque, T., Lor{\'e}n-Aguilar, P., Garc{\'\i}a-Berro, E., \&
  Fisher, R. 2018, The Astrophysical Journal, 869, 140

\bibitem[{Kawabata {et~al.}(2018)Kawabata, Kawabata, Maeda, Yamanaka, Nakaoka,
  Takaki, Fukushima, Kojiguchi, Masumoto, Matsumoto, {et~al.}}]{kawabata2018a}
Kawabata, M., Kawabata, K.~S., Maeda, K., {et~al.} 2018, Publications of the
  Astronomical Society of Japan, 70, 111

\bibitem[{Kawabata {et~al.}(2021)Kawabata, Maeda, Yamanaka, Nakaoka, Kawabata,
  Aoki, Anupama, Burgaz, Dutta, Isogai, Kino, Kojiguchi, Kota, Kumar, Kuroda,
  Maehara, Matsubayashi, Morihana, Murata, Ohshima, Otsuka, Sahu, Singh,
  Sugitani, Takahashi, \& Takagi}]{kawabata2021a}
Kawabata, M., Maeda, K., Yamanaka, M., {et~al.} 2021, Intermediate Luminosity
  Type Iax SN 2019muj With Narrow Absorption Lines: Long-Lasting Radiation
  Associated With a Possible Bound Remnant Predicted by the Weak Deflagration
  Model

\bibitem[{{Khokhlov}(1989)}]{khokhlov1989a}
{Khokhlov}, A.~M. 1989, \mnras, 239, 785

\bibitem[{{Khokhlov}(1991)}]{khokhlov1991a}
{Khokhlov}, A.~M. 1991, \aap, 245, 114

\bibitem[{{Khokhlov}(2000)}]{khokhlov2000a}
{Khokhlov}, A.~M. 2000, preprint: astro-ph/0008463

\bibitem[{Kobayashi {et~al.}(2020)Kobayashi, Leung, \& Nomoto}]{kobayashi2020a}
Kobayashi, C., Leung, S.-C., \& Nomoto, K. 2020, The Astrophysical Journal,
  895, 138

\bibitem[{{Kobayashi} {et~al.}(2006){Kobayashi}, {Umeda}, {Nomoto}, {Tominaga},
  \& {Ohkubo}}]{kobayashi2006a}
{Kobayashi}, C., {Umeda}, H., {Nomoto}, K., {Tominaga}, N., \& {Ohkubo}, T.
  2006, \apj, 653, 1145

\bibitem[{{Kromer} {et~al.}(2013){Kromer}, {Fink}, {Stanishev}, {Taubenberger},
  {Ciaraldi-Schoolman}, {Pakmor}, {R{\"o}pke}, {Ruiter}, {Seitenzahl}, {Sim},
  {Blanc}, {Elias-Rosa}, \& {Hillebrandt}}]{kromer2013a}
{Kromer}, M., {Fink}, M., {Stanishev}, V., {et~al.} 2013, \mnras, 429, 2287

\bibitem[{{Kromer} {et~al.}(2015){Kromer}, {Ohlmann}, {Pakmor}, {Ruiter},
  {Hillebrandt}, {Marquardt}, {R{\"o}pke}, {Seitenzahl}, {Sim}, \&
  {Taubenberger}}]{kromer2015a}
{Kromer}, M., {Ohlmann}, S.~T., {Pakmor}, R., {et~al.} 2015, \mnras, 450, 3045

\bibitem[{{Kromer} {et~al.}(2017){Kromer}, {Ohlmann}, \&
  {Roepke}}]{kromer2017a}
{Kromer}, M., {Ohlmann}, S.~T., \& {Roepke}, F.~K. 2017, ArXiv e-prints
  [\eprint[arXiv]{1706.09879}]

\bibitem[{{Kromer} \& {Sim}(2009)}]{kromer2009a}
{Kromer}, M. \& {Sim}, S.~A. 2009, \mnras, 398, 1809

\bibitem[{{Kuhlen} {et~al.}(2006){Kuhlen}, {Woosley}, \&
  {Glatzmaier}}]{kuhlen2006a}
{Kuhlen}, M., {Woosley}, S.~E., \& {Glatzmaier}, G.~A. 2006, \apj, 640, 407

\bibitem[{{Lach} {et~al.}(2020){Lach}, {R{\"o}pke}, {Seitenzahl}, {Cot{\'e}},
  {Gronow}, \& {Ruiter}}]{lach2020a}
{Lach}, F., {R{\"o}pke}, F.~K., {Seitenzahl}, I.~R., {et~al.} 2020, \aap, 644,
  A118

\bibitem[{{Langanke} \& {Mart{\'{\i}}nez-Pinedo}(2001)}]{langanke2001a}
{Langanke}, K. \& {Mart{\'{\i}}nez-Pinedo}, G. 2001, Atomic Data and Nuclear
  Data Tables, 79, 1

\bibitem[{{Lesaffre} {et~al.}(2006){Lesaffre}, {Han}, {Tout}, {Podsiadlowski},
  \& {Martin}}]{lesaffre2006a}
{Lesaffre}, P., {Han}, Z., {Tout}, C.~A., {Podsiadlowski}, P., \& {Martin},
  R.~G. 2006, \mnras, 368, 187

\bibitem[{Leung \& Nomoto(2020)}]{leung2020a}
Leung, S.-C. \& Nomoto, K. 2020, The Astrophysical Journal, 900, 54

\bibitem[{{Li} {et~al.}(2018){Li}, {Wang}, {Zhang}, {Arcavi}, {Zhang}, {Rui},
  {Hosseinzadeh}, {Howell}, {McCully}, {Zhang}, {Valenti}, {Mo}, {Li}, {Huang},
  {Xiang}, {Wang}, \& {Zhou}}]{li2018a}
{Li}, L., {Wang}, X., {Zhang}, J., {et~al.} 2018, \mnras, 478, 4575

\bibitem[{{Li} {et~al.}(2003){Li}, {Filippenko}, {Chornock}, {Berger},
  {Berlind}, {Calkins}, {Challis}, {Fassnacht}, {Jha}, {Kirshner}, {Matheson},
  {Sargent}, {Simcoe}, {Smith}, \& {Squires}}]{li2003a}
{Li}, W., {Filippenko}, A.~V., {Chornock}, R., {et~al.} 2003, \pasp, 115, 453

\bibitem[{{Li} {et~al.}(2011){Li}, {Leaman}, {Chornock}, {Filippenko},
  {Poznanski}, {Ganeshalingam}, {Wang}, {Modjaz}, {Jha}, {Foley}, \&
  {Smith}}]{li2011a}
{Li}, W., {Leaman}, J., {Chornock}, R., {et~al.} 2011, \mnras, 412, 1441

\bibitem[{Li {et~al.}(2006)Li, Van~Dyk, Filippenko, Cuillandre, Jha, Bloom,
  Riess, \& Livio}]{li2006a}
Li, W., Van~Dyk, S.~D., Filippenko, A.~V., {et~al.} 2006, The Astrophysical
  Journal, 641, 1060

\bibitem[{{Liu} {et~al.}(2013){Liu}, {Kromer}, {Fink}, {Pakmor}, {R{\"o}pke},
  {Chen}, {Wang}, \& {Han}}]{liu2013c}
{Liu}, Z.-W., {Kromer}, M., {Fink}, M., {et~al.} 2013, \apj, 778, 121

\bibitem[{Livio \& Mazzali(2018)}]{livio2018a}
Livio, M. \& Mazzali, P. 2018, Physics Reports, 736, 1

\bibitem[{Long {et~al.}(2014)Long, Jordan~IV, Van~Rossum, Diemer, Graziani,
  Kessler, Meyer, Rich, \& Lamb}]{long2014a}
Long, M., Jordan~IV, G.~C., Van~Rossum, D.~R., {et~al.} 2014, The Astrophysical
  Journal, 789, 103

\bibitem[{Lundqvist {et~al.}(2013)Lundqvist, Mattila, Sollerman, Kozma, Baron,
  Cox, Fransson, Leibundgut, \& Spyromilio}]{lundqvist2013a}
Lundqvist, P., Mattila, S., Sollerman, J., {et~al.} 2013, Monthly Notices of
  the Royal Astronomical Society, 435, 329

\bibitem[{Lyman {et~al.}(2018)Lyman, Taddia, Stritzinger, Galbany, Leloudas,
  Anderson, Eldridge, James, Kr{\"u}hler, Levan, {et~al.}}]{lyman2018a}
Lyman, J., Taddia, F., Stritzinger, M., {et~al.} 2018, Monthly Notices of the
  Royal Astronomical Society, 473, 1359

\bibitem[{Magee {et~al.}(2019)Magee, Sim, Kotak, Maguire, \&
  Boyle}]{magee2019a}
Magee, M., Sim, S., Kotak, R., Maguire, K., \& Boyle, A. 2019, Astronomy \&
  Astrophysics, 622, A102

\bibitem[{{Magee} {et~al.}(2016){Magee}, {Kotak}, {Sim}, {Kromer},
  {Rabinowitz}, {Smartt}, {Baltay}, {Campbell}, {Chen}, {Fink}, {Gal-Yam},
  {Galbany}, {Hillebrandt}, {Inserra}, {Kankare}, {Le Guillou}, {Lyman},
  {Maguire}, {Pakmor}, {R{\"o}pke}, {Ruiter}, {Seitenzahl}, {Sullivan},
  {Valenti}, \& {Young}}]{magee2016a}
{Magee}, M.~R., {Kotak}, R., {Sim}, S.~A., {et~al.} 2016, \aap, 589, A89

\bibitem[{{Marquardt} {et~al.}(2015){Marquardt}, {Sim}, {Ruiter}, {Seitenzahl},
  {Ohlmann}, {Kromer}, {Pakmor}, \& {R{\"o}pke}}]{marquardt2015a}
{Marquardt}, K.~S., {Sim}, S.~A., {Ruiter}, A.~J., {et~al.} 2015, \aap, 580,
  A118

\bibitem[{{McClelland} {et~al.}(2010){McClelland}, {Garnavich}, {Galbany},
  {Miquel}, {Foley}, {Filippenko}, {Bassett}, {Wheeler}, {Goobar}, {Jha},
  {Sako}, {Frieman}, {Sollerman}, {Vinko}, \& {Schneider}}]{mcclelland2010a}
{McClelland}, C.~M., {Garnavich}, P.~M., {Galbany}, L., {et~al.} 2010, \apj,
  720, 704

\bibitem[{{McCully} {et~al.}(2014){McCully}, {Jha}, {Foley}, {Bildsten},
  {Fong}, {Kirshner}, {Marion}, {Riess}, \& {Stritzinger}}]{mccully2014a}
{McCully}, C., {Jha}, S.~W., {Foley}, R.~J., {et~al.} 2014, \nat, 512, 54

\bibitem[{McCully {et~al.}(2014)McCully, Jha, Foley, Chornock, Holtzman, Balam,
  Branch, Filippenko, Frieman, Fynbo, {et~al.}}]{mccully2014b}
McCully, C., Jha, S.~W., Foley, R.~J., {et~al.} 2014, The Astrophysical
  Journal, 786, 134

\bibitem[{McCully {et~al.}(2021)McCully, Jha, Scalzo, Howell, Foley, Zeng, Liu,
  Hosseinzadeh, Bildsten, Riess, {et~al.}}]{mccully2021a}
McCully, C., Jha, S.~W., Scalzo, R.~A., {et~al.} 2021, arXiv preprint
  arXiv:2106.04602

\bibitem[{{Narayan} {et~al.}(2011){Narayan}, {Foley}, {Berger}, {Botticella},
  {Chornock}, {Huber}, {Rest}, {Scolnic}, {Smartt}, {Valenti}, {Soderberg},
  {Burgett}, {Chambers}, {Flewelling}, {Gates}, {Grav}, {Kaiser}, {Kirshner},
  {Magnier}, {Morgan}, {Price}, {Riess}, {Stubbs}, {Sweeney}, {Tonry},
  {Wainscoat}, {Waters}, \& {Wood-Vasey}}]{narayan2011a}
{Narayan}, G., {Foley}, R.~J., {Berger}, E., {et~al.} 2011, \apjl, 731, L11

\bibitem[{Neunteufel(2020)}]{neunteufel2020a}
Neunteufel, P. 2020, Astronomy \& Astrophysics, 641, A52

\bibitem[{{Nomoto} {et~al.}(2013){Nomoto}, {Kobayashi}, \&
  {Tominaga}}]{nomoto2013a}
{Nomoto}, K., {Kobayashi}, C., \& {Tominaga}, N. 2013, \araa, 51, 457

\bibitem[{{Nonaka} {et~al.}(2012){Nonaka}, {Aspden}, {Zingale}, {Almgren},
  {Bell}, \& {Woosley}}]{nonaka2012a}
{Nonaka}, A., {Aspden}, A.~J., {Zingale}, M., {et~al.} 2012, \apj, 745, 73

\bibitem[{{Ohlmann} {et~al.}(2014){Ohlmann}, {Kromer}, {Fink}, {Pakmor},
  {Seitenzahl}, {Sim}, \& {R{\"o}pke}}]{ohlmann2014a}
{Ohlmann}, S.~T., {Kromer}, M., {Fink}, M., {et~al.} 2014, \aap, 572, A57

\bibitem[{Oliphant(2007)}]{oliphant2007a}
Oliphant, T.~E. 2007, Computing in Science \& Engineering, 9, 10

\bibitem[{{Osher} \& {Sethian}(1988)}]{osher1988a}
{Osher}, S. \& {Sethian}, J.~A. 1988, Journal of Computational Physics, 79, 12

\bibitem[{Oskinova {et~al.}(2020)Oskinova, Gvaramadze, Gr{\"a}fener, Langer, \&
  Todt}]{oskinova2020a}
Oskinova, L.~M., Gvaramadze, V.~V., Gr{\"a}fener, G., Langer, N., \& Todt, H.
  2020, Astronomy \& Astrophysics, 644, L8

\bibitem[{{Pakmor} {et~al.}(2012){Pakmor}, {Edelmann}, {R{\"o}pke}, \&
  {Hillebrandt}}]{pakmor2012b}
{Pakmor}, R., {Edelmann}, P., {R{\"o}pke}, F.~K., \& {Hillebrandt}, W. 2012,
  \mnras, 424, 2222

\bibitem[{{Pakmor} {et~al.}(2008){Pakmor}, {R{\"o}pke}, {Weiss}, \&
  {Hillebrandt}}]{pakmor2008a}
{Pakmor}, R., {R{\"o}pke}, F.~K., {Weiss}, A., \& {Hillebrandt}, W. 2008, \aap,
  489, 943

\bibitem[{Pastorello {et~al.}(2009)Pastorello, Valenti, Zampieri, Navasardyan,
  Taubenberger, Smartt, Arkharov, B{\"a}rnbantner, Barwig, Benetti,
  {et~al.}}]{pastorello2009a}
Pastorello, A., Valenti, S., Zampieri, L., {et~al.} 2009, Monthly Notices of
  the Royal Astronomical Society, 394, 2266

\bibitem[{P\'{e}rez \& Granger(2007)}]{perez2007a}
P\'{e}rez, F. \& Granger, B.~E. 2007, Computing in Science \& Engineering, 9,
  21

\bibitem[{{Perlmutter} {et~al.}(1999){Perlmutter}, {Aldering}, {Goldhaber},
  {Knop}, {Nugent}, {Castro}, {Deustua}, {Fabbro}, {Goobar}, {Groom}, {Hook},
  {Kim}, {Kim}, {Lee}, {Nunes}, {Pain}, {Pennypacker}, {Quimby}, {Lidman},
  {Ellis}, {Irwin}, {McMahon}, {Ruiz-Lapuente}, {Walton}, {Schaefer}, {Boyle},
  {Filippenko}, {Matheson}, {Fruchter}, {Panagia}, {Newberg}, {Couch}, \& {The
  Supernova Cosmology Project}}]{perlmutter1999a}
{Perlmutter}, S., {Aldering}, G., {Goldhaber}, G., {et~al.} 1999, \apj, 517,
  565

\bibitem[{{Pfannes} {et~al.}(2010){Pfannes}, {Niemeyer}, {Schmidt}, \&
  {Klingenberg}}]{pfannes2010a}
{Pfannes}, J.~M.~M., {Niemeyer}, J.~C., {Schmidt}, W., \& {Klingenberg}, C.
  2010, \aap, 509, A74

\bibitem[{{Phillips}(1993)}]{phillips1993a}
{Phillips}, M.~M. 1993, \apjl, 413, L105

\bibitem[{{Phillips} {et~al.}(2007){Phillips}, {Li}, {Frieman}, {Blinnikov},
  {DePoy}, {Prieto}, {Milne}, {Contreras}, {Folatelli}, {Morrell}, {Hamuy},
  {Suntzeff}, {Roth}, {Gonz{\'a}lez}, {Krzeminski}, {Filippenko}, {Freedman},
  {Chornock}, {Jha}, {Madore}, {Persson}, {Burns}, {Wyatt}, {Murphy}, {Foley},
  {Ganeshalingam}, {Serduke}, {Krisciunas}, {Bassett}, {Becker}, {Dilday},
  {Eastman}, {Garnavich}, {Holtzman}, {Kessler}, {Lampeitl}, {Marriner},
  {Frank}, {Marshall}, {Miknaitis}, {Sako}, {Schneider}, {van der Heyden}, \&
  {Yasuda}}]{phillips2007a}
{Phillips}, M.~M., {Li}, W., {Frieman}, J.~A., {et~al.} 2007, \pasp, 119, 360

\bibitem[{{Piro} \& {Bildsten}(2008)}]{piro2008c}
{Piro}, A.~L. \& {Bildsten}, L. 2008, \apj, 673, 1009

\bibitem[{Prantzos {et~al.}(2018)Prantzos, Abia, Limongi, Chieffi, \&
  Cristallo}]{prantzos2018a}
Prantzos, N., Abia, C., Limongi, M., Chieffi, A., \& Cristallo, S. 2018,
  Monthly Notices of the Royal Astronomical Society, 476, 3432

\bibitem[{Provencal {et~al.}(1998)Provencal, Shipman, H{\o}g, \&
  Thejll}]{provencal1998a}
Provencal, J.~L., Shipman, H., H{\o}g, E., \& Thejll, P. 1998, The
  Astrophysical Journal, 494, 759

\bibitem[{Raddi {et~al.}(2018{\natexlab{a}})Raddi, Hollands, G{\"a}nsicke,
  Townsley, Hermes, Gentile~Fusillo, \& Koester}]{raddi2018b}
Raddi, R., Hollands, M., G{\"a}nsicke, B., {et~al.} 2018{\natexlab{a}}, Monthly
  Notices of the Royal Astronomical Society: Letters, 479, L96

\bibitem[{Raddi {et~al.}(2018{\natexlab{b}})Raddi, Hollands, Koester,
  G{\"a}nsicke, Fusillo, Hermes, \& Townsley}]{raddi2018a}
Raddi, R., Hollands, M., Koester, D., {et~al.} 2018{\natexlab{b}}, The
  Astrophysical Journal, 858, 3

\bibitem[{Raddi {et~al.}(2019)Raddi, Hollands, Koester, Hermes, G{\"a}nsicke,
  Heber, Shen, Townsley, Pala, Reding, {et~al.}}]{raddi2019a}
Raddi, R., Hollands, M., Koester, D., {et~al.} 2019, Monthly Notices of the
  Royal Astronomical Society, 489, 1489

\bibitem[{{Rauscher} \& {Thielemann}(2000)}]{rauscher2000a}
{Rauscher}, T. \& {Thielemann}, F.-K. 2000, Atomic Data and Nuclear Data
  Tables, 75, 1

\bibitem[{{Reinecke} {et~al.}(1999{\natexlab{a}}){Reinecke}, {Hillebrandt}, \&
  {Niemeyer}}]{reinecke1999b}
{Reinecke}, M., {Hillebrandt}, W., \& {Niemeyer}, J.~C. 1999{\natexlab{a}},
  \aap, 347, 739

\bibitem[{{Reinecke} {et~al.}(1999{\natexlab{b}}){Reinecke}, {Hillebrandt},
  {Niemeyer}, {Klein}, \& {Gr{\"o}bl}}]{reinecke1999a}
{Reinecke}, M., {Hillebrandt}, W., {Niemeyer}, J.~C., {Klein}, R., \&
  {Gr{\"o}bl}, A. 1999{\natexlab{b}}, \aap, 347, 724

\bibitem[{{Riess} {et~al.}(1996){Riess}, {Press}, \& {Kirshner}}]{riess1996a}
{Riess}, A.~G., {Press}, W.~H., \& {Kirshner}, R.~P. 1996, \apj, 473, 88

\bibitem[{Ritter {et~al.}(2021)Ritter, Parker, Lykou, Zijlstra, Guerrero, \&
  le~Du}]{ritter2021a}
Ritter, A., Parker, Q., Lykou, F., {et~al.} 2021, The remnant and origin of the
  historical supernova 1181AD

\bibitem[{{R{\"o}pke}(2005)}]{roepke2005c}
{R{\"o}pke}, F.~K. 2005, \aap, 432, 969

\bibitem[{{R{\"o}pke}(2007)}]{roepke2007d}
{R{\"o}pke}, F.~K. 2007, \apj, 668, 1103

\bibitem[{{R{\"o}pke} {et~al.}(2006{\natexlab{a}}){R{\"o}pke}, {Gieseler},
  {Reinecke}, {Travaglio}, \& {Hillebrandt}}]{roepke2006b}
{R{\"o}pke}, F.~K., {Gieseler}, M., {Reinecke}, M., {Travaglio}, C., \&
  {Hillebrandt}, W. 2006{\natexlab{a}}, \aap, 453, 203

\bibitem[{{R{\"o}pke} \& {Hillebrandt}(2004)}]{roepke2004c}
{R{\"o}pke}, F.~K. \& {Hillebrandt}, W. 2004, \aap, 420, L1

\bibitem[{{R{\"o}pke} \& {Hillebrandt}(2005)}]{roepke2005b}
{R{\"o}pke}, F.~K. \& {Hillebrandt}, W. 2005, \aap, 431, 635

\bibitem[{{R{\"o}pke} \& {Hillebrandt}(2006)}]{roepke2006d}
{R{\"o}pke}, F.~K. \& {Hillebrandt}, W. 2006, in American Institute of Physics
  Conference Series, Vol. 847, Origin of Matter and Evolution of Galaxies, ed.
  S.~{Kubono}, W.~{Aoki}, T.~{Kajino}, T.~{Motobayashi}, \& K.~{Nomoto},
  190--195

\bibitem[{{R{\"o}pke} {et~al.}(2006{\natexlab{b}}){R{\"o}pke}, {Hillebrandt},
  {Niemeyer}, \& {Woosley}}]{roepke2006a}
{R{\"o}pke}, F.~K., {Hillebrandt}, W., {Niemeyer}, J.~C., \& {Woosley}, S.~E.
  2006{\natexlab{b}}, \aap, 448, 1

\bibitem[{{R{\"o}pke} {et~al.}(2007){R{\"o}pke}, {Hillebrandt}, {Schmidt},
  {Niemeyer}, {Blinnikov}, \& {Mazzali}}]{roepke2007c}
{R{\"o}pke}, F.~K., {Hillebrandt}, W., {Schmidt}, W., {et~al.} 2007, \apj, 668,
  1132

\bibitem[{R{\"o}pke \& Sim(2018)}]{roepke2018a}
R{\"o}pke, F.~K. \& Sim, S.~A. 2018, Space Science Reviews, 214, 72

\bibitem[{{Schmidt} {et~al.}(1998){Schmidt}, {Suntzeff}, {Phillips},
  {Schommer}, {Clocchiatti}, {Kirshner}, {Garnavich}, {Challis}, {Leibundgut},
  {Spyromilio}, {Riess}, {Filippenko}, {Hamuy}, {Smith}, {Hogan}, {Stubbs},
  {Diercks}, {Reiss}, {Gilliland}, {Tonry}, {Maza}, {Dressler}, {Walsh}, \&
  {Ciardullo}}]{schmidt1998a}
{Schmidt}, B.~P., {Suntzeff}, N.~B., {Phillips}, M.~M., {et~al.} 1998, \apj,
  507, 46

\bibitem[{{Seitenzahl} {et~al.}(2013){Seitenzahl}, {Ciaraldi-Schoolmann},
  {R{\"o}pke}, {Fink}, {Hillebrandt}, {Kromer}, {Pakmor}, {Ruiter}, {Sim}, \&
  {Taubenberger}}]{seitenzahl2013a}
{Seitenzahl}, I.~R., {Ciaraldi-Schoolmann}, F., {R{\"o}pke}, F.~K., {et~al.}
  2013, \mnras, 429, 1156

\bibitem[{{Seitenzahl} {et~al.}(2009){Seitenzahl}, {Townsley}, {Peng}, \&
  {Truran}}]{seitenzahl2009a}
{Seitenzahl}, I.~R., {Townsley}, D.~M., {Peng}, F., \& {Truran}, J.~W. 2009,
  Atomic Data and Nuclear Data Tables, 95, 96

\bibitem[{{Shen} {et~al.}(2021){Shen}, {Blondin}, {Kasen}, {Dessart},
  {Townsley}, {Boos}, \& {Hillier}}]{shen2021a}
{Shen}, K.~J., {Blondin}, S., {Kasen}, D., {et~al.} 2021, \apjl, 909, L18

\bibitem[{Shen \& Schwab(2017)}]{shen2017a}
Shen, K.~J. \& Schwab, J. 2017, The Astrophysical Journal, 834, 180

\bibitem[{{Shingles} {et~al.}(2020){Shingles}, {Sim}, {Kromer}, {Maguire},
  {Bulla}, {Collins}, {Ballance}, {Michel}, {Ramsbottom}, {R{\"o}pke},
  {Seitenzahl}, \& {Tyndall}}]{shingles2020a}
{Shingles}, L.~J., {Sim}, S.~A., {Kromer}, M., {et~al.} 2020, \mnras, 492, 2029

\bibitem[{{Sim}(2007)}]{sim2007b}
{Sim}, S.~A. 2007, \mnras, 375, 154

\bibitem[{Soker(2019)}]{soker2019a}
Soker, N. 2019, New Astronomy Reviews, 87, 101535

\bibitem[{{Srivastav} {et~al.}(2020){Srivastav}, {Smartt}, {Leloudas}, {Huber},
  {Chambers}, {Malesani}, {Hjorth}, {Gillanders}, {Schultz}, {Sim}, {Auchettl},
  {Fynbo}, {Gall}, {McBrien}, {Rest}, {Smith}, {Wojtak}, \&
  {Young}}]{srivastav2020a}
{Srivastav}, S., {Smartt}, S.~J., {Leloudas}, G., {et~al.} 2020, \apjl, 892,
  L24

\bibitem[{{Stritzinger} {et~al.}(2014){Stritzinger}, {Hsiao}, {Valenti},
  {Taddia}, {Rivera-Thorsen}, {Leloudas}, {Maeda}, {Pastorello}, {Phillips},
  {Pignata}, {Baron}, {Burns}, {Contreras}, {Folatelli}, {Hamuy},
  {H{\"o}flich}, {Morrell}, {Prieto}, {Benetti}, {Campillay}, {Haislip},
  {LaClutze}, {Moore}, \& {Reichart}}]{stritzinger2014a}
{Stritzinger}, M.~D., {Hsiao}, E., {Valenti}, S., {et~al.} 2014, \aap, 561,
  A146

\bibitem[{{Stritzinger} {et~al.}(2015){Stritzinger}, {Valenti}, {Hoeflich},
  {Baron}, {Phillips}, {Taddia}, {Foley}, {Hsiao}, {Jha}, {McCully}, {Pandya},
  {Simon}, {Benetti}, {Brown}, {Burns}, {Campillay}, {Contreras},
  {F{\"o}rster}, {Holmbo}, {Marion}, {Morrell}, \&
  {Pignata}}]{stritzinger2015a}
{Stritzinger}, M.~D., {Valenti}, S., {Hoeflich}, P., {et~al.} 2015, \aap, 573,
  A2

\bibitem[{Taubenberger(2017)}]{taubenberger2017a}
Taubenberger, S. 2017, Handbook of Supernovae, 317

\bibitem[{{Thielemann} {et~al.}(2018){Thielemann}, {Isern}, {Perego}, \& {von
  Ballmoos}}]{thielemann2018a}
{Thielemann}, F.-K., {Isern}, J., {Perego}, A., \& {von Ballmoos}, P. 2018,
  \ssr, 214, 62

\bibitem[{{Thielemann} {et~al.}(1996){Thielemann}, {Nomoto}, \&
  {Hashimoto}}]{thielemann1996a}
{Thielemann}, F.-K., {Nomoto}, K., \& {Hashimoto}, M.-A. 1996, \apj, 460, 408

\bibitem[{{Timmes} \& {Arnett}(1999)}]{timmes1999a}
{Timmes}, F.~X. \& {Arnett}, D. 1999, \apjs, 125, 277

\bibitem[{{Timmes} {et~al.}(2003){Timmes}, {Brown}, \& {Truran}}]{timmes2003a}
{Timmes}, F.~X., {Brown}, E.~F., \& {Truran}, J.~W. 2003, \apjl, 590, L83

\bibitem[{{Timmes} \& {Woosley}(1992)}]{timmes1992a}
{Timmes}, F.~X. \& {Woosley}, S.~E. 1992, \apj, 396, 649

\bibitem[{{Tomasella} {et~al.}(2016){Tomasella}, {Cappellaro}, {Benetti},
  {Pastorello}, {Hsiao}, {Sand}, {Stritzinger}, {Valenti}, {McCully}, {Arcavi},
  {Elias-Rosa}, {Harmanen}, {Harutyunyan}, {Hosseinzadeh}, {Howell}, {Kankare},
  {Morales-Garoffolo}, {Taddia}, {Tartaglia}, {Terreran}, \&
  {Turatto}}]{tomasella2016a}
{Tomasella}, L., {Cappellaro}, E., {Benetti}, S., {et~al.} 2016, \mnras, 459,
  1018

\bibitem[{{Tomasella} {et~al.}(2020){Tomasella}, {Stritzinger}, {Benetti},
  {Elias-Rosa}, {Cappellaro}, {Kankare}, {Lundqvist}, {Magee}, {Maguire},
  {Pastorello}, {Prentice}, \& {Reguitti}}]{tomasella2020a}
{Tomasella}, L., {Stritzinger}, M., {Benetti}, S., {et~al.} 2020, \mnras, 496,
  1132

\bibitem[{{Townsley} {et~al.}(2009){Townsley}, {Jackson}, {Calder}, {Chamulak},
  {Brown}, \& {Timmes}}]{townsley2009a}
{Townsley}, D.~M., {Jackson}, A.~P., {Calder}, A.~C., {et~al.} 2009, \apj, 701,
  1582

\bibitem[{{Travaglio} {et~al.}(2004){Travaglio}, {Hillebrandt}, {Reinecke}, \&
  {Thielemann}}]{travaglio2004a}
{Travaglio}, C., {Hillebrandt}, W., {Reinecke}, M., \& {Thielemann}, F.-K.
  2004, \aap, 425, 1029

\bibitem[{{Umeda} {et~al.}(1999){Umeda}, {Nomoto}, {Kobayashi}, {Hachisu}, \&
  {Kato}}]{umeda1999a}
{Umeda}, H., {Nomoto}, K., {Kobayashi}, C., {Hachisu}, I., \& {Kato}, M. 1999,
  \apjl, 522, L43

\bibitem[{{Valenti} {et~al.}(2009){Valenti}, {Pastorello}, {Cappellaro},
  {Benetti}, {Mazzali}, {Manteca}, {Taubenberger}, {Elias-Rosa}, {Ferrando},
  {Harutyunyan}, {Hentunen}, {Nissinen}, {Pian}, {Turatto}, {Zampieri}, \&
  {Smartt}}]{valenti2009a}
{Valenti}, S., {Pastorello}, A., {Cappellaro}, E., {et~al.} 2009, \nat, 459,
  674

\bibitem[{Vennes {et~al.}(2017)Vennes, Nemeth, Kawka, Thorstensen, Khalack,
  Ferrario, \& Alper}]{vennes2017a}
Vennes, S., Nemeth, P., Kawka, A., {et~al.} 2017, Science, 357, 680

\bibitem[{{Wang} \& {Han}(2012)}]{wang2012b}
{Wang}, B. \& {Han}, Z. 2012, New Astronomy Review, 56, 122

\bibitem[{{Whelan} \& {Iben}(1973)}]{whelan1973a}
{Whelan}, J. \& {Iben}, I.~J. 1973, \apj, 186, 1007

\bibitem[{{White} {et~al.}(2015){White}, {Kasliwal}, {Nugent}, {Gal-Yam},
  {Howell}, {Sullivan}, {Goobar}, {Piro}, {Bloom}, {Kulkarni}, {Laher},
  {Masci}, {Ofek}, {Surace}, {Ben-Ami}, {Cao}, {Cenko}, {Hook}, {J{\"o}nsson},
  {Matheson}, {Sternberg}, {Quimby}, \& {Yaron}}]{white2015a}
{White}, C.~J., {Kasliwal}, M.~M., {Nugent}, P.~E., {et~al.} 2015, \apj, 799,
  52

\bibitem[{{Woosley} {et~al.}(1973){Woosley}, {Arnett}, \&
  {Clayton}}]{woosley1973a}
{Woosley}, S.~E., {Arnett}, W.~D., \& {Clayton}, D.~D. 1973, \apjs, 26, 231

\bibitem[{{Yamanaka} {et~al.}(2015){Yamanaka}, {Maeda}, {Kawabata}, {Tanaka},
  {Tominaga}, {Akitaya}, {Nagayama}, {Kuroda}, {Takahashi}, {Saito},
  {Yanagisawa}, {Fukui}, {Miyanoshita}, {Watanabe}, {Arai}, {Isogai},
  {Hattori}, {Hanayama}, {Itoh}, {Ui}, {Takaki}, {Ueno}, {Yoshida}, {Ali},
  {Essam}, {Ozaki}, {Nakao}, {Hamamoto}, {Nogami}, {Morokuma}, {Oasa},
  {Izumiura}, \& {Sekiguchi}}]{yamanaka2015a}
{Yamanaka}, M., {Maeda}, K., {Kawabata}, K.~S., {et~al.} 2015, \apj, 806, 191

\bibitem[{{Yao} {et~al.}(2019){Yao}, {Miller}, {Kulkarni}, {Bulla}, {Masci},
  {Goldstein}, {Goobar}, {Nugent}, {Dugas}, {Blagorodnova}, {Neill}, {Rigault},
  {Sollerman}, {Nordin}, {Bellm}, {Cenko}, {De}, {Dhawan}, {Feindt},
  {Fremling}, {Gatkine}, {Graham}, {Graham}, {Ho}, {Hung}, {Kasliwal},
  {Kupfer}, {Laher}, {Perley}, {Rusholme}, {Shupe}, {Soumagnac}, {Taggart},
  {Walters}, \& {Yan}}]{yao2019a}
{Yao}, Y., {Miller}, A.~A., {Kulkarni}, S.~R., {et~al.} 2019, \apj, 886, 152

\bibitem[{{Yaron} \& {Gal-Yam}(2012)}]{yaron2012a}
{Yaron}, O. \& {Gal-Yam}, A. 2012, \pasp, 124, 668

\bibitem[{{Yoon} \& {Langer}(2005)}]{yoon2005b}
{Yoon}, S. \& {Langer}, N. 2005, \aap, 435, 967

\bibitem[{Zeng {et~al.}(2020)Zeng, Liu, \& Han}]{zeng2020a}
Zeng, Y., Liu, Z.-W., \& Han, Z. 2020, The Astrophysical Journal, 898, 12

\bibitem[{Zhang {et~al.}(2019)Zhang, Fuller, Schwab, \& Foley}]{zhang2019a}
Zhang, M., Fuller, J., Schwab, J., \& Foley, R.~J. 2019, The Astrophysical
  Journal, 872, 29

\bibitem[{Zhou {et~al.}(2021)Zhou, Leung, Li, Nomoto, Vink, \&
  Chen}]{zhou2021a}
Zhou, P., Leung, S.-C., Li, Z., {et~al.} 2021, The Astrophysical Journal, 908,
  31

\bibitem[{{Zingale} {et~al.}(2009){Zingale}, {Almgren}, {Bell}, {Nonaka}, \&
  {Woosley}}]{zingale2009a}
{Zingale}, M., {Almgren}, A.~S., {Bell}, J.~B., {Nonaka}, A., \& {Woosley},
  S.~E. 2009, \apj, 704, 196

\bibitem[{{Zingale} {et~al.}(2011){Zingale}, {Nonaka}, {Almgren}, {Bell},
  {Malone}, \& {Woosley}}]{zingale2011a}
{Zingale}, M., {Nonaka}, A., {Almgren}, A.~S., {et~al.} 2011, \apj, 740, 8

\end{thebibliography}
